\pdfoutput=1
\documentclass[epjc3,twocolumn,a4paper,fleqn]{svjour3} 
\usepackage[fleqn]{amsmath}
\usepackage{graphicx}
\usepackage{epstopdf}
\usepackage{fix-cm}
\usepackage{txfonts}
\usepackage{lipsum}
\usepackage{mathtools}
\usepackage{cuted}
\usepackage{multirow}
\usepackage{comment}
\usepackage{empheq}
\usepackage[titletoc,title]{appendix}

\usepackage{listings}
\usepackage{lineno}
\usepackage{mdwlist}
\usepackage{pennames}
\usepackage[british]{babel}
\usepackage{color}
\usepackage{subfigure}
\usepackage{siunitx}
\usepackage{import}
\usepackage{amssymb}
\usepackage{epsfig}
\usepackage{cite}
\usepackage{hyperref}
\hypersetup{
colorlinks=true,
linkcolor=blue,
citecolor=blue,
urlcolor=blue,
}

\DeclareSIUnit \clight  {\textit{c}}
\RequirePackage[normalem]{ulem} 
\RequirePackage{color}\definecolor{RED}{rgb}{1,0,0}\definecolor{BLUE}{rgb}{0,0,1} 
\RequirePackage[normalem]{ulem} 
\RequirePackage{color}\definecolor{RED}{rgb}{1,0,0}\definecolor{BLUE}{rgb}{0,0,1} 

\DeclareGraphicsExtensions{.eps,.pdf,.jpeg,.jpg,.png}

\newcommand{\bea}{\begin{eqnarray}}
\newcommand{\eea}{\end{eqnarray}}
\newcommand{\be}{\begin{equation}}
\newcommand{\ee}{\end{equation}}

\newcommand*{\muonp}          {\ifmmode\mathrm{\muup^+}\else$\mathrm{\muup^+}$\fi}
\newcommand*{\muon}           {\ifmmode\mathrm{\muup}\else$\mathrm{\muup}$\fi}
\newcommand*{\tauon}          {\ifmmode\mathrm{\tauup}\else$\mathrm{\tauup}$\fi}

\newcommand*{\photon}         {\ifmmode{\gammaup}\else${\gammaup}$\fi}
\newcommand*{\positron}       {\ifmmode{\mathrm{e}^+}\else${\mathrm{e}^+}$\fi}
\newcommand*{\electron}       {\ifmmode{\mathrm{e}}\else${\mathrm{e}}$\fi}

\newcommand{\tg}{{\ifmmode t_{\gammaup_1\mathrm{e}^+}\else$t_{\gammaup_1\mathrm{e}^+}$\fi}}
\newcommand*{\tgg}{{\ifmmode t_{\gammaup\gammaup}\else$t_{\gammaup\gammaup}$\fi}}

\newcommand{\meg}{\ifmmode{\muup \to e \gammaup}\else$\mathrm{\muup \to e \gammaup}$\fi}
\newcommand{\megp}{\ifmmode{\muup^+ \to \mathrm{e}^+ \gammaup}\else$\mathrm{\muup^+ \to e^+ \gammaup}$\fi}
\newcommand{\michel}{\ifmmode{\muup^+ \to e^+ \nuup\bar{\nuup}}\else$\mathrm{\muup^+ \to e^+ \nuup\bar{\nuup}}$\fi}
\newcommand{\radiative}{\ifmmode{\muup^+ \to e^+\nuup\bar{\nuup}\gammaup} \else$\mathrm{\muup^+ \to e^+ \nuup\bar{\nuup}\gammaup}$\fi}
\newcommand{\conv}{\ifmmode{\muup^- \to e^-}\else$\mathrm{\muup^- \to e^-}$\fi}
\newcommand{\convN}{\ifmmode{\muup^-N \to e^-N}\else$\mathrm{\muup^-N \to e^-N}$\fi}
\newcommand{\mute}{\ifmmode{\muup \to 3e}\else $\mathrm{\muup \to 3e}$\fi}
\newcommand{\mutec}{\ifmmode{\muup^+ \to e^+e^+e^-}\else $\mathrm{\muup^+ \to e^+e^+e^-}$\fi}
\newcommand{\aif}{\ifmmode\mathrm{e}^+ \mathrm{e}^- \to \gammaup\gammaup \else$\mathrm{e}^+ \mathrm{e}^- \to \gammaup \gammaup$\fi}
\newcommand{\teg}{\ifmmode{\tauup \to e \gammaup} \else$\mathrm{\tauup \to e \gammaup}$\fi}
\newcommand{\tmg}{\ifmmode{\tauup \to \gammaup} \else$\mathrm{\tauup \to \muup \gammaup}$\fi}
\newcommand{\tmueg}{\ifmmode{\mathrm\tauup \to \ell \gammaup}\else$\mathrm{\tauup \to \ell \gammaup}$\fi}
\newcommand{\tautl}{\ifmmode{\mathrm\tauup \to 3\ell} \else$\mathrm\tauup \to 3\ell$\fi}

\newcommand*{\mathtentative}{}
\def\mathtentative#1#{\mathcoloraux{#1}}
\newcommand*{\mathcoloraux}[3]{%
  \protect\leavevmode
  \begingroup
    \color#1{#2}#3%
  \endgroup
}

\journalname{Eur. Phys. J. C} 

\begin{document}

\title{Performances of a new generation tracking detector: the MEG~II cylindrical drift chamber}

\newcommand*{\INFNPi}{INFN Sezione di Pisa$^{a}$; Dipartimento di Fisica$^{b}$ dell'Universit\`a, Largo B.~Pontecorvo~3, 56127 Pisa, Italy}
\newcommand*{\INFNGe}{INFN Sezione di Genova$^{a}$; Dipartimento di Fisica$^{b}$ dell'Universit\`a, Via Dodecaneso 33, 16146 Genova, Italy}
\newcommand*{\INFNPv}{INFN Sezione di Pavia$^{a}$; Dipartimento di Fisica$^{b}$ dell'Universit\`a, Via Bassi 6, 27100 Pavia, Italy}
\newcommand*{\INFNRm}{INFN Sezione di Roma$^{a}$; Dipartimento di Fisica$^{b}$ dell'Universit\`a ``Sapienza'', Piazzale A.~Moro, 00185 Roma, Italy}
\newcommand*{\INFNNa}{INFN Sezione di Napoli, Via Cintia, 80126 Napoli, Italy}
\newcommand*{\INFNLe}{INFN Sezione di Lecce$^{a}$; Dipartimento di Matematica e Fisica$^{b}$ dell'Universit\`a del Salento, Via per Arnesano, 73100 Lecce, Italy}
\newcommand*{\ICEPP} {ICEPP, The University of Tokyo, 7-3-1 Hongo, Bunkyo-ku, Tokyo 113-0033, Japan }
\newcommand*{\Kobe} {Kobe University, 1-1 Rokkodai-cho, Nada-ku, Kobe, Hyogo 657-8501, Japan}
\newcommand*{\UCI}   {University of California, Irvine, CA 92697, USA}
\newcommand*{\KEK}   {KEK, High Energy Accelerator Research Organization, 1-1 Oho, Tsukuba, Ibaraki 305-0801, Japan}
\newcommand*{\PSI}   {Paul Scherrer Institut PSI, 5232 Villigen, Switzerland}
\newcommand*{\Waseda}{Research Institute for Science and Engineering, Waseda~University, 3-4-1 Okubo, Shinjuku-ku, Tokyo 169-8555, Japan}
\newcommand*{\BINP}  {Budker Institute of Nuclear Physics of Siberian Branch of Russian Academy of Sciences, 630090 Novosibirsk, Russia}
\newcommand*{\JINR}  {Joint Institute for Nuclear Research, 141980 Dubna, Russia}
\newcommand*{\ETHZ}  {Institute for Particle Physics and Astrophysics, ETH Z\" urich, 
Otto-Stern-Weg 5, 8093 Z\" urich, Switzerland}
\newcommand*{\NOVS}  {Novosibirsk State University, 630090 Novosibirsk, Russia}
\newcommand*{\NOVST} {Novosibirsk State Technical University, 630092 Novosibirsk, Russia}
\newcommand*{\ScuolaPi}{Scuola Normale Superiore, Piazza dei Cavalieri 7, 56126 Pisa, Italy}
\newcommand*{\INFNLNF}{\textit{Present Address}: INFN, Laboratori Nazionali di Frascati, Via 
E. Fermi, 40-00044 Frascati, Rome, Italy}
\newcommand*{\Liverpool}{Oliver Lodge Laboratory, University of Liverpool, Liverpool, L69 7ZE, United Kingdom}

\date{Received: date / Accepted: date}

\author{
        A.~M.~Baldini~\thanksref{addr1}$^{a}$ \and
        H.~Benmansour~\thanksref{addr1}$^{ab}$ \and
        G.~Boca~\thanksref{addr4}$^{ab}$ \and   
        G.~Cavoto~\thanksref{addr5}$^{ab}$ \and
        F.~Cei~\thanksref{addr1}$^{ab}$\thanksref{e1} \and
        M.~Chiappini~\thanksref{addr1}$^{ab}$ \and
        G.~Chiarello~\thanksref{addr6}$^{a}$\thanksref{e2} \and
        A.~Corvaglia~\thanksref{addr6}$^{a}$ \and
        F.~Cuna~\thanksref{addr6}$^{ab}$\thanksref{e3} \and
        M.~Francesconi~\thanksref{addr17} \and 
        L.~Galli~\thanksref{addr1}$^{a}$ \and
        F.~Grancagnolo~\thanksref{addr6}$^{a}$ \and
        E.~G.~Grandoni~\thanksref{addr1}$^{ab}$ \and 
        M.~Grassi~\thanksref{addr1}$^{a}$ \and 
        M.~Hildebrandt~\thanksref{addr2} \and
        F.~Ignatov~\thanksref{addr15} \and
        M.~Meucci~\thanksref{addr5}$^{ab}$ \and   
        W.~Molzon~\thanksref{addr11} \and
        D.~Nicol\`o~\thanksref{addr1}$^{ab}$ \and
        A.~Oya~\thanksref{addr10} \and
        D.~Palo~\thanksref{addr11} \and
        M.~Panareo~\thanksref{addr6}$^{ab}$ \and
        A.~Papa~\thanksref{addr1}$^{ab}$\thanksref{addr2} \and
        F.~Raffaelli~\thanksref{addr1}$^{a}$ \and
        F.~Renga~\thanksref{addr5}$^{a}$ \and
        G.~Signorelli~\thanksref{addr1}$^{a}$ \and
        G.F.~Tassielli~\thanksref{addr6}$^{ab}$\thanksref{e4} \and
        Y.~Uchiyama~\thanksref{addr10,addr14} \and
        A.~Venturini~\thanksref{addr1}$^{ab}$ \and        
        B.~Vitali~\thanksref{addr1}$^{a,}$\thanksref{addr5}$^{b}$ \and
        C.~Voena~\thanksref{addr5}$^{ab}$ \and   
}

\institute{  \INFNPi \label{addr1}
           \and
             \INFNPv \label{addr4}
           \and
             \INFNRm \label{addr5}
           \and
             \INFNLe \label{addr6} 
           \and
             \INFNNa \label{addr17}
            \and
             \PSI \label{addr2}
            \and
             \Liverpool  \label{addr15}
           \and
             \UCI    \label{addr11}
           \and
             \ICEPP \label{addr10}
           \and
             \Kobe    \label{addr14}
}

\thankstext[*]{e1}{Corresponding author: fabrizio.cei@pi.infn.it} 
\thankstext[**]{e2}{Presently at Department of Engineering, University of Palermo, Viale delle Scienze,
Building 9, 90128 Palermo, Italy} 
\thankstext[***]{e3}{Presently at INFN Sezione di Bari, Via Giovanni Amendola, 173, 70126, Bari, Italy} 
\thankstext[****]{e4}{Presently at Dipartimento di Medicina e Chirurgia, Università LUM “Giuseppe
Degennaro”, 70010, Casamassima, Bari, Italy} 

\maketitle 

\begin{abstract}
The cylindrical drift chamber is the most innovative part of the MEG~II detector, the upgraded version of the MEG experiment. The MEG~II chamber differs from the MEG one because it is a single volume cylindrical structure, instead of a segmented one, chosen to improve its resolutions and efficiency in detecting low energy positrons from muon decays at rest. In this paper, we show the characteristics and performances of this fundamental part of the MEG~II apparatus and we discuss the impact of its higher resolution and efficiency on the sensitivity of the MEG~II experiment. Because of its innovative structure and high quality resolution and efficiency the MEG~II cylindrical drift chamber will be a cornerstone in the development of an ideal tracking detector for future positron-electron collider machines.   
\end{abstract}

\tableofcontents 

\section{Introduction}
\label{sec:intro}
The goal of the MEG~II experiment~\cite{baldini_2018} is to improve by one order of magnitude the sensitivity of the MEG experiment~\cite{megdet}, which established the world best upper limit on the branching ratio of the Standard Model (almost) forbidden decay 
$\mu^{+} \rightarrow {\rm e^{+}} \gamma$~\cite{Baldini_2016}:
\begin{equation}
{\rm BR} \left( \mu^{+} \rightarrow {\rm e^{+}} \gamma \right) < 4.2 \times 10^{-13}  
\label{eq:MEGlimit}   
\end{equation}    
at $90~\%$ confidence level. 

The upper limit and the sensitivity of the experiment depend on the number of background events which is determined by the muon stopping rate and by the resolutions on the positron and photon kinematic variables~\cite{kuno_2001} according to the formula:
\begin{equation}
N_\mathrm{bkg} \propto R_{\mu}^2 \Delta T_{{\rm e}\gamma} \Delta E_{\rm e} \Delta E_{\gamma}^2 \Delta \Theta_{{\rm e}\gamma}^2  
\label{eq:nBKG}   
\end{equation}    
In this expression $N_\mathrm{bkg}$ is the number of accidental background events, $R_{\mu}$ is the muon stopping rate and $\Delta T_{{\rm e}\gamma}$, $\Delta E_{\rm e}$, $\Delta E_{\gamma}$ and $\Delta \Theta_{{\rm e}\gamma}$ are the positron--photon relative timing, positron energy, photon energy and positron--photon relative angle resolutions, respectively. 

Therefore, a significant improvement in the experimental sensitivity requires substantial modifications and refurbishments of the old detector, extensively discussed in~\cite{baldini_2018}. Particularly, the new Cylindrical Drift CHamber (from now on: CDCH) is an innovative and challenging detector, based on a geometrical configuration completely different from the old one. 

In the MEG experiment the positron tracking detector was formed by sixteen independent drift chambers each at fixed azimuth and with a collective $\approx\pi$ coverage on the azimuthal angle around the beam line axis. This configuration ensured a minimum number of track points (\num{>10}) usually sufficient for a good quality track reconstruction, but strong efficiency loss and resolution degradation were caused by the presence along the positron  trajectories of mechanical structures and electronic boards. 

\begin{figure*}[htb]
\centering  \includegraphics[width=0.99\textwidth,angle=0] {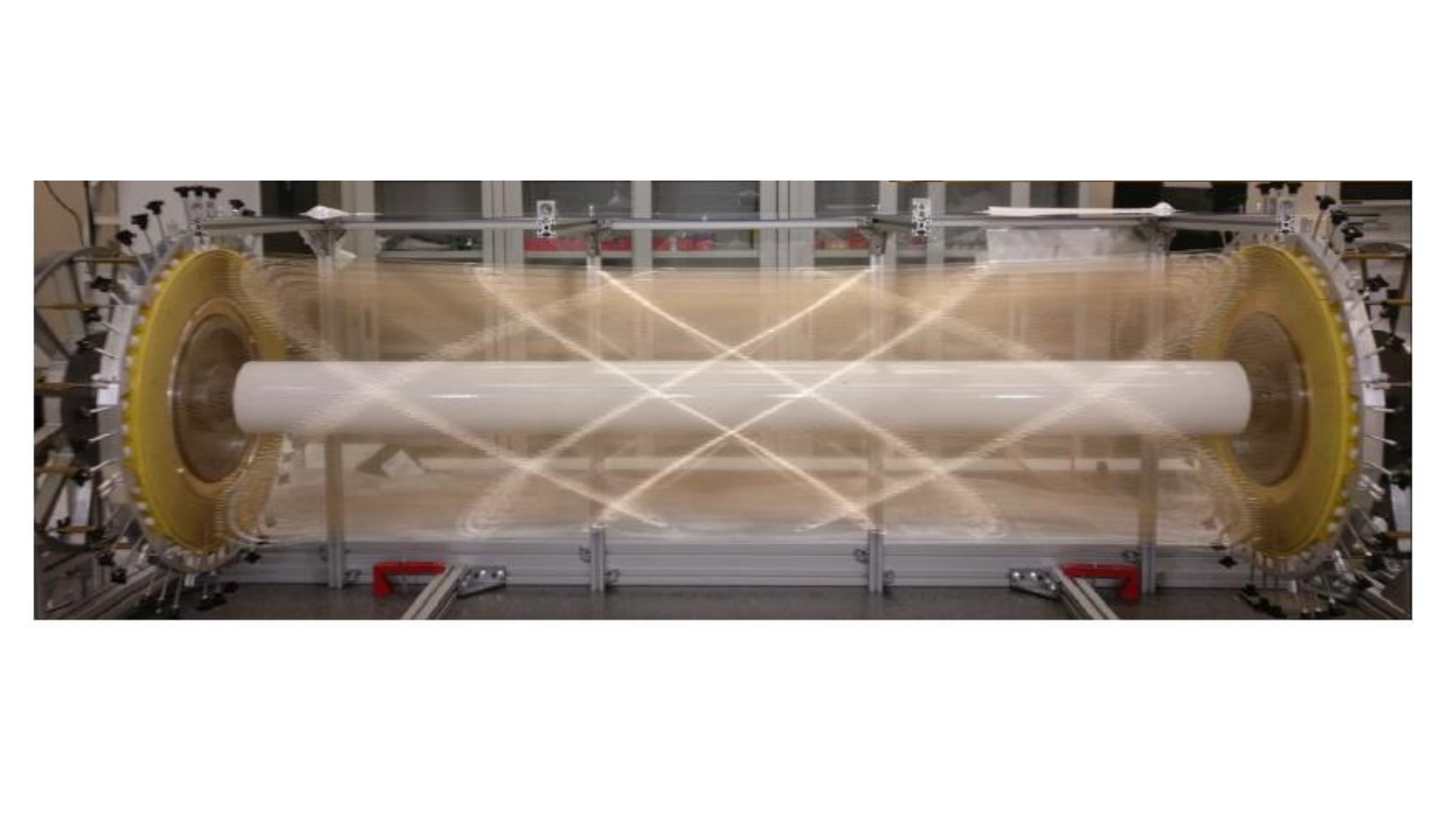}
  \caption{Picture of MEG~II cylindrical drift chamber CDCH.}
 \label{fig:cyldch}
\end{figure*}

The MEG~II chamber, on the opposite, is a unique volume detector, formed by a \SI{1.93}{\meter} length and \SI{35}{\centi\meter} external diameter cylinder, filled with a $90:10$ helium-isobutane gas mixture; small percentages of oxygen and isopropyl alcohol have been added to avoid current spikes during the transient phases of the detector start-up (details on the gas mixture are provided in Section~\ref{sec:Gas_Mixture}). A picture of the CDCH is shown in Fig.~\ref{fig:cyldch}.
The volume is equipped with a dense array of gold plated tungsten sense wires and silver plated aluminum cathode wires, arranged in a two views (conventionally called \lq\lq U\rq\rq~and \lq\lq V\rq\rq) stereo configuration. The two views are separated on average by \SI{\approx 7.5}{\degree} and this arrangement allows a high resolution three-dimensional reconstruction of the coordinates of the positron track points. 
The total number of sense wires is \num{1728}, distributed in nine concentric layers, and that of cathode wires is $\approx$ \num{10000}; \num{7680} cathode wires have a \SI{40}{\micro\meter} diameter, the remaining ones and the guard wires (needed to define the electric field at the boundaries of the active volume) have a \SI{50}{\micro\meter} diameter. The diameter of the gold plated tungsten wires is \SI{20}{\micro\meter}.
Only a part of the sense wires (\num{1152} wires in the 2021 run and \num{1216} in the following runs) are equipped with readout electronics since the remaining ones are located in chamber regions outside the acceptance for positron tracks coming from the target. 
Cathode wires are placed around the sense wires to form almost squared
drift cells, with side's dimensions ranging from \SIrange{5.8}{7.5}{\mm} at center and from \SIrange{6.7}{8.7}{\mm} at the end-plates.   

The external mechanical structure is formed by two wheel shaped end caps, where the wires are anchored, and a carbon fiber structure which encloses the whole active volume. This arrangement ensures the rigidity of the structure at a level of few microns. In the inner part the chamber is closed by an aluminated mylar cylindrical foil of \SI{20}{\micro\meter} thickness. Another aluminum foil of \SI{100}{\micro\meter} thickness is placed in the inner surface of the carbon fiber to provide the ground connection to the guard wires. 

A superconducting magnet COBRA (COstant Bending RAdius) produces a longitudinally varying magnetic field in the whole volume of the CDCH, with its maximum intensity of \SI{1.26}{\tesla} at the chamber center. The field is arranged to ensure a positron bending radius almost independent of the emission angle from the target (hence the name COBRA) and an efficient extraction of positrons with small longitudinal momentum from the chamber volume. A positron emitted from a muon decay at rest, whose maximum energy is \SI{52.83}{\mega\eV}, traverses on average a total amount of material (gas and wires) corresponding to \num{1.6e-3} radiation lengths $X_{0}$ for each chamber crossing (\lq\lq turn\rq\rq). 

The other components of the MEG~II experiment are a liquid xenon photon detector (from now on: LXe) 
to measure energy, arrival time and interaction point of photons emitted in $\mu^{+} \rightarrow {\rm e^{+}} \gamma$ decay, a pixelated scintillation timing counter (from now on: pTC)
to measure the positron timing, and an auxiliary detector, the Radiative Decay Counter, 
useful to reduce the high energy photon background. 
A complete description of the MEG~II detector is given elsewhere \cite{megiidetector}.  

Here, we discuss the performances reached by the CDCH in the resolution on positron trajectory reconstruction and in the efficiency on positron tracking. We show that the CDCH momentum and angular resolutions and efficiency are in good agreement with the Monte Carlo predictions and represent a substantial improvement with respect to what was obtained in MEG. 
\section{Reference frame and positron kinematic variables}
\label{sec:refframe}
\begin{figure*}[htb]
\centering  \includegraphics[width=0.99\textwidth,angle=0] {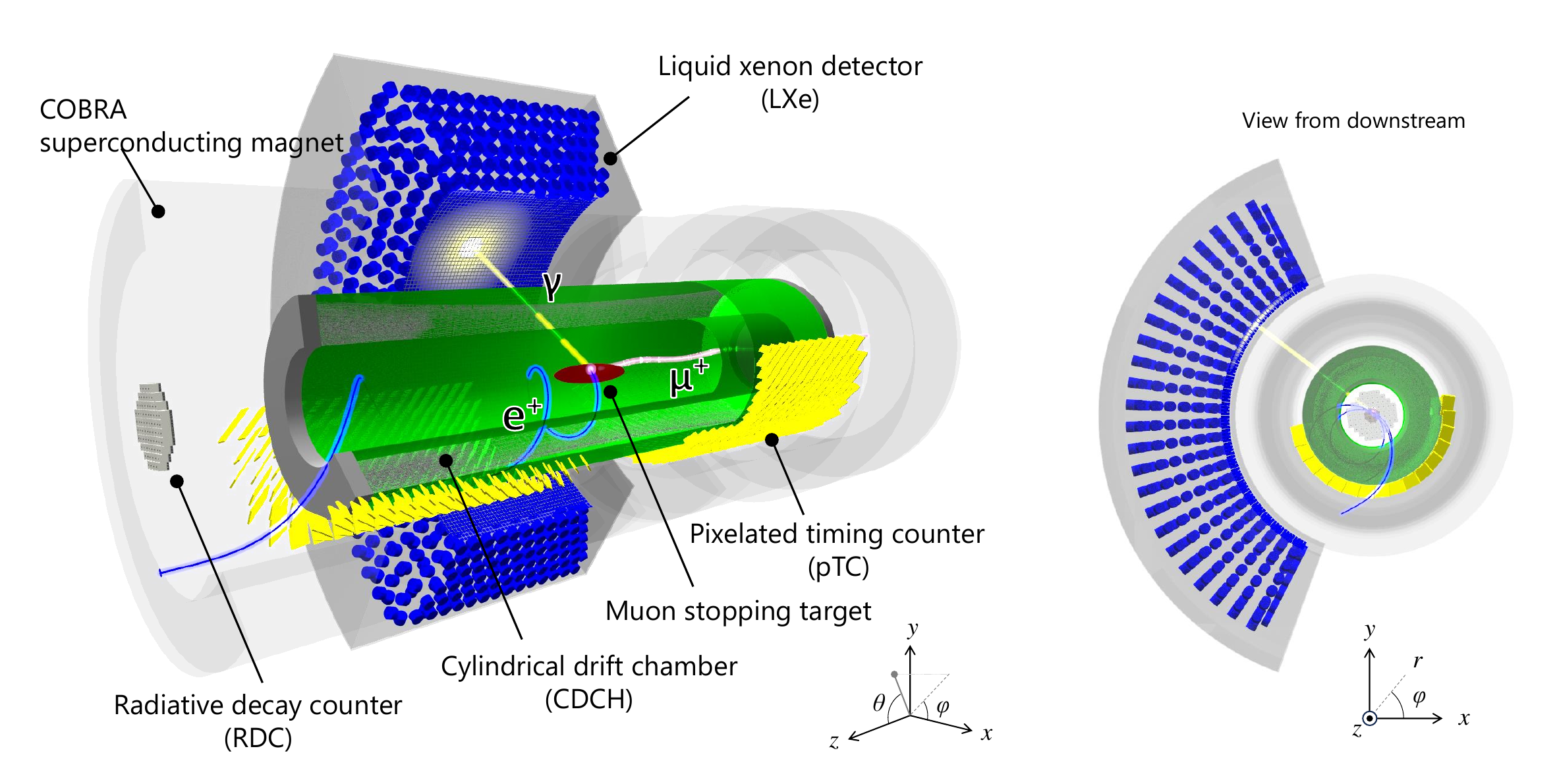}
  \caption{Schematic drawing of the MEG~II detector showing the reference frame and a simulated $\mu^{+} \rightarrow {\rm e^{+}} \gamma$ event.}
 \label{fig:meg2}
\end{figure*}

The MEG~II reference frame is defined starting from the center of the COBRA magnet, used as origin of the coordinate axes. The beam direction is the $z$ axis, the $x$ axis is chosen to have the LXe detector in $x<0$ half-space and the $y$ axis is directed upwards. The half-space with $z>0$ is called downstream and that with $z<0$ is called upstream. Physical quantities (like timing or charges) collected on electronics at the two chamber ends are labelled with \lq\lq 0\rq\rq~on the upstream side and with \lq\lq 1\rq\rq~on the downstream side. A cylindrical and a spherical coordinate system are frequently used to define the positron kinematics: these two systems have the $z$ axis in common with the Cartesian system discussed above, while the $\left(r, \phi\right)$ (for cylindrical system) and $\left( \theta, \phi \right)$ (for spherical system) pairs of coordinates are defined as usually with respect to the $z$ axis and in the plane orthogonal to it. The $\phi = 0$ direction corresponds to the positive $x$ axis. A schematic drawing of the MEG~II experiment showing the various components of the detector and the reference frame, together with a simulated $\mu^{+} \rightarrow {\rm e^{+}} \gamma$ event, is shown in Fig.~\ref{fig:meg2}.
%

The positron trajectory parameters are the momentum vector components, the timing and the coordinates at the positron production point. The momentum vector is described in polar form using its module $P_{\rm e}$ and the polar and azimuthal angles $\theta_{\rm e}$ and $\phi_{\rm e}$ at the target, where the positron track begins; these variables are reconstructed by a tracking algorithm, together with the intersection point of the track with the target (used as positron production point), described by $y_{\rm e}$ and $z_{\rm e}$ (the $x_{\rm e}$ coordinate is determined by $z_{\rm e}$ and by the target inclination). The positron timing is measured by the pTC detector and the timing at the target is computed taking into account the trajectory length from the target to the pTC. The positron information is matched with that of the photon provided by the LXe detector in order to identify possible $\mu^{+} \rightarrow {\rm e^{+}} \gamma$ candidates. Using the experimentally measured positron vertex as hypothetical photon production point, one can check if the energies and relative timing and directions associated to any positron--photon pair are compatible with a $\mu^{+} \rightarrow {\rm e^{+}} \gamma$ decay  originated in the muon stopping target.  
\section{Gas Mixture}
\label{sec:Gas_Mixture}
The CDCH uses a helium based gas mixture, continuously flowed through the detector by a dedicated gas system~\cite{Baldini:2018ing}. Because of its large radiation length ($X_0 \sim \SI{5300}{\meter}$ at STP), the choice of the helium ensures a small contribution in terms of Coulomb multiple scattering, a very important feature in low momentum measurements. A small amount (\SI{10}{\percent}) of isobutane is required as a quencher to avoid self-sustained discharge. Such a percentage is sufficient as it raises the number of primary ionization pairs to \SI{\sim 12.3}{\per\cm}~\cite{Adinolfi:2002} for a minimum ionization particle (from now on: m.i.p.) though lowers the mixture radiation length to $X_0 \sim \SI{1300}{\meter}$. On the other hand, the use of an organic quencher involves specific problems relating to exposure to high radiation fluxes, since the recombination of dissociated organic molecules leads to the formation of solid or liquid polymers which, by accumulating on the anodes and cathodes, contribute to the aging of the detector.

The fairly constant drift velocity in helium based gas mixtures ensures a linear time--distance relation, up to very close distance to the sense wire. Moreover, the high helium ionization potential of \SI{24.6}{\eV} is such that a crossing particle produces only a small number of primary electron--ion pairs in helium based gas mixture. The average number of total clusters per cell is about \num{13} for a m.i.p. and the average number of ionization electrons per cluster is \num{1.6}. In combination with the small size of the drift cells, it enhances the contribution to the spatial resolutions coming from the statistical fluctuation of the primary ionization along the track, if only the first arriving electrons are timed. An improvement can be obtained using the cluster timing technique, i.e.\ by timing all the arriving ionization clusters to the anode wire and so reconstructing their distribution along the ionization track~\cite{Cascella:2014}.


To compensate for the variable dimensions of the CDCH drift cells, the HV bias of the anode wires is made variable by \SI{10}{\V}/layer, ranging from \SI{1400}{\V} for the cells of the innermost layer, up to \SI{1480}{\V} for the cells of the outermost layer. In this way, a uniform gas gain of about \num{2.5e5} is obtained, as measured by comparing the signal amplitude distributions in data and Monte Carlo simulations. The amplification in the avalanche process has large fluctuations following a Polya distribution, having a standard deviation comparable to the mean value~\cite{BlumRolandi}.

\begin{figure*}[htb]
\centering 
\includegraphics[width=0.95\textwidth]{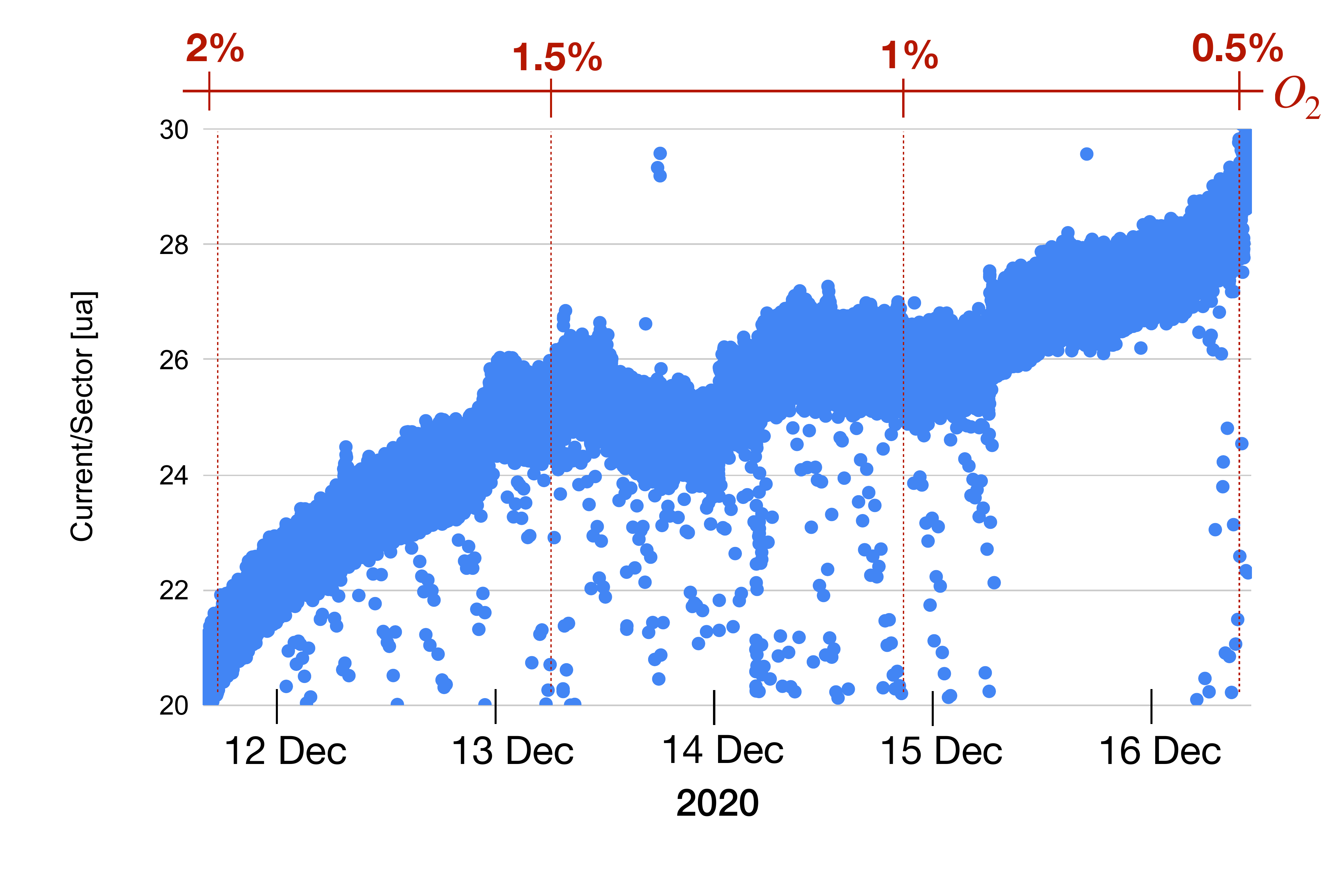}
\caption{\label{CurrentOxygen} Current in a sector as the oxygen percentage is lowered from $\SI{2}{\percent}\rightarrow \SI{1.5}{\percent} \rightarrow \SI{1}{\percent} \rightarrow \SI{0.5}{\percent}$. The abrupt reductions of the current 
were due to instabilities of the PSI particle beam.}
\end{figure*}

During nominal operations an abrupt increase in current up to \SI{400}{\micro\A} was observed, while the normal level is around
\SIrange[range-phrase=--,range-units = single]{10}{20}{\micro\A}. A deep investigation revealed the formation of corona-like discharges in correspondence of some regions along some wires. Several additives to the gas mixture were tried to recover the normal detector operation. The reduction of the high currents was achieved with an oxygen level up to \SI{2}{\percent}, then gradually lowered to avoid electron attachment effects. The effect of oxygen on the current per sector is shown in Fig.~\ref{CurrentOxygen}; the lowering of the oxygen content is indicated by 
the red scale and vertical lines on the top of the figure; the current is lowered from oxygen capture and some loss of gas gain. In addition, a small percentage of isopropyl alcohol has been added to keep the current level stable from the beginning of the stabilizing procedure.
The CDCH is now operated at the HV working point in stable conditions at full MEG~II beam intensity with the standard gas mixture + isopropyl alcohol (\SI{1.5}{\percent}) + O$_2$ (\SI{0.5}{\percent}). 

\section{Electric parameters of the drift cells}
\label{sec:Drift cells}
Each drift cell is approximately squared, \SI{6.7}{\milli\m} (at the lowest radius) to \SI{8.7}{\milli\m} (at the highest radius) wide, with a \SI{20}{\micro\m} diameter gold plated tungsten sense wire, surrounded by \SI{40}{\micro\m} and \SI{50}{\micro\m} diameter silver plated aluminium field wires, in a ratio of 5:1. Since the mean active length of the wires is \SI{1.93}{\m}, the mean distributed resistance is \SI{140}{\ohm}/$\si{\m}$, the mean distributed inductance is \SI{1.2}{\micro\henry}/$\si{\m}$ and the mean distributed capacitance is \SI{9.4}{\pico\farad}/$\si{\m}$; the distributed conductance is negligible. The characteristic impedance of the drift cell depends on the frequency, since the drift cell behaves as a lossy coaxial transmission line, with a  significant resistance due to the wires. Nevertheless its variation is less then \SI{10}{\percent} from the mean value of \SI{354}{\ohm}, for frequencies higher than \SI{200}{\mega\hertz}. 

\begin{figure}[htb]
\centering 
\includegraphics[width=0.48\textwidth]{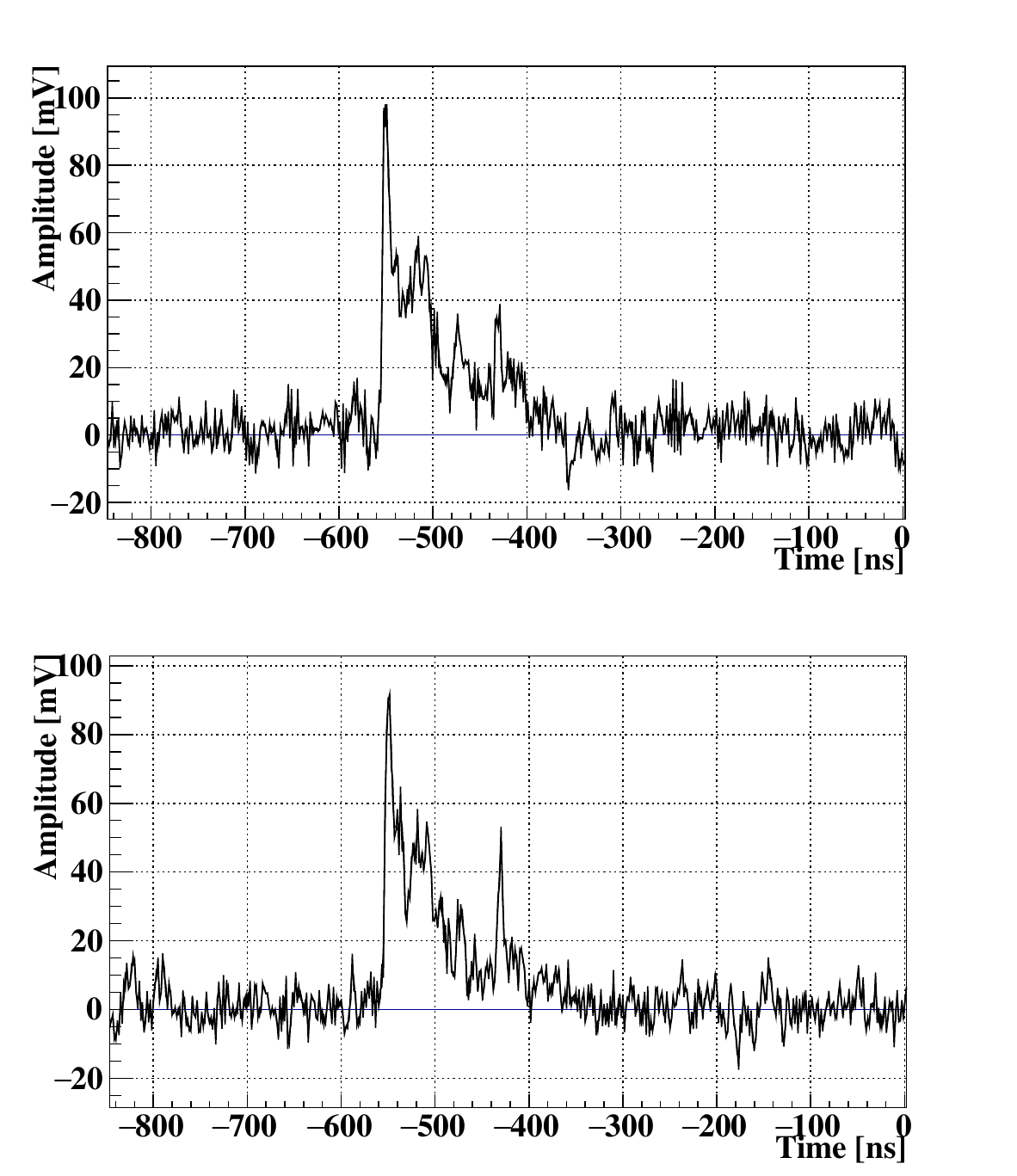}
\caption{\label{pulse} Signal measured at both ends of a drift cell.}
\end{figure}

The typical signal waveform is a pulses train (as shown in Fig.~\ref{pulse}): the time separation between different pulses goes from few nanoseconds to a few tens of nanoseconds. Main signal information is contained within a bandwidth less than \SI{1}{\giga\hertz}.
\section{CDCH front end electronics}
\label{sec:FE}
A two stages 8-channel front end amplifier based on commercial active components has been developed to readout the signals. The amplifier is characterized by a high linearity, low distortion, and a bandwidth adequate to the expected spectral density of the signal. The same board is used for the distribution of the high voltage bias to the anode wires~\cite{Panareo:2020}.
\begin{figure}[htb]
\centering 
\includegraphics[width=0.48\textwidth]{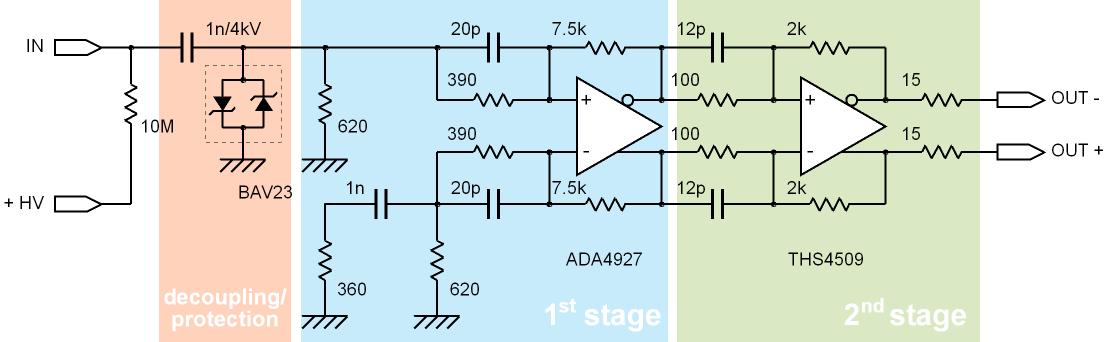}
\caption{\label{schematic} The schematic of the front end electronics.}
\end{figure}

The schematic of the amplifier is shown in Fig.~\ref{schematic}. The input network provides HV decoupling, voltage spike protection and matching to the mean characteristic impedance of the drift cell. The first gain stage is based on Analog Device\textquotesingle s low noise, ultra-low distortion, high speed, current feedback differential amplifier ADA4927. The second gain stage uses the Texas Instruments\textquotesingle~ wide-band, very low noise, fully differential operational amplifier THS4509; this stage is used as output driver as well. The differential output of the amplifier is connected to the digitizing unit (see later) through a multiwire, low attenuation, custom made \SI{5}{\m} long cable, designed to have a stable, flat frequency response. This cable is also used for powering the front-end board.

In order to balance the attenuation of the output cable, a pre-emphasis on both gain stages of the front end amplifier has been implemented. The pre-emphasis introduces a high frequency peak that compensates the output cable losses, resulting in a total bandwidth larger than \SI{500}{\mega\hertz}. The gain at middle bandwidth is about $\sim$\SI{30}{\decibel} on \SI{120}{\ohm} load. The average non-linearity is less than $0.1~\%$ for input short pulses (rise time on the order of \SI{1}{\nano\s}) in the range of amplitudes \SIrange[range-phrase=--,range-units = single]{15}{75}{\milli\volt}. The noise level is less than \SI{2}{\milli\volt} (RMS), after the output cable, on \SI{120}{\ohm} load. The cross talk between adjacent channels of the front end board is \SI{\sim1}{\percent}, that to the next channel is negligible (\SI{< 0.5}{\percent}).

The amplified differential signals are successively digitized by the WaveDREAM board, one of the board of the WaveDAQ system~\cite{Galli:2019,francesconi2023wavedaq} at a programmable sampling speed of \SI{1.2}{GSPS} with an analogue bandwidth of about \SI{1}{\giga\hertz}.

The current consumption for each channel of the front end board is \SI{60}{\milli\ampere} at a voltage supply of $\pm$\SI{2.5}{\volt}, corresponding to a total power dissipation per endcap of approximately \SI{300}{\watt}. Therefore a dedicated cooling system based on a \SI{1}{\kilo\watt} chiller and a cold water distribution system with piping embedded in the front end board holders is used. 
Moreover, a system of perforated pipes, glued to the external faces of the end-plates, flushing dry air
is used to avoid water condensation and dangerous temperature gradients.

\section{Waveform processing}
\label{sec:waveform}

The analogue signal from each wire end is digitized at \SI{1.2}{GSPS} for \num{1024} points by the DRS4 chip~\cite{ritt_2004_nim} on the WaveDREAM board when a trigger signal is issued, and the digital waveform is collected to a DAQ machine. To reduce the data size, the digital waveforms are averaged and quantized every two points (rebinning) for \SI{16}{\bit} at \SI{0.1}{\mV} steps when written to disk. Therefore, \num{2304} channels (or \num{2432} channels from 2022 run) of digital waveforms with \SI{850}{\ns} depth at an effective sampling speed of \SI{0.6}{GSPS} are saved and analyzed for each event.

When a charged particle (likely a positron) passes through a drift cell, it ionises the gas and creates several ionization clusters.
A \lq\lq hit\rq\rq~is defined as the group of such ionization clusters by a single passage of the particle in a cell. Different clusters in a hit go along different drift lines, resulting in stretching in time. Therefore, the signal of a hit usually consists with discrete pulses spanning up to \SI{400}{\ns}. 
The main goal of the waveform processing is to detect the pulse signals from hits, especially the earliest arrival pulse of each hit.  
The hit rate per cell can be \SI{1.2}{\mega\hertz} at the innermost wires for a stopping muon intensity of \SI{5e7}{\per\second}. Under the high intensity beam, multiple hits can overlap, making the detection of the first cluster difficult.

\begin{figure}[htb]
\centering 
\includegraphics[width=0.5\textwidth]{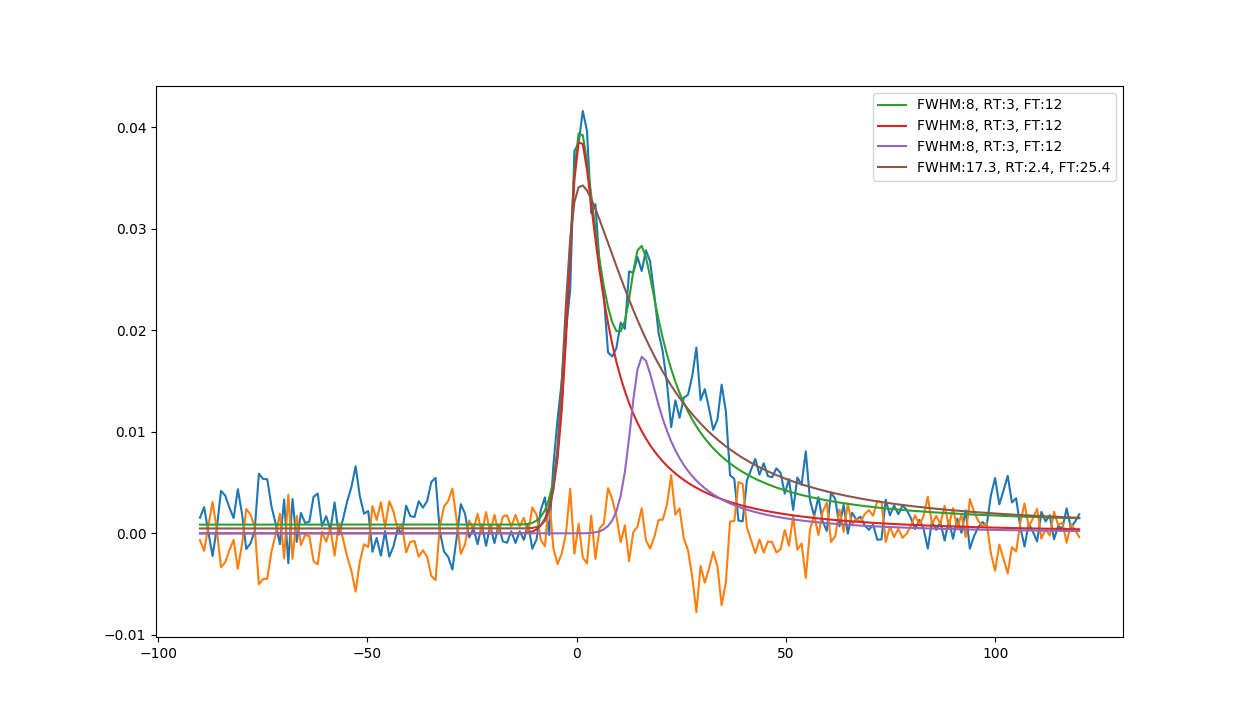}
\caption{An example hit with a two-pulse fit (green),  the two individual pulses in the two-pulse fit (red, purple), the one-pulse fit (brown), and the residuals to the two-pulse fit (orange). RT is the 20\%--80\% rise time and FT is the 80\%--20\% fall time; both are listed in nanoseconds.  \label{fig:ExamplePulseFit} }
\end{figure}

\begin{figure}[htb]
\centering 
\includegraphics[width=0.48\textwidth]{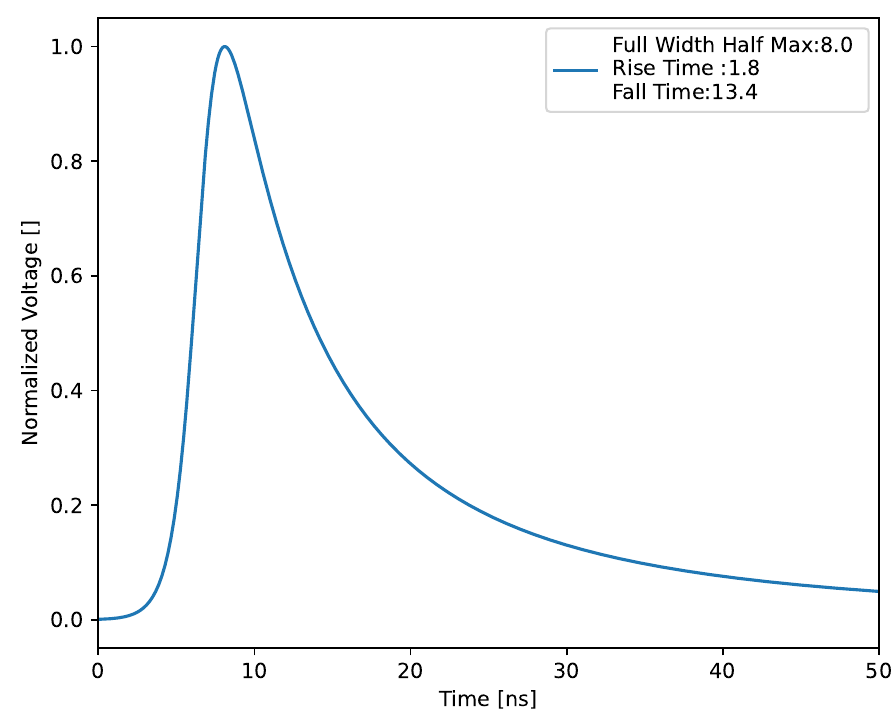} 
\caption{The estimated pulse shape of a single ionization cluster based on many fits to narrow waveform signals. RT is the 20\%--80\% rise time and FT is the 80\%--20\% fall time; both are listed in nanoseconds.\label{fig:PulseShape} }
\end{figure}

The digitized signal pulse shapes of single ionization clusters were studied to aid in optimizing the signal detection. Instances of narrow pulses were fitted to estimate the signal shape; an example is shown in Fig.~\ref{fig:ExamplePulseFit}. Here, the figure contains a fit to two pulses with the same shape; the fit results in flat residuals.  The average shape of the pulse is shown in Fig.~\ref{fig:PulseShape}. The average shape has a very short rise time (\SI{<2}{\ns}) and loses nearly all of its amplitude after 20 ns.

On each side of the CDCH (US and DS), the signals coming from one end of \num{16} adjacent wires are processed with two Front-End (FE) boards (\num{8} wire ends per FE board). The signals from each FE board are transmitted through a bundle of cables and digitized by one DRS4 chip. Each WaveDREAM has two DRS4 chips and digitizes \num{16} signals. Thus, two FE boards are connected to a single WaveDREAM board.  A coherent, low frequency noise was observed over the \num{16} wires connected to the same pair (one on US and one on DS end) of WaveDREAM boards. This noise was studied in detail and several noise sources were identified and removed: we upgraded all the unshielded Ethernet in the area cables to shielded versions and replaced the DCDC conversion unit in the electronics crate with a low-noise emission model. We also investigated other possible noise sources in the apparatus, such as circulation pumps and slow control nodes, none of them was successfully identified. At the same time the PiE5 experimental hall is an intrinsically noisy environment with many possible sources that we cannot control and even test; as a result a residual noise is still present and probably originates on a single side of the CDCH by the  WaveDREAM board on this side which picks up an external disturbance; from that board the noise propagates also to the WaveDREAM board on the other side of the wires. 

\begin{figure}[htb]
\centering 
\includegraphics[width=0.48\textwidth]{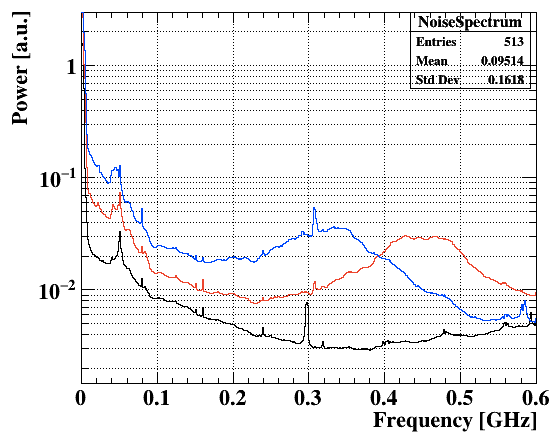}
\caption{\label{fig:FFTElectronicGain} Power spectra of pedestal data with varying the pre-amplifier gain. The gains $\times1$,$\times2$,$\times4$ are shown in black, red, and blue, respectively. Here, the standard rebinning by a factor of 2 was not yet applied.}
\end{figure}

This noise was investigated by running with different pre-amplifier gains. As expected, the signal amplitude scaled with the pre-amplifier gain. The RMS of the noise was measured using pedestal runs without the particle beam; the noise scaled at slightly less than the electronic gain. Thus the highest gain ($\times4$) had the highest signal to noise ratio. The power spectra with a discrete Fourier transform of pedestal data for $\times1$,$\times2$,$\times4$ gains are shown in Fig.~\ref{fig:FFTElectronicGain}.

\begin{figure}[htb]
\centering 
\includegraphics[width=0.48\textwidth]{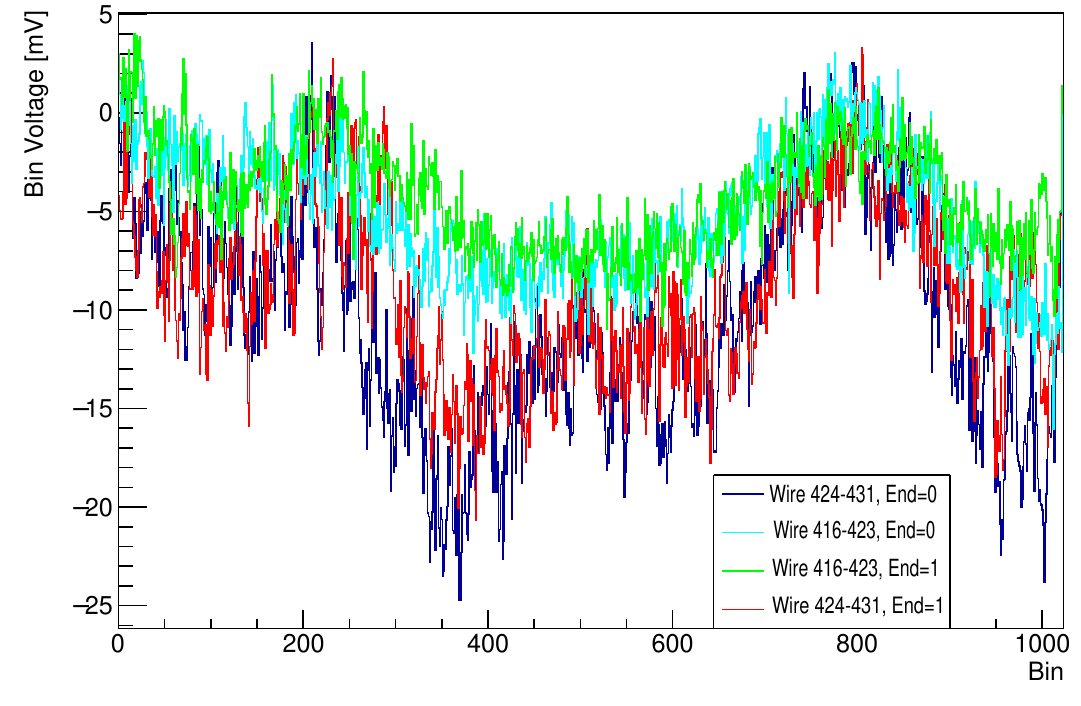}
\caption{\label{fig:wfmpednofilter} Example of waveform pedestals without any filter. Wires 416-423 and 424-431 have a factor 2 and 4 pre-amplifier gain applied, respectively. }
\end{figure}

An example of the average voltage on a pair of upstream, downstream DRS4 boards (four DRS4 chips, each with eight channels) in pedestal data is shown in Fig.~\ref{fig:wfmpednofilter}.  A vast majority of the low frequency noise phase is coherent on all four DRS4 chips; the amplitude is also coherent, but scaled by the relative pre-amplifier gains. Some small fraction of the noise is coherent only to a single DRS4 chip. 

Two algorithms were developed for the hit detection from the complicated waveform data: one uses a conventional waveform processing and the other one a 
deep-learning based algorithm. 

\subsection{Hit detection with a conventional waveform processing}
\label{sec:waveform_conventional}
The conventional algorithm begins with a coherent noise subtraction. We estimate the average noise coherent over each DRS4 chip (eight channels). The noise evaluation calculates the average voltage, bin-by-bin while excluding waveform bins associated with signals wire-by-wire. 
Possible biases introduced by the presence of signals are eliminated by requiring that the bin where the calculation is performed is surrounded by bins on the same waveform whose voltage is lower than a preset threshold, usually set to $3.5~\sigma_{RMS}$. If in some bins all channels are above threshold, the average voltage is replaced by the corresponding one computed on surrounding bins. 

The coherent noise calculation procedure was optimized on the data at standard beam intensity by maximizing the resulting tracking efficiency, number of hits per track, and minimizing the fitted $\chi^{2}/N_\mathrm{dof}$. This technique resulted in a large improvement in the tracking efficiency over usual digital filter approaches like a moving average high-pass filter or a discrete Fourier transform low frequency cut off. Such approaches significantly suppress the signal peak amplitude and integrated charge, which are critical for hit detection and for estimating the hit arrival time. The technique was verified using Michel positron events in the Monte Carlo simulation at high beam rates, simulating event out-of-time tracks, digitizing the signal shapes and simulating the coherent and non-coherent noise spectra.   

\begin{figure}[ht!]
\centering 
\includegraphics[width=0.4\textwidth]{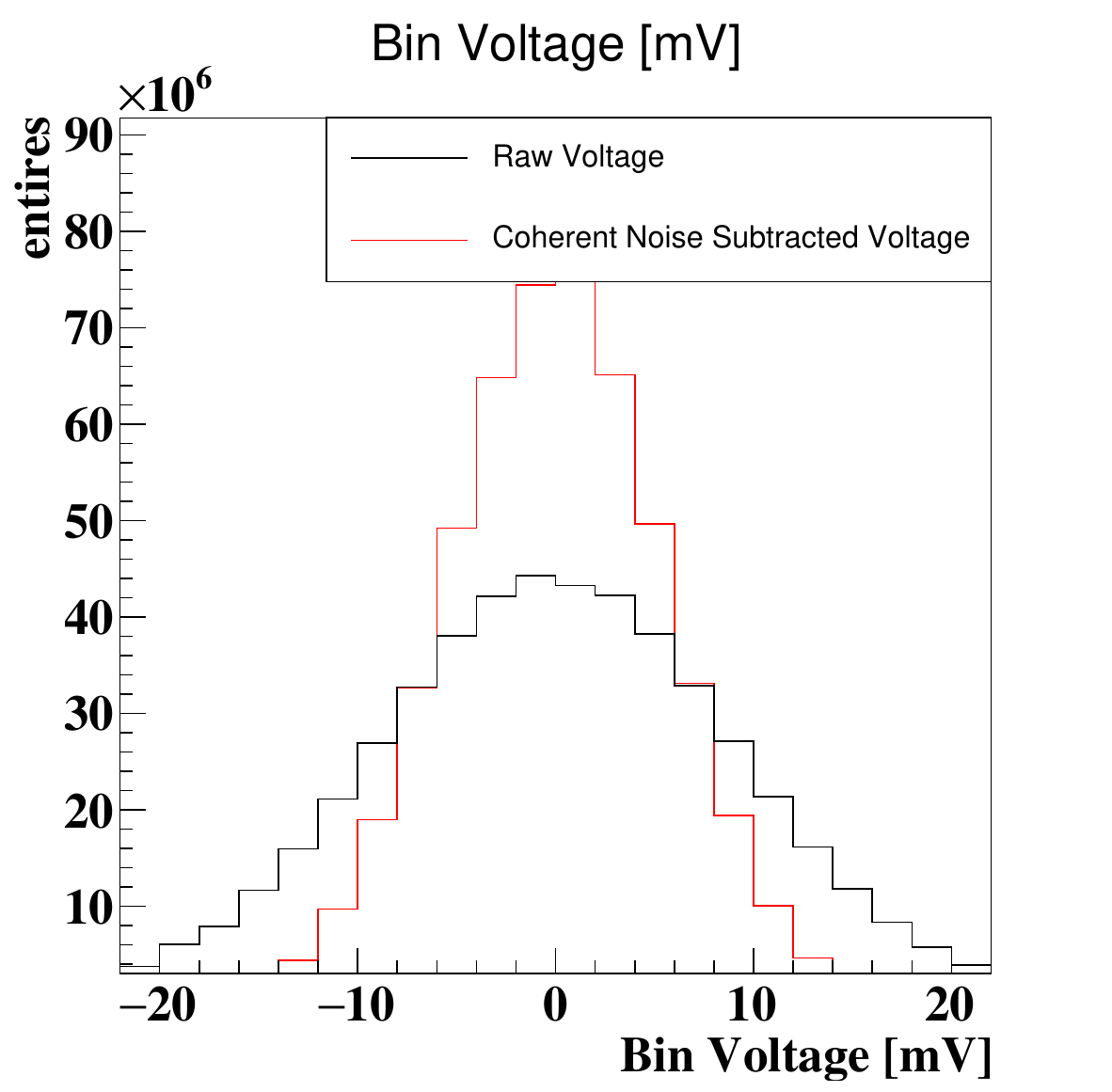}
\caption{\label{fig:wfmpedcomp} Effects of the coherent noise subtraction on the voltage distribution in Michel data.}  
\end{figure}
\begin{figure}[htb]
\centering 
\includegraphics[width=0.48\textwidth]{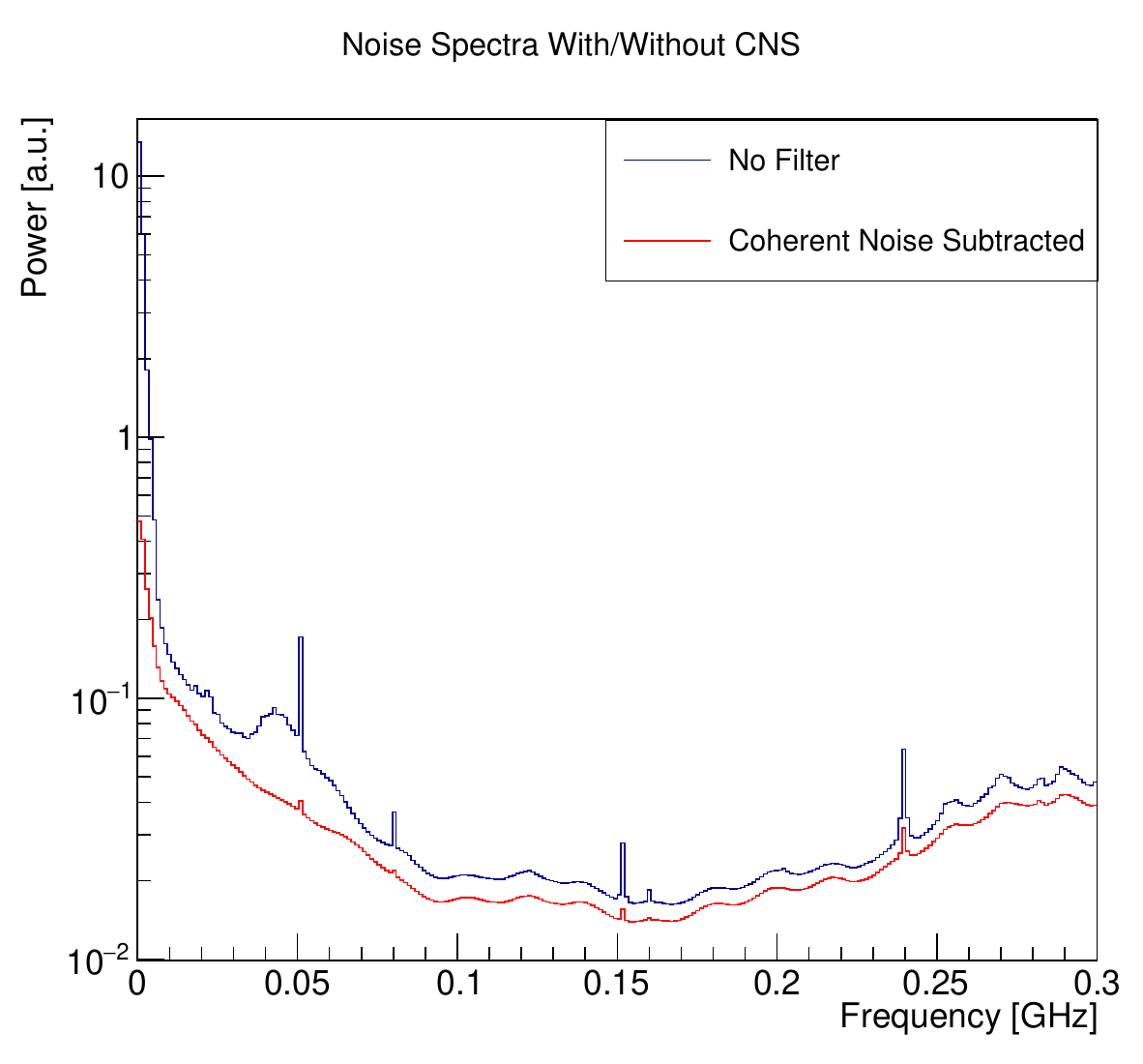}
\caption{\label{fig:FFTCNS} Power spectrum of pedestal data with (red) and without (blue) coherent noise subtraction with pre-amplifier gain 4.}
\end{figure}

Integrating over all wires and bins in Michel data, the bin voltage with and without the coherent noise subtraction is shown in Fig.~\ref{fig:wfmpedcomp}. The FWHM is suppressed from $\sim$ \SI{23}{\milli \V} $\rightarrow$ \SI{13}{\milli \V}. In Fig.~\ref{fig:FFTCNS}, the power spectrum of pedestal data is shown for before and after the coherent noise subtraction. It is clear that the power of low frequency noise is significantly suppressed. The discrete peaks (e.g. 50 MHz, 150 MHz) and the wide peak at \SI{\sim 45}{\MHz} are almost completely suppressed. 

\begin{figure}[htb]
\centering 
\includegraphics[width=0.44\textwidth]{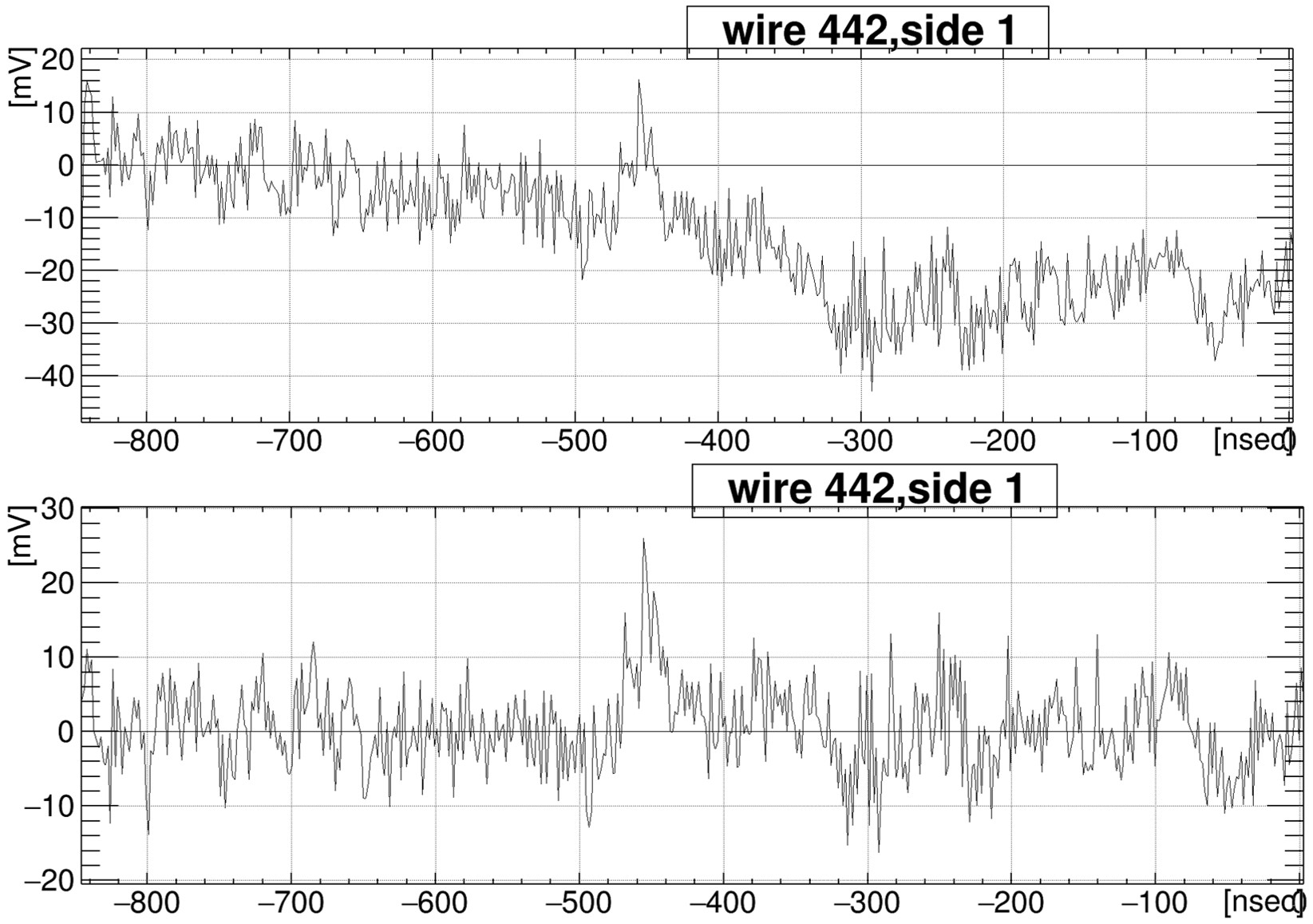} 
\caption{The top and bottom sub-figures show an example signal waveform without and with the coherent noise subtraction applied.\label{fig:CNSWaveform} }
\end{figure}

An example of a low amplitude signal on top of a coherent low frequency noise is shown in Fig.~\ref{fig:CNSWaveform}. 
When the coherent noise subtraction is applied, the signal maintains its amplitude while the baseline is flat and well-centered at zero. 

The signal pulse shape (Fig.~\ref{fig:PulseShape}) in the simulation indicates minimal signal power with frequency above \SI{200}{\MHz}. Since we still observe significant incoherent high frequency noise in this frequency region, we apply a discrete Fourier transform high frequency cut off at \SI{225}{\MHz}.  That is, transforming the waveforms into frequency domain, setting the power of all frequencies above the high frequency cut off to zero, and then transforming the waveform back to time domain. Applying this filter after the coherent noise subtraction resulted in higher number of hits per track and tracking efficiency when compared to a moving average or no low pass filtering technique. 

The waveform analysis proceeds by searching for hits on the filtered waveform. The hit detection implements a fixed threshold (in Volt) on two adjacent bins and a wide “pulse-shape” fixed threshold (again in Volt) integrating over 20 ns after the two initial adjacent bins. The discriminator is roughly based on studies fitting signals assumed to be a single or a pair of clusters in data.

After the hit is detected, the thresholds are lowered to search for a low amplitude cluster before the detected hit. The two-level (hit detection and then hit arrival time) algorithm results in a high efficiency of detecting low amplitude clusters without a high fake hit rate. 

\subsection{Hit detection with a deep-learning algorithm}
\label{sec:waveform_deep}
\begin{figure}[htb]
\centering 
\includegraphics[width=1\linewidth]{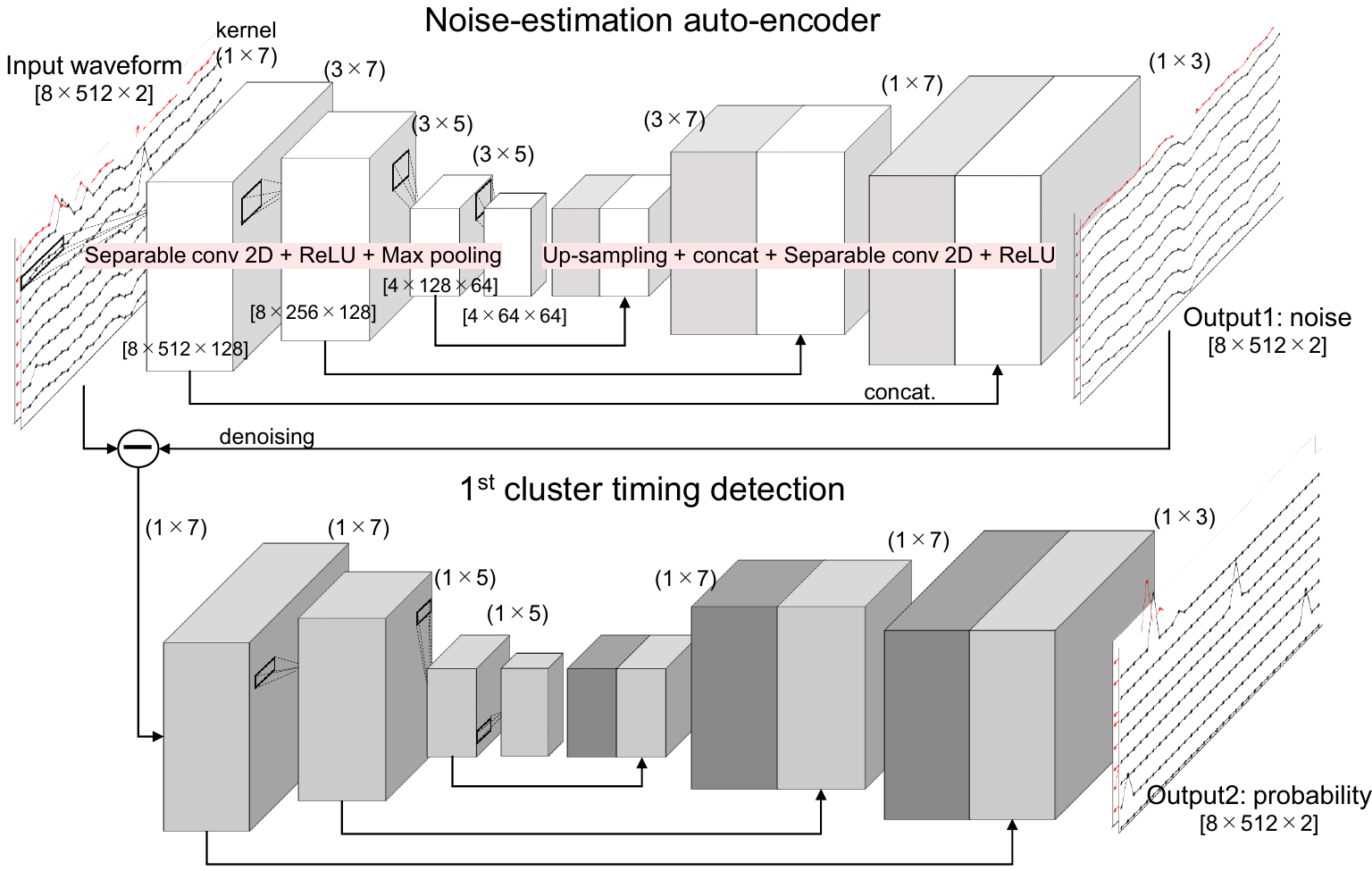}
\caption{Architecture of the hit detection CNN model.\label{fig:CNNArchitecture} }
\end{figure}

The other method uses a deep-learning algorithm based on a convolutional neural network (CNN). 
The input is a set of waveform data (512 points each) from adjacent eight wires, in the form of \numproduct{8x512x2}, where the last dimension corresponds to the two ends of wires.
The output shape is the same \numproduct{8x512x2}, where each point represents the probability of having the first cluster timing at this sample point.

Fig.~\ref{fig:CNNArchitecture} shows the architecture of the model, which consists of two cascades of similar architecture based on a 2D-CNN-based auto-encoder with U-Net like structure~\cite{unet}. The first stage is designed to estimate and remove noise. 
The kernels in each CNN layer work as digital filters for the dimension of time in waveforms, and the activation layers (rectified linear unit) introduce non-linearity.
Therefore, the CNN is a multi-stage parallel non-linear digital filter with optimized filter parameters.
Convolving over the different waveforms is to learn the pattern of the coherent noise components.

The output of the first stage is passed to the second stage for hit detection.
In this stage, the kernels of CNN layers do not convolve points from different wires because the signals are not correlated with different wires. 
The hit is identified at the local maximum points above 0.05 in the output.

\begin{figure}[htb]
\centering 
\includegraphics[width=1\linewidth]{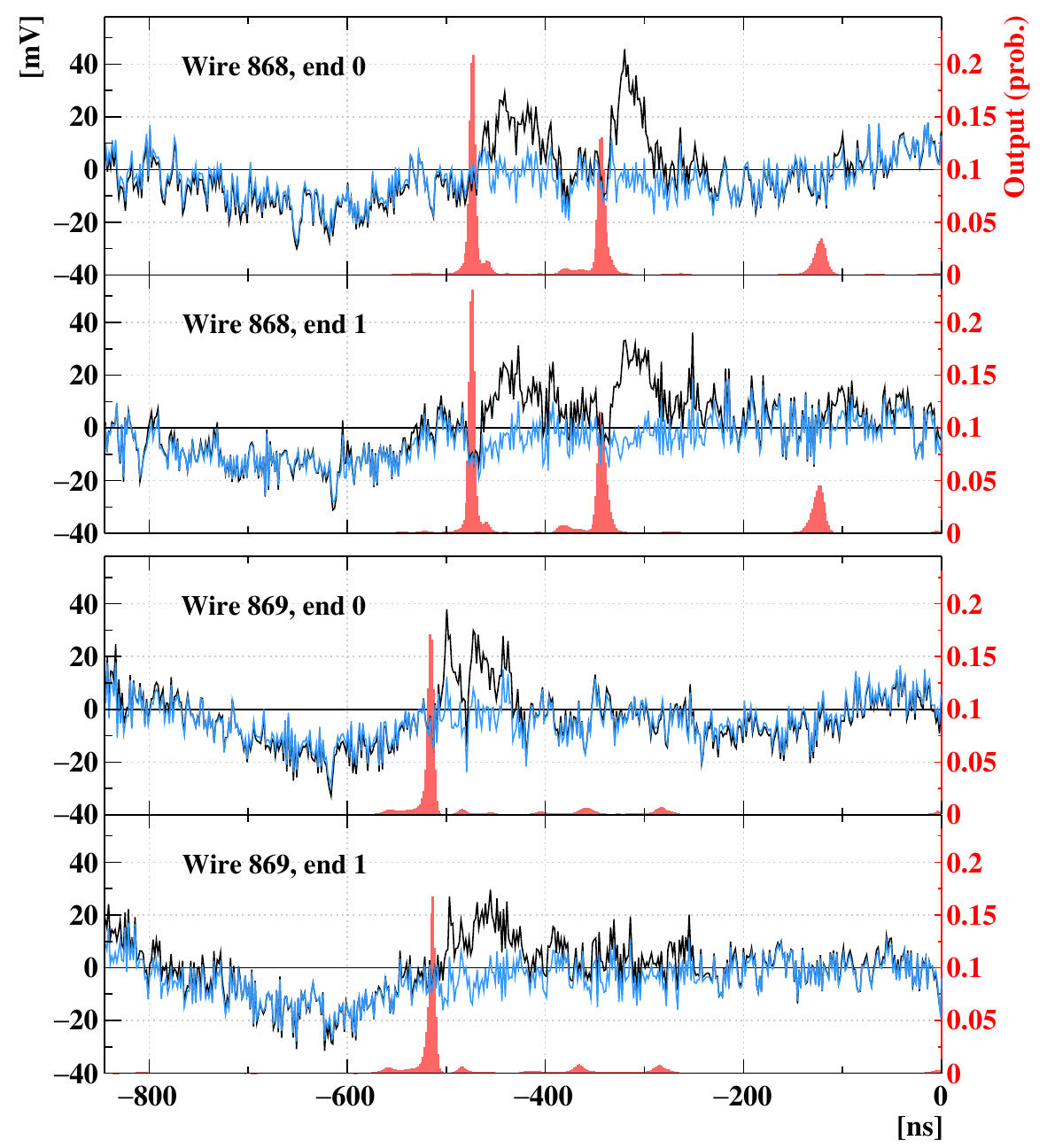}
\caption{Example of the inputs  and outputs of the hit detection CNN (4 out of 16 channels are shown). The black waveforms are the input raw waveforms. The light blue waveforms are the intermediate output of the CNN (``output1'' in Fig.~\ref{fig:CNNArchitecture}), showing the estimated noise. The red bar graphs are the final output of the CNN (``output2''), showing the probability of the first cluster arrival time of a hit at each sampling point.\label{fig:WF_MLProb} }
\end{figure}

The network was trained with samples of the simulated waveform data superposed on the pedestal data taken without beam.  The hits in the simulated waveform are randomly added in time at the expected rate for $R_{\mu} = \SI{3e7}{\per\second}$.
Fig.~\ref{fig:WF_MLProb} illustrates how the CNN works for the real data.

The maximum hit detection efficiency is given by combining the hits detected by the two algorithms; 
the lists of hits by the two are merged and if two hits are within \SI{2}{\ns} they are merged and the earlier timing is adopted. 
However, the combination also results in a higher fake hit rate.
Therefore, to make the best use of the results by the two methods, the following reconstruction is repeated twice,
one with only the hits detected with the conventional processing method and the other with the combined hits.
These results are combined after the track reconstruction is completed (see Section~\ref{sec:selection}).
This approach improves the final tracking efficiency by a factor of 1.17 at $R_{\mu} = \SI{2e7}{\per\second}$ and 1.35 at $R_{\mu} = \SI{5e7}{\per\second}$ compared to the result with only the
conventional method.

\subsection{The cross-fitting}

A hit always induces signal on both ends of the wire with similar shapes. As described in the next section, the hit position is computed from the relative time difference and the relative size of the signal on the two ends of the wire. 

The measurements are made by a ``cross-fitting" algorithm. In this algorithm, one end of a waveform is used as the fitting function of the other waveform. The fitting is performed by minimizing the chi-squared with three floating parameters: baseline voltage, relative amplitude scale, and the time difference. 

The fitting is performed when a pulse is detected on one end regardless of the detection on the other end. Therefore, the hit detection is given by the \lq\lq or\rq\rq~logic of the two ends. If pulses are detected on both ends, fitting is performed in both directions and the mean is taken. 
The relative size of the signal is conventionally measured by integrating the charge on the two ends; this already uses many bins and thus results in a comparable resolution to the cross-fitting approach. 

The cross-fitting gives only relative values, while the absolute values are calculated from the summed waveform of the two pulses after adjusting their relative timing. 

\section{Hit Reconstruction}
\label{sec:hitrec}
\subsection{Hit $z$ reconstruction}
The difference in arrival times and the ratio between the total charges collected on the two ends of the cell sense wire allows a preliminary determination of the longitudinal ($z$) coordinate of the hit. The $z_{w}$ coordinate along the wire is calculated as:
 \begin{equation}
    z_{w,t} = \frac{v}{2} \left(t_{0} - t_{1} -t_{\mathrm{wireends}} \right) \label{eq:zrawtimes}
\end{equation}
using the arrival times and as:
\begin{equation}
    z_{w,q} = L_{wire} \left(  \frac{R_{G} Q_{1} - Q_{0}}
     {R_{G} Q_{1} + Q_{0}} \right), 
\label{eq:zrawcharges}
\end{equation}
using the charges and is then transformed into $z$ in the MEG~II coordinate system by using the individual wire position and direction within the chamber. In Eq.~(\ref{eq:zrawtimes}) $v$ is the signal propagation velocity inside the wire, $t_{0}$ and $t_{1}$ are the signal arrival times on the two wire ends and $t_{\mathrm{wireends}}$ is the calibrated timing offset between the two wire ends, while in Eq.~(\ref{eq:zrawcharges}) $R_{G}$ is the calibrated ratio of the gains at the two wire ends and $Q_{0}$ and $Q_{1}$ are the corresponding charges. All the parameters were obtained by calibration procedures and iteratively optimized to get the best single hit resolution, as discussed in Section~\ref{sec:qtcalib}.

\subsection{Hit distance of closest approach measurement}

\begin{figure}[htb]
\centering 
\includegraphics[width=0.48\textwidth]{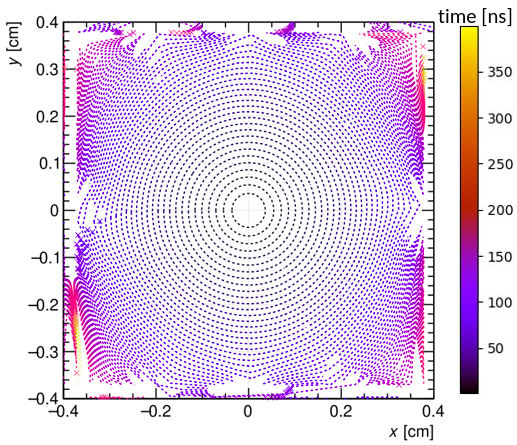}
\caption{\label{fig:isochrone} An example of the lines connecting points with the same drift time towards the sense wire (\lq\lq isochrones\rq\rq).}
\end{figure}

The first detected ionization cluster usually (but not always) corresponds to the energy release nearest to the wire on the analyzed waveform. The time of the first cluster is converted into the particle’s distance of closest approach (DOCA) to the anode wire using the time--distance relationship (TXY). The lines at fixed time (\lq\lq isochrones\rq\rq) are approximately circles around the sense wire for a time \SIrange[range-phrase=--,range-units = single]{150}{200}{ns} and exhibit important deviations from the circular symmetry when one approaches the cathode wires and because of the magnetic field effects. An example of isochrone curves from Garfield++~\cite{garfield++} is shown in Fig.~\ref{fig:isochrone}.

\begin{figure}[htbp]
\centering 
\includegraphics[width=0.48\textwidth]{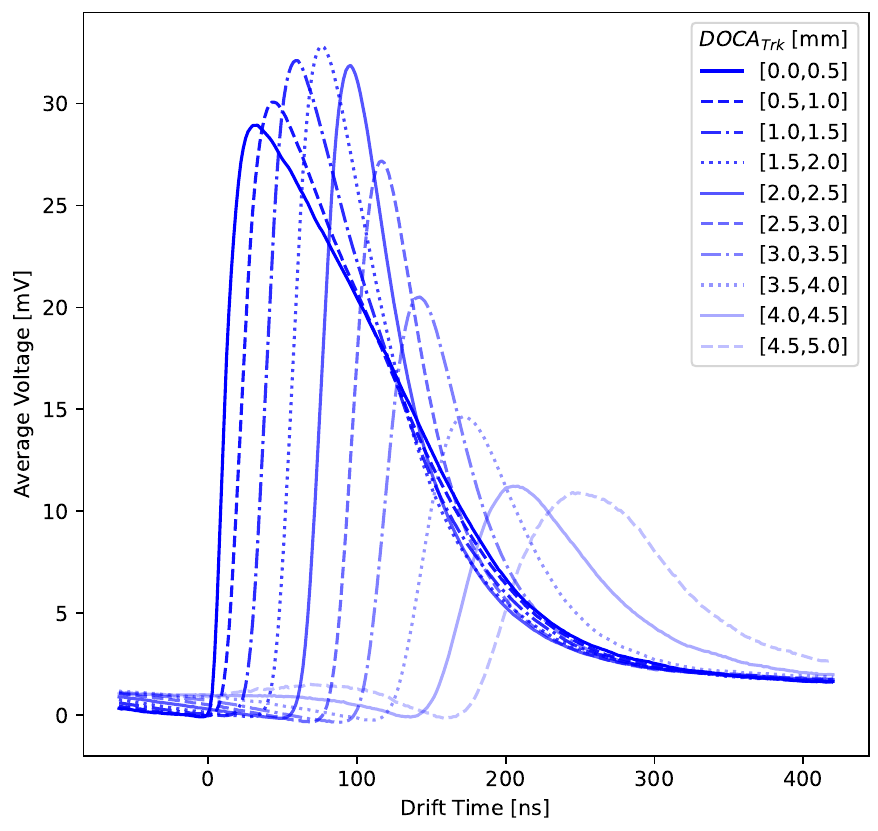}
\caption{\label{fig:AverageVoltageDriftTime} The average voltage per drift time as a function of the track DOCA.  }
\end{figure}

A first look at the cluster statistics and drift velocity in the \num{2021} data can be shown by plotting the average voltage per drift time summed over many waveforms for hits with a range in track DOCA ($D_\mathrm{Trk}$); this is shown in Fig.~\ref{fig:AverageVoltageDriftTime}. Here, we integrate over all cell sizes. 
The time corresponding to the peak value of the distribution in a given track DOCA interval yields the local drift velocity value, e.g. \SI{1.5}{\milli \meter} $< D_\mathrm{Trk} <$ \SI{2}{\milli \meter} at \SI{80}{\nano\second} gives an average drift velocity of \SI{22}{\micro\metre\per\ns} at \SI{1.75}{\milli \meter} from the sense wire, which remains constant for most of the cell width, in agreement with measured values in literature. The peak value of each distribution, proportional to the charge collected, shows, for larger drift times, clear indication of effects due to electron recombination, most likely due to the oxygen content in the gas mixture. The fall of the peak values at lower track DOCA intervals suggests non negligible space charge effects for tracks close to the sense wire.

The cluster's arrival time is composed of a track time $T_{0}$, measured by the pTC detector with a resolution $\sigma_{T_{0}} \sim \SI{40}{\ps}$, and by the cluster's drift time, which can reach \SI{\sim 400}{\ns} if the track crosses a cell near one edge. Two techniques are used to determine the DOCA from the hit arrival time: the conventional and the neural network based approaches. The DOCA estimates is done iteratively: a first evaluation is done by using an angle-averaged time--distance relationship; then, the DOCA estimate is refined at the tracking stage taking into account the track direction and the new DOCA estimates are inputs for tracking refinements; and finally, the DOCA from the neural network approach is used. 

In the conventional approach the time--distance relationship is stored in TXY tables, estimated using Garfield++~\cite{garfield++} simulations. 
Since the cell size and shape as well as the magnetic field change along the longitudinal, radial and azimuth coordinates of the chamber, a bi-dimensional sampling of the CDCH volume in $r$ and $\phi$ is performed at fixed $z$ and repeated in $z$ bins of $10~{\rm cm}$ width, forming arrays of TXY tables. The effective DOCA corresponding to a measured drift time is determined by interpolation using the tables computed in the location nearest to the hit wire. Intrinsic uncertainties in the DOCA determination come from the use of tables obtained with simulations. Additional uncertainty can potentially by the result of the sampling and interpolation procedure, however in the Monte Carlo this was shown to be negligible.

The conventional determination of DOCA is affected by a bias coming from the low density of the ionization clusters along the track, which causes a systematic overestimate of the DOCA at a level of \SIrange[range-phrase=--,range-units = single]{50}{100}{\micro\meter}, depending on the track angle with respect to the wire direction.
Since the conventional DOCA estimate only uses the drift time of the first electron, an improved estimate can be achieved by taking into account also other clusters with the cluster counting technique discussed in Section~\ref{sec:Gas_Mixture}. However, the presence of the additives (oxygen and alcohol) in gas mixture reduced 
the number of primary clusters and the total signal charge, making difficult the practical application of this method on \num{2021} and \num{2022} MEG~II data. Then, even if all ionization cluster information is present in digitized waveforms, a specific cluster counting algorithm was not yet developed. 

The neural network approach trains on fitted tracks created using the conventional time--distance relationship, to create an improved DOCA estimator that overcomes the intrinsic limitations of the conventional approach. The neural network inputs a series of variables (hit arrival time, $T_{0}$, layer, $z$, etc.) and trains on the track DOCA to create a model that maps the hit properties to a final DOCA estimate.  The purpose of this training is to produce a data-driven DOCA estimator, which is free from possible systematic differences between the \lq\lq real\rq\rq~TXY and that based on Garfield++ simulations. Being trained on experimental data, the network can also learn the bias due to the ionization statistics. The optimal neural network approach is a convolutional neural network (CNN) model that additionally inputs waveform voltages to map the signals from all ionization clusters to the DOCA estimate. The details of the neural network approach are presented in ~\cite{CNN}.

\begin{figure}[htbp]
\centering 
\includegraphics[width=0.4\textwidth]{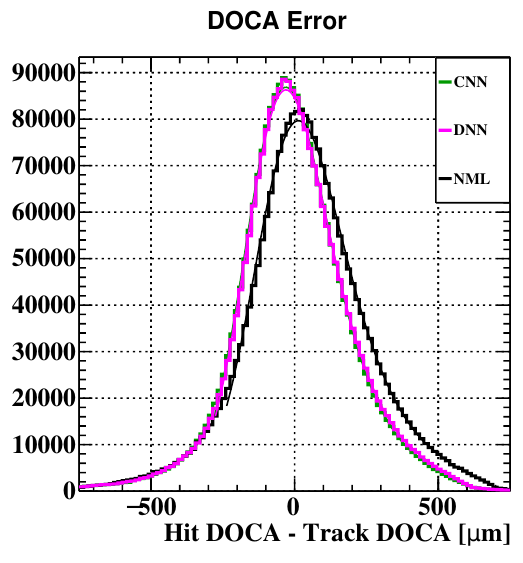}
\caption{\label{fig:hitres} The hit residuals for the three methods: conventional approach is in black, convolutional neural network is in green and dense neural network is in red.}
\end{figure}

We evaluate the performance of the DOCA evaluation in the data by looking at the DOCA hit residual, defined as the difference between the final DOCA estimate ($D_\mathrm{hit}$) and the DOCA determined by the track reconstruction. Fig.~\ref{fig:hitres} shows the distribution of the DOCA residuals ($D_\mathrm{hit} - D_\mathrm{Trk}$). The residual distribution with a conventional approach for DOCA reconstruction is approximately Gaussian, with a central core of $\sigma \sim$ \SI{160}{\nano\second} and a small non Gaussian tail for positive values due to the ionization statistics. The main benefit of the neural network approaches (especially of CNN) is a reduction of the right tail and an increased number of entries in the peak; this implies a suppression of the ionization statistics bias. 

The widths of the hit residual distributions represent the resolutions in the single hit coordinate reconstruction and are not far from  the expected values cited in the MEG~II proposal of \SI{100}{\micro\meter}~\cite{baldini_2018}. Some improvements are possible by further optimizations of the alignment and reconstruction algorithms. This neural network technique has been verified using data-driven kinematic resolution estimating techniques to improve all kinematic resolutions by $\sim 5-13\%$~\cite{CNN}; an example of the kinematic improvement is given in Section \ref{sec:trackres}.
\section{Track finding and track fit} 
\label{sec:tracking}

The reconstruction of a positron track starting from the reconstructed hits is performed in two steps: the track finding, which combines hits produced by the same positron into a track candidate, and the track fit, which extracts the best estimate of the positron's kinematics at the target.

\subsection{Track finding}
\label{sec:trackfinding}

\begin{figure}[htbp]
  \centering
  \includegraphics[width=1\linewidth]{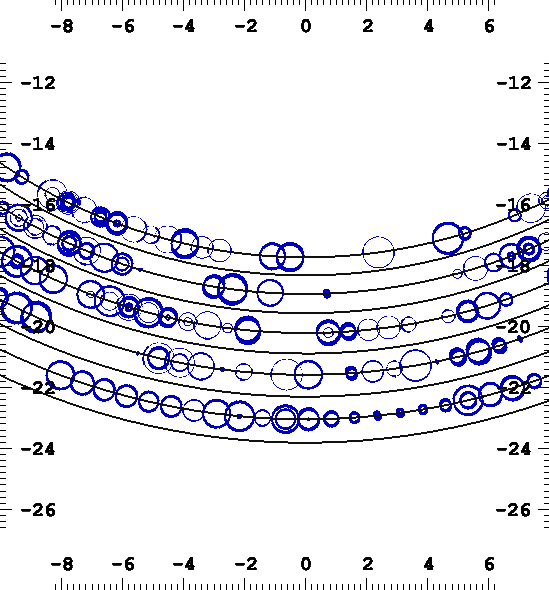}
  \includegraphics[width=1\linewidth]{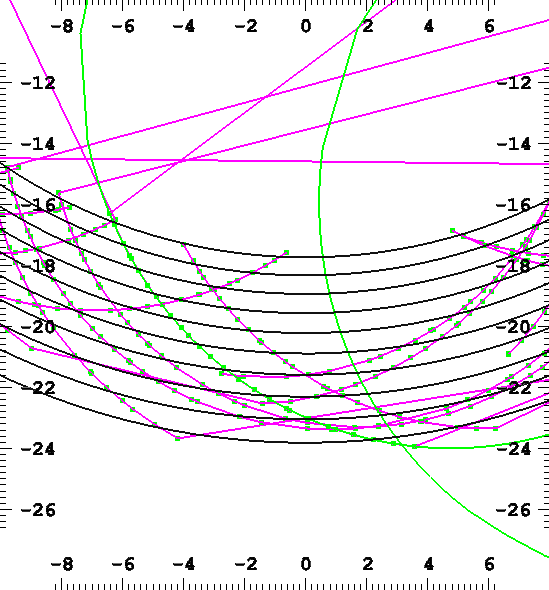}
  \caption{Example of event in one stereo view projection at $z=0$. Top:
    circles correspond to the drift distance
    of hits relative to $T_{0}$ of the signal track; hits are selected within the
    $[\SI{-50}{\ns},\SI{200}{\ns}]$ time window around $T_{0}$. Bottom:
    reconstructed tracks with hit points in both stereo view, the green curve is the trajectory of the signal positron. }
  \label{fig:prevent}
\end{figure}

The CDCH operates in the high-rate environment, which makes the positron reconstruction a very challenging task. The corresponding cell occupancy is at the level of \SIrange[range-phrase=--,range-units = single]{15}{25}{\percent} per event in time window of maximum drift time.
For example, this is on the level of the channels occupancy of the former ALICE TPC\footnote{The present ALICE TPC uses a MPGD-based readout which replaced the wire chamber} in Pb--Pb events or even higher than in the Belle~II drift chamber~\cite{Alme:2010ke,BelleIITrackingGroup:2020hpx}.
Taking in the fact that the CDCH has only nine layers in comparison
with 156 and 56 layers of Alice and Belle~II trackers respectively,
this makes a pattern recognition more complicated for MEG~II, as shorter tracks can be more easily hidden under overlapping
background events. An example of event is shown in Fig.~\ref{fig:prevent}. 

The track finding is a pattern recognition algorithm that starts from the reconstructed hits and tries to put them together to form a track candidate, i.e.\ a list of hits supposedly produced by the same positron and a preliminary estimate of the particle's kinematics based on the hit positions.
The implemented pattern recognition algorithm is based on the track following with the Kalman filter method~\cite{Fruhwirth:1987fm}. 

To reconstruct the distance to wires of found hits, as described in the previous section, an initial estimation of the track time $T_0$ is required. 
For this purpose, times of all pTC hit clusters within \SI{50}{\ns} of the
trigger time are taken for consideration. 
To take into account the time of flight from CDCH region to the pTC, additional average corrections are applied,
which correspond to cases when the trajectory leaves the CDCH immediately or do after one or two full 
turns before crossing the pTC.
The average time of flight of one positron between different turns is around \SI{4}{\ns}. 
In addition, the $T_0$ seed can be evaluated based on three consecutive
cells in a same layer, assuming they are crossed by a straight line. 
This gives the time resolution of the tracklet about \SI{6}{\ns}. 
For each initial $T_0$, only hits within the time window slice up to the maximum possible drift time in a cell are considered for the tracking.

\begin{figure}[htbp]
  \centering
  \includegraphics[width=0.8\linewidth]{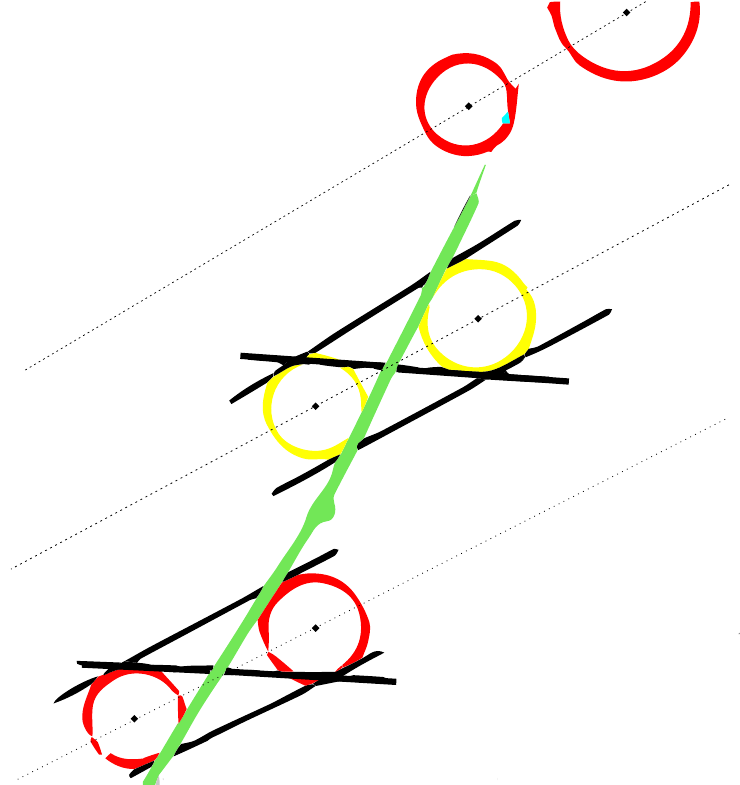}
  \caption{Pairs of hits from two stereo views projected to same $z$
    coordinate. Black lines: all combinations of tracklets, green
    line: constructed track seed from compatible pairs.}
  \label{fig:seedpairs}
\end{figure}

The tracking starts with a seed construction, where the procedure starts from the external layers with lowest occupancy.
For this purpose, two (or three for the self $T_0$ without pTC timing) hits from spatially adjacent wires in the same layer
are combined together into a tracklet, which has the position and transverse direction. 
The hits in a tracklet should be compatible with the measured $z$ coordinate within resolution,
and the position of this tracklet should be loosely compatible with the position of the initial pTC hit cluster as to be roughly in the same quarter of the CDCH. 
The first tracklet is matched with other pairs of hits in next two lower layers with opposite stereo view, where they should be compatible with the $z$ position and the projection of transversal direction.
An example of the constructed seed is shown in Fig.~\ref{fig:seedpairs}.
These two pairs together with an additional constraint on the beam axis crossing give fully defined track parameters for further tracking. 
All compatible combinations of two pairs between different layers, including possible different left--right side of wires in the pair, produce the set of track seeds.

Since the CDCH has the stereo geometry, then the
$z$-coordinate measurements over single wires are not strictly necessary to reconstruct the track parameters. 
However, the additional $z$ hit information is greatly helpful to reduce seeding
combinatorics and to prevent improper attaching of hits from pileup
tracks during tracking.
This is very important especially in the high-rate environment of the MEG~II experiment.

The constructed seed trajectory is then propagated backward and forward to the contiguous layers, according to the expected motion in the magnetic field, and accounting for the average energy loss in material. If hits are found in these layers, consistent with the expected trajectory by means of $\chi^2$, they are added to the track candidate and the estimate of the trajectory is updated using the Kalman filter algorithm. The procedure is iterated until the innermost layers are reached.

Several quality checks of followed tracks are performed at the intermediate
steps to discard the seeds at earliest as possible stage; these include to find some hits in
nearest to seed layers, checks of number of found hits and their density, the track $\chi^2$ of shorter track segment.

The seeding is repeated starting from lower layers, where hits already
used by found good tracks at the previous tracking iteration are excluded from consideration. 
From innermost layers all tracks are propagated again to the forward direction with attempt to find additional compatible hits and to form the full single turn track candidate.
Additional seeding is performed from found tracks when they
propagated to next turns in both direction from first and last points 
and their track parameters are used to find new track segments.

Since different initial seeds can result in near same or strongly
overlapped track candidates, a cleanup of the output list is performed at the final stage.
All obtained track candidates are sorted according to the quality factor, which include number of attached hits, hits density over reconstructed trajectory and 
track $\chi^2$. Tracks from this list are selected if the number of shared hits with tracks of higher quality is below some threshold.

As it is customary in wire detectors, ambiguities affect the determination of the hit position, which, for given track angle and drift time, can be either on the left or on the right of the wire (left--right ambiguity). 
The track finder procedure also provides a preliminary resolution of these ambiguities, as the one minimizing the hit--track residuals. 

The pattern recognition results in single and half turn track segments inside the CDCH (\numproduct{2x9} and 9 layers intersections, respectively). About \SI{85}{\percent} of tracks contain three full crossing of the CDCH layers (\numproduct{3 x 9} layer intersections or a 1.5 full turn track). The remaining fraction is 2.5 turn and even more rare 3.5 turn tracks. The 2.5 and 3.5 turn tracks are a consequence of the graded magnetic field; the larger number of turn tracks correspond to tracks with $\theta_\mathrm{e} \sim \SI{90}{\degree}$. 

The currently implemented pattern recognition by using the track
following method is of type of ``local" approach, where local continuity of trajectory is important.
Such methods can be spoiled by inefficiency of the hit reconstruction
discussed in the previous section and the high rate of overlapped hits. 
Other pattern recognition strategies are under initial discussion as well, which can be based on global
hit dealing approaches, such as Hough transform method (with a proper dealing of the CDCH stereo configuration)
or trying to exploit neural nets in track reconstructions
algorithms. Combination of different approaches can potentially further improve the track finding efficiency.

\subsection{Track fit}
The track fit is developed using the GenFit toolkit~\cite{genfit1} and allows to include an optimal treatment of the material effects (energy loss and multiple scattering). For this purpose, the CDCH is modeled as a uniform volume of gas. In reality, there are wires in the volume. However, considering individual wires is not good for tracking because small changes in the trajectory change whether the track crosses the wire volumes or not, making the calculated material budget in the track propagation unstable and unreliable. Another possibility is replacing the volume of gas with an average medium, diluting the wire material over the gas volume. When this method was applied to the data, it resulted in a deterioration of the tracking performance. Therefore, the wire material was omitted in the model. 
The measurements entering the $\chi^2$ minimized with the Kalman algorithm are the  drift distance of the hits and their longitudinal position $z$, estimated from the weighted average of the charge-division and time-difference estimates. It should be noticed that the drift distance would be enough, by itself, to fit the trajectory, thanks to the combination of the different stereo angles of the wires. The hit self-estimate of the longitudinal position adds poor statistical information, but it is helpful to numerically stabilize the fit.

The Kalman filter is complemented with a deterministic annealing filter (DAF)~\cite{daf} to identify and reject hits not really produced by the particle under consideration, and to improve the resolution of the left--right ambiguities. The inclusion of the $z$ self-estimate in the measurement model is also important to improve the bad hit rejection power of the DAF. 

In the first stage of track fitting, we fit the track segments output by the track finder. We then merge the fitted track segments to get the full, multi-turn trajectory of the positron. This is done by propagating the first and last hit of each segment to the point of closest approach to the $z$ axis, and comparing two by two the resulting positions and directions, looking for a good match (quantified by a $\chi^2$ based on the Kalman covariance matrix).

\begin{figure}[htb]
  \centering
  \includegraphics[width=1\linewidth]{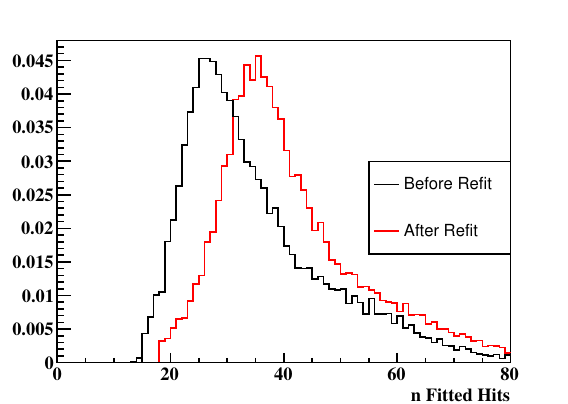}
  \caption{Number of hits on the fitted tracks with and without the \lq\lq refit\rq\rq~algorithm.}
  \label{fig:refithits}
\end{figure}

\begin{figure}[htb]
  \centering
  \includegraphics[width=\linewidth]{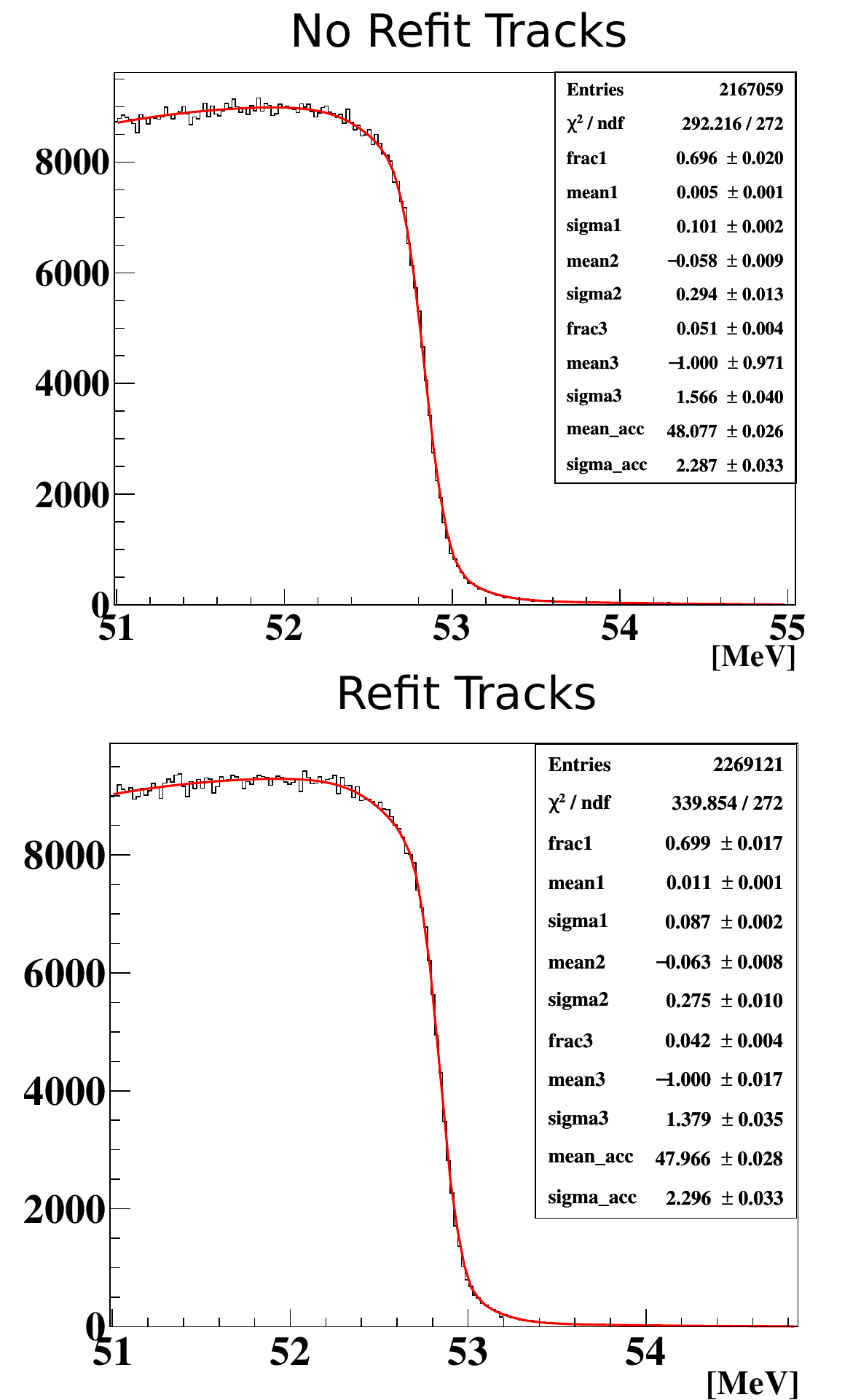}
  \caption{Michel edge fit with and without the track \lq\lq refit\rq\rq~algorithm applied. Core resolution is improved by $\sim 14\%$. Details on the determination of positron momentum resolution will be explained in section \ref{sec:trackres}.}
  \label{fig:refitmichel}
\end{figure}

Once the track segment merge has been tried, the fitted tracks are propagated through the CDCH on its full trajectory from the target to the pTC to search for missing hits that the track finder, with its lower accuracy, was unable to associate to the track. Frequently, the final half turn output from the track finder results in a low quality fit due to a low number of hits and thus is difficult to merge. Adding hits on the final half turn to single full turns results in improved momentum resolution. 
The impact of repeating the fit after this algorithm (the procedure of adding hits and fit again the track is called \lq\lq refit\rq\rq~algorithm) is applied is shown in Figures~\ref{fig:refithits} and~\ref{fig:refitmichel}. An increase of about \SI{30}{\percent} in the typical number of hits and an improvement of about \SI{14}{\percent} in the core momentum resolution are observed.

The precise trajectory obtained from the track fit is propagated to the pTC, so that the association of the pTC hit cluster to the track can be verified (or included if the track was built without a reference pTC hit cluster), and the best possible estimate of the track time can be extracted. It includes the time of flight from each hit to the pTC itself. In this way, the drift time (and consequently the drift distance) of each single hit can be updated. 

Finally, the track is propagated from its first hit to the target. Deformations of the target with respect to an ideal plane, down to a few hundred micrometers, can introduce significant biases in the measurement of the track angles. For this reason, the unflatness of the target foil is modeled by a triangle mesh, and the propagation is performed in two steps.  
In the first step, the track is propagated to a virtual plane, a few mm in front of the real target. In the second step, the track is propagated to the nearest triangle of the mesh. For the energy loss and multiple scattering calculations, the positron is assumed to be produced at half-thickness depth inside the target foil. 

The propagation to the target provides the best estimate of the muon decay point and time, and of the positron kinematics at production, including an estimate of the uncertainty and of the correlations among these quantities, in the form of a covariance matrix.

\section{Quality cuts and track selection}
\label{sec:selection}

Even though several quality requirements are imposed on the track candidates and a cleanup procedure is applied to them in the track finding stage as described in Section~\ref{sec:trackfinding}, 
poor quality tracks and duplicated tracks of a single positron (ghost tracks) exist in the fitted tracks.
We apply the following selection criteria to the fitted tracks to select the best measured track per physical positron and to guarantee the reconstruction quality to be usable in physics analyses:
\begin{enumerate}
\item Cut on the quality of the track fit,
\item Cut on the condition of propagation to the target and the pTC,
\item Identification and grouping of ghost tracks,
\item Rating the ghost tracks and selecting the best one.
\end{enumerate}
\begin{table}[tb]
\caption{Criteria for the track quality cut.}
\label{tab:criteria}
\centering
  \begin{minipage}{1\linewidth}
   \renewcommand{\thefootnote}{\alph{footnote})}	
   \renewcommand{\thempfootnote}{\alph{mpfootnote})}	
\centering
\small
\begin{tabular}{@{}lr}
\hline
Parameter  & Condition \\
\hline
\textbf{Track quality} & \\
Number of fitted hits  & $N_\mathrm{hit} \geq 18$ \\
Those in the first half turn & $N_\mathrm{hit,first} \geq 5$ \\
Chisquare of the fit & $\chi^2_\mathrm{fit}/N_\mathrm{dof} < 4.33-0.0167N_\mathrm{hit}$ \\
Energy fit uncertainty & $\mathrm{cov}(E_\mathrm{e},E_\mathrm{e}) < (\SI{300}{\keV})^2$ \\
Angular fit uncertainties & $\mathrm{cov}(\theta_\mathrm{e},\theta_\mathrm{e}) < (\SI{50}{\milli rad})^2$ \\
                                    & $\mathrm{cov}(\phi_\mathrm{e},\phi_\mathrm{e}) < (\SI{12}{\milli rad})^2$ \\
Position fit uncertainties & $\mathrm{cov}(y_\mathrm{e},y_\mathrm{e}) < (\SI{5}{\mm})^2$ \\
                                    & $\mathrm{cov}(z_\mathrm{e},z_\mathrm{e}) < (\SI{5}{\mm})^2$ \\      
\hline
\textbf{Matching with pTC}\footnotemark[1] & \\
Timing & $|\Delta T|<\SI{15}{\ns}$ \\          
Distance & $\Delta v^2 +\Delta w^2  < (\SI{10}{\cm})^2$ \\
 &  $|\Delta w|  < 5\sigma_{w,\mathrm{pTC}} \approx \SI{6}{\cm}$ \\
Fiducial volume & $+\SI{3}{\cm}$\\
Extrapolation length & $l_\mathrm{pTC} < \SI{80}{\cm}$ \\
\hline
\textbf{Extrapolation to target} & \\
Fiducial volume & $-2\sigma_{y,z}$\\
Extrapolation length & $l_\mathrm{target} < \SI{45}{\cm}$\\
\hline
\textbf{Multivariate analysis} & \\
NN output & $O_\mathrm{NN} < 0.1$ \\
\hline
\end{tabular}
\footnotetext[1]{Compared at the plane of the matched pTC counter. $w$ ($v$) is the local coordinate along the long (short) side of the counter.}
\end{minipage}
\end{table}

The criteria for the first and second items are listed in Table~\ref{tab:criteria}. 
The quality cuts are basically based on the track-fit outputs and geometrical consistency.
Among them, the number of fitted hits and the fit $\chi^{2}$ are especially important while the others are practically to reject outliers.

\begin{figure}[!htb]
  \centering
  \includegraphics[width=0.9\linewidth]{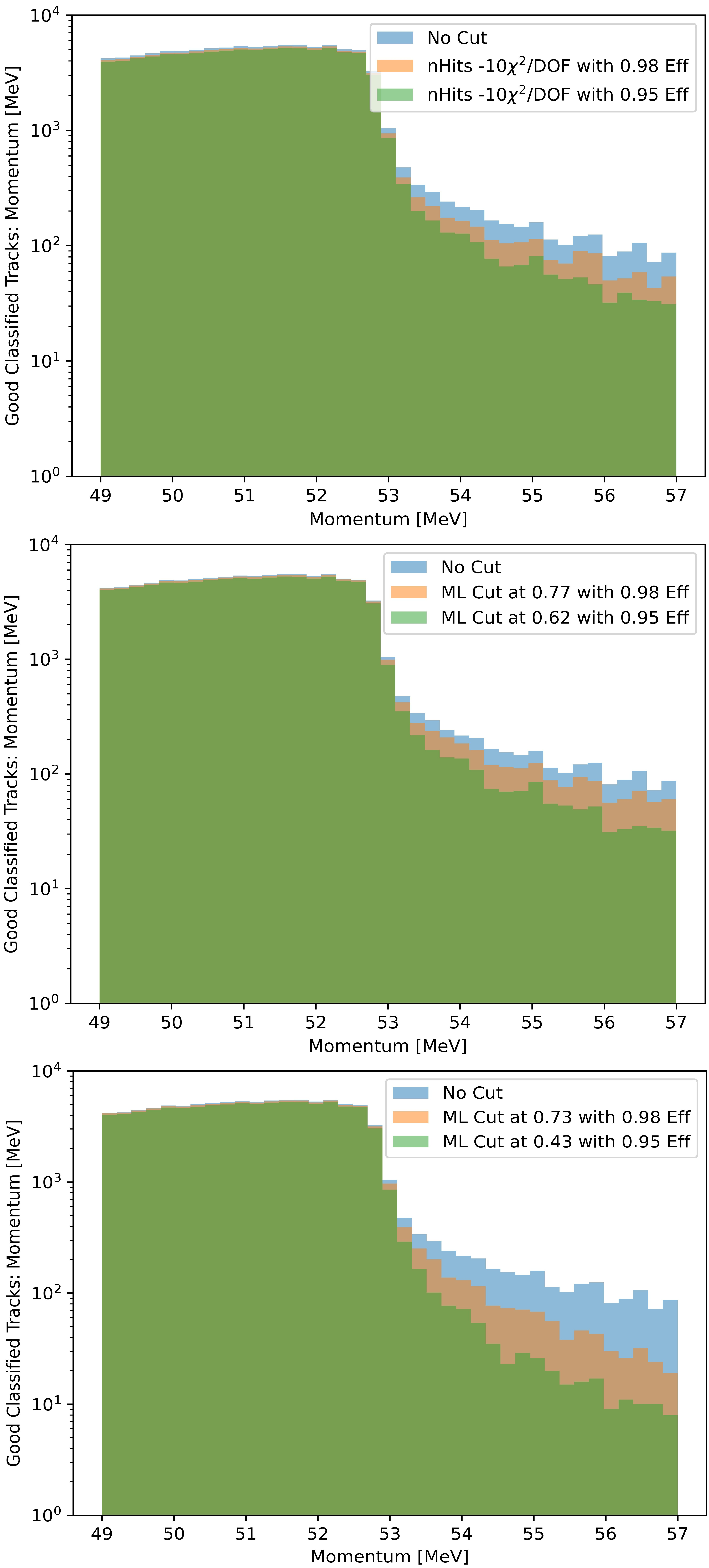}
  \caption{Effects of selection criteria discussed in the text on Michel momentum distribution. See text for details.}
  \label{fig:MomentumML}
\end{figure}

On top of these standard criteria, a multivariate approach is adopted to efficiently apply a stricter cut.  
A neural network with dense layers is formed with the input variables listed in Table~\ref{tab:NNselection} and 
trained on samples for ``good'' and ``bad'' labeled tracks from the data to output $O_\mathrm{NN} \in [0,1]$, a larger value for a higher probability of being a mismeasured track.
We used Michel tracks reconstructed in $49.0 < E_\mathrm{e} < \SI{53.5}{\MeV}$ as the good track sample and those in $53.5 < E_\mathrm{e} < \SI{57.0}{\MeV}$ (a tail region) for the bad one. In the ``bad" track sample, tracks are mismeasured by at least 670 keV whereas the ``good" track sample is dominated by well-measured tracks with a small fraction of mismeasured tracks. In this way, we can train the network without help of MC simulation.
This neural network incorporated information from the Kalman covariance matrix, number of fitted hits, $\chi^{2}/DOF$, information from the comparison with the timing counter (described below), axnd other measurables such as hit efficiency estimates, energy loss, etc. Some measurables were removed as they were correlated with the reconstructed momentum and therefore the neural network output was rejecting tracks based on momentum, not quality. This "bias" towards high momentum was verified to be highly suppressed by measuring the rate of tracks as a function of momentum that were rejected/accepted by the neural network algorithm. We found that including more input variables into the neural network algorithm improved the classification accuracy. In Figure \ref{fig:MomentumML}, the Michel momentum distribution is shown using conventional selection criteria tuned on the Monte Carlo simulation using the track $\chi^{2}/DOF$ and the number of fitted hits, a neural network inputting the same two measurables, and the final neural network selection. The optimal network incorporates 17 input quantities. This selection was verified in the MC (although using the same neural network weights) to improve all kinematic resolutions. The selection required tracks to have a probability of being "bad" of less than $0.1$, resulting in a significant reduction of the tail events and a small improvement in the core energy resolution while keeping the relative efficiency \num{>0.93}.

The identification and grouping of ghost tracks are based on the matched pTC hit cluster.
We refer the reader to \cite{megiidetector} for the pTC detector and its analysis. Basically, a single positron hits multiple counters of pTC (typically 9 counters for a signal positron) in a turn, and
these series of hits are grouped into a hit cluster.
Tracks matched with the same pTC cluster is judged as ghost tracks.
Since the pTC has much faster time response and its resolving power against pileup is much higher than CDCH, 
the probability of loosing a well-reconstructed positron by accidental coincidence to the same pTC cluster is $\lesssim \SI{1}{\percent}$.

\begin{table}[tb]
\caption{Inputs for the track rating neural network.}
\label{tab:NNselection}
\centering
  \begin{minipage}{1\linewidth}
   \renewcommand{\thefootnote}{\alph{footnote})}	
   \renewcommand{\thempfootnote}{\alph{mpfootnote})}	
\centering
\small
\begin{tabular}{@{}l}
\hline
$N_\mathrm{hit}$, $N_\mathrm{hit,first}$, $N_\mathrm{hit,last}$ (last half turn), $N_\mathrm{turn}$, \\
$\chi^2_\mathrm{fit}/N_\mathrm{dof}$, $\mathrm{cov}(E_\mathrm{e},E_\mathrm{e})$, \\
$l_\mathrm{target}$, $l_\mathrm{pTC}$, $\chi^2_\mathrm{CDCH-pTC}$,\\
Energy loss from target to CDCH, \\
 \qquad first turn to second turn, and CDCH to pTC, \\
Ratio of $N_\mathrm{hit}$ to the expected one from the trajectory,\\
Ratio of number of layers with hits\\ 
  \qquad to those passed by the trajectory.\\
\hline
\end{tabular}
\end{minipage}
\end{table}

Tracks of the same positron are ranked according to 
\begin{equation}
R_\mathrm{ghost} = -O_\mathrm{NN} + 2\times N_\mathrm{turn} - B_\mathrm{hit\_det},
\end{equation}
where $N_\mathrm{turn}$ is the number of turns of the track and $B_\mathrm{hit\_det}$ is a binary flag for the algorithm used in the hit detection described in Section~\ref{sec:waveform}. If a track is reconstructed with only the hits detected by the conventional algorithm (Section~\ref{sec:waveform_conventional}), $B_\mathrm{hit\_det} = 0$; if it is also with the hits by the deep-learning algorithm (Section~\ref{sec:waveform_deep}), $B_\mathrm{hit\_det} = 1$.  This term is introduced to preferentially select tracks by the conventional algorithm because the resolutions are usually worse when using the hits by the deep-learning algorithm due to the higher fake rate.
The track with largest $R_\mathrm{ghost}$ is selected for each positron.

\section{Calibrations}
\label{sec:calib}

\subsection{Time and charge calibrations}
\label{sec:qtcalib}

\begin{figure}[htb]
  \centering
  \includegraphics[width=\linewidth]{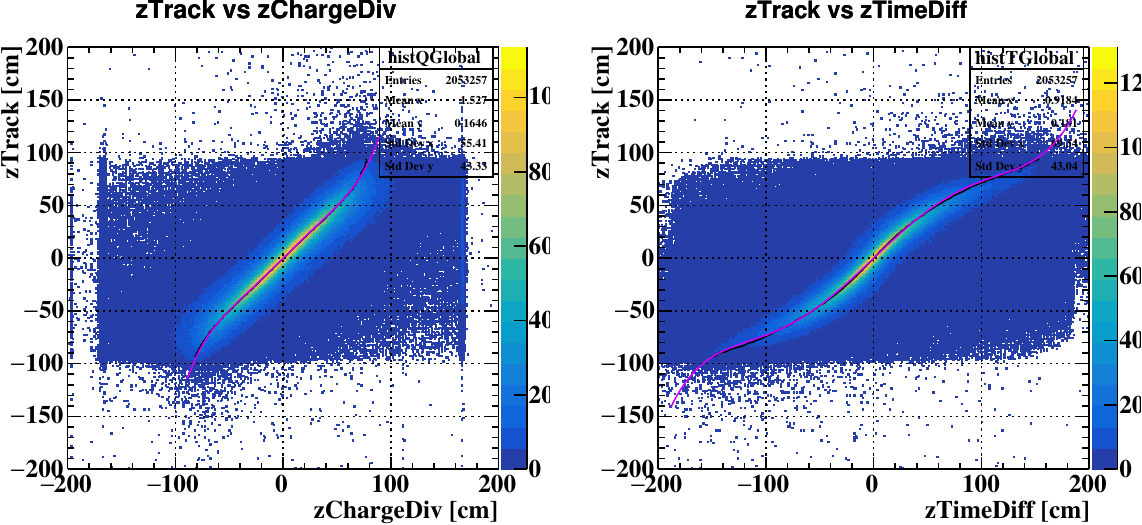}
  \caption{Non linearity correction of the $z$ measurement in CDCH hit reconstrunction.}
  \label{fig:CDCHZCalibGLBCorrection}
\end{figure}

\begin{figure}[htb]
  \centering
  \includegraphics[width=\linewidth]{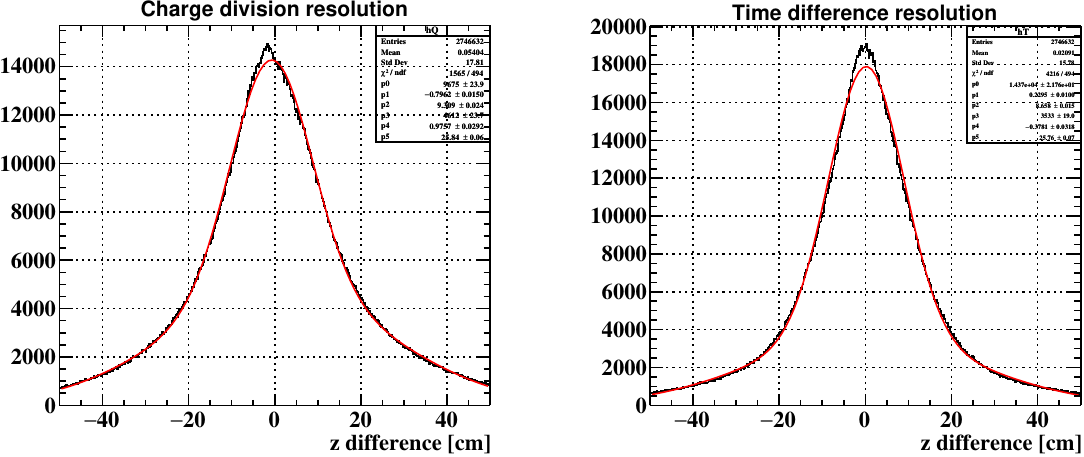}
  \caption{Precision of $z$ measurement in hit reconstruction with positron tracks using charge division (left plot) and time difference method (right plot).}
  \label{fig:CDCHHitReconstructionZResolution}
\end{figure}

The wire end to end charge ratio and time difference ($R_G$ and $t_\mathrm{wireends}$) are calibrated by comparing the reconstructed track $z$ position with the calculation in Eqs.~(\ref{eq:zrawtimes}) and (\ref{eq:zrawcharges}).
In addition, non-linearities of the $z$ reconstruction observed in the data (Fig.~\ref{fig:CDCHZCalibGLBCorrection}) are corrected with a polynomial modeling. 
The nominal signal propagation speed and the wire resistivity are also corrected here so that the slope around $z=0$ becomes one.
The non-linearity calibration uses cosmic ray events that have a high coverage in the end regions, while the $R_G$ and $t_\mathrm{wireends}$ calibration mainly use the Michel positron events.
The $z$ measurement resolution in the hit reconstruction is evaluated with positron tracks to be \SI{9}{cm} for both methods as shown in Fig.~\ref{fig:CDCHHitReconstructionZResolution}.
These two measurements are combined as $z=0.7 z_{w,t} + 0.3 z_{w,q}$, where a larger weight is put on the time difference method because of the larger tail component in the charge division method.
As a result, the combined resolution of $z$ measurement is \SI{7.5}{cm}.

The wire to wire offsets are calibrated by comparing the reconstructed hit time ($t_\mathrm{hit}$) with the expected hit time, which is the sum of the drift time ($t_\mathrm{drift}$) and the referenced track crossing time ($t_\mathrm{reference}$).
The drift time, $t_\mathrm{drift}$, is estimated from the TXY table based on the Garfield++ simulation.
The track crossing time is estimated from the reference time at the matched pTC counter with time-of-flight corrections.
\begin{figure}[htb]
  \centering
  \includegraphics[width=\linewidth]{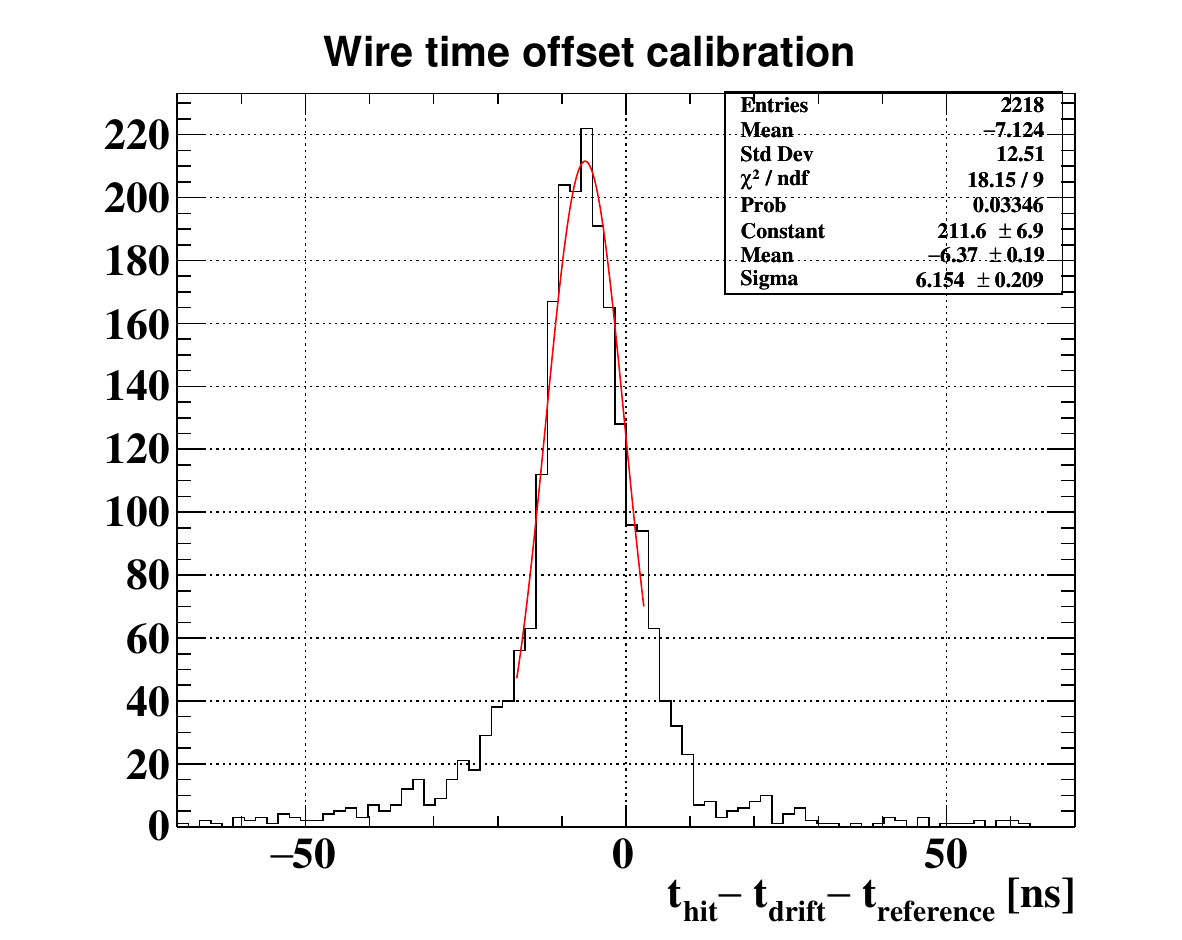}
  \caption{CDCH wire time offset calibration.}
  \label{fig:CDCHWireTimeOffsetCalibration}
\end{figure}

\subsection{Wire alignment and wire sag correction}
\label{sec:align}

The CDCH global position was measured in both \num{2021} and \num{2022} using an optical survey; the  relative wire-by-wire alignment was refined by taking into account multiple measurements taken during the chamber construction and all geometrical information regarding CDCH was incorporated in the general MEG~II database.

\begin{figure}[htb]
\centering 
\includegraphics[width=0.49\textwidth]{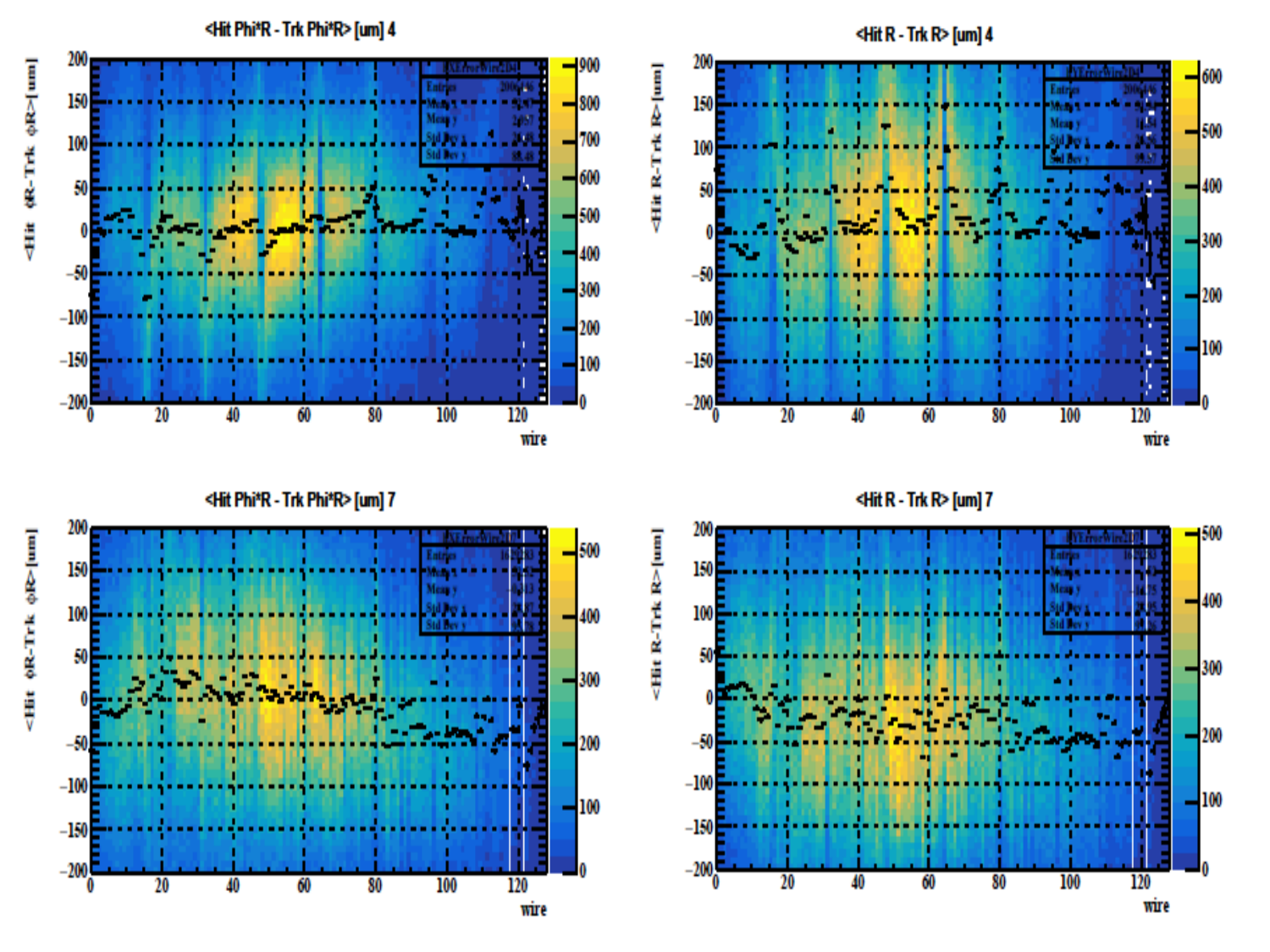}
  \caption{Examples of hit residuals for two representative layers, $4$ and $7$,in the $X$ and $y$ axes of the MEG~II reference frame due to wire-by-wire misalignments.}
 \label{fig:dchxymisalign}
\end{figure}

As observed before, at the end of the tracking stage, the DOCA from the track to each hit wire is determined. Combined with the fitted track angle and the left/right ambiguity solution, we have 2D estimates of the track’s position at the DOCA from the measured DOCA in the cell and from the fitted track (unbiased by the hit itself). Averaged over many tracks, the systematic difference between the hit position estimate and the track position estimate yields the difference between the nominal position of the wire and wire position observed by the fitted tracks. The \lq\lq hit residual\rq\rq~typical size is \SI{\sim 100}{\micro\meter}: an example of hit position residual distribution is shown in Fig.~\ref{fig:dchxymisalign} for two  representative layers ($4$ and $7$) along the $x$ (left) and $y$ (right) axes of MEG~II reference frame. Here, we show the average hit position residual in a cell integrating over the position along the wire axis. A change in the systematic residual along the wire axis corresponds to an angular misalignment or a wire sagitta.

The presence of systematic differences in the hit residual distributions is due to relative wire-by-wire misalignments and is an important degradation factor of the CDCH performances, together with the uncertainty on time-distance relationship, the precision in the knowledge of the magnetic field etc. Such differences must be corrected by appropriate calibration procedures based on particle tracks in experimental data. 

The more abundant tracks present in the CDCH data stream are that of Michel positrons, collected during the normal data taking in a huge sample. Michel positron tracks are the closest possible to a signal track and are reconstructed by the same procedure. Therefore, using Michel tracks for the alignment is a natural choice, but with a couple of disadvantages, coming from being curved tracks: first, a complex tracking algorithm is needed; second, possible uncertainties in the knowledge of magnetic field affect the hit residual distribution and cannot be unfolded. A possible alternative is based on the use of cosmic ray induced events, which have the advantage of producing straight tracks, but the disadvantage of requiring a dedicated data taking, separated from the normal acquisition; moreover, since cosmic ray tracks come from above and not from the target, the coverage of the tracking volume for cosmic ray events is not uniform and different from what expected for particles produced inside the CDCH. We discuss here the official alignment based on Michel tracks, since the cosmic ray alignment is still under development; the latter one will be presented at the end of the paper as a promising calibration tool. However, for both methods we will show present performances and possible improvements 

\subsubsection{Alignment with Michel positron tracks}  
\label{sec:alimich}
\begin{figure}[htb]
\centering 
\includegraphics[width=0.47\textwidth]{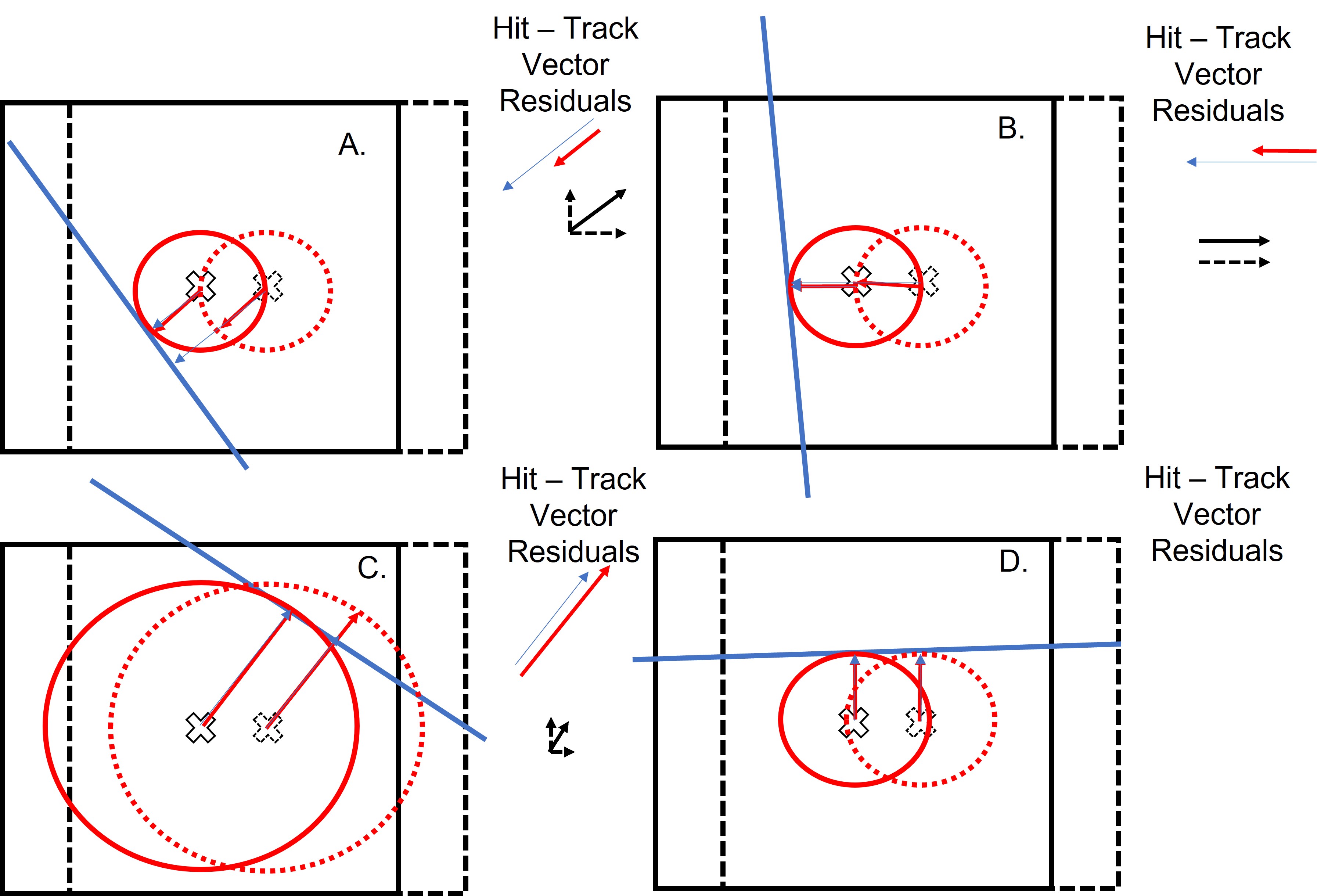}
  \caption{An example of four hypothetical tracks with the same misalignment. It’s clear that in all cases except when the track is parallel to the misalignment (plot on the bottom right), the misalignment creates a mean residual in the direction of the misalignment itself. This mean residual is the handle used to perform the track alignment.}
 \label{fig:dchalignscheme}
\end{figure}

In this section, we describe the procedure used to improve the alignment of the drift chamber by using the Michel positron tracks. This procedures works only for the relative wire-by-wire alignment, i.e. it is insensitive to a global displacement of the whole CDCH with respect to the other subdetectors of MEG~II experiment. First, Fig.~\ref{fig:dchalignscheme} illustrates schematically how the positron tracks can be used to align the CDCH.
In this figure the solid and dotted $X$ represent the true and misaligned wire position respectively and the solid and dotted circles represent the (approximately circular) isochrone curves corresponding to the hit DOCA. The misalignment creates a systematic residual in the direction of the misalignment for all track angles except that parallel with the misalignment direction (case D, plot on the bottom right). Therefore, by iteratively adjusting the wire coordinates on the basis of the mean residual, we iteratively improve the mean residuals and thus the wire alignment. 

\begin{figure}[htb]
\centering 
\includegraphics[width=0.48\textwidth]{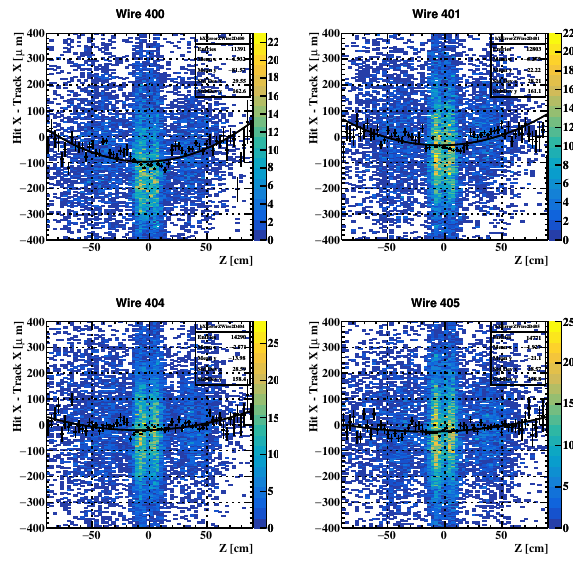}
  \caption{Examples of mean residual on the $x$ axis as a function of wire longitudinal coordinate ($z$) for four wires.}
 \label{fig:Xresid}
\end{figure}

The general approach is to fit the mean residual per wire as a function of the position along the wire axis ($z$). An example of the mean residual on the $x$ axis is shown in Fig.~\ref{fig:Xresid}. 
The residual plots in $x$,$y$ and $z$ global coordinate system are fit using the equation below:
\begin{align}
R \left( z \right) = p_{0} + p_{1} z + 
p_{2} \left[ \left( \frac{2z}{L} \right)^{2} - 1 \right]
\label{eq:dchalign}
\end{align}
The fit parameters $p_{0}$, $p_{1}$, and $p_{2}$ correspond to a translation, rotation, and a wire sagitta respectively; $L$ is the wire length. This is the sagitta equation used in the MEG~II software. The sagitta angle is defined with respect to the plane defined by the wire axis and the radial vector pointing outwards from the CDCH center: if the sagitta lies in this plane, the sagitta angle is zero. The sagitta equation conveniently sets the sagitta to $p_{2}$ at $z=0$ and to $0$ at the wire edges $\left(z = \pm L/2 \right)$. 
The sagitta is calculated in the $x$ and $y$ global coordinate system and converted into the local coordinate system; this has a small and negligible error by not taking into account the global $z$ error due to the sagitta. The sagitta can be due to both electrostatics or gravity and can be as large as \SI{100}{\micro \meter}. 

As already discussed, this method works well if the tracks are not parallel to the misalignment vector (case D of Fig.~\ref{fig:dchalignscheme}). Therefore, in an ideal situation one should use isotropic tracks, coming from all directions with the same probability, but this is clearly not the case since the Michel positron tracks are normally collected using the MEG~II trigger, which is best suited to acquire almost backward going positron-photon pairs. Since the photon detector has a defined position within the volume of the apparatus and a defined angular acceptance, the distribution of the wire crossing angles of the tracks is not uniform, but somewhat biased, depending on the wire position and direction in space. We previously observed that in most cases the positron tracks cross the CDCH active volume three times, corresponding to three distinct segments of tracks. This particular pattern is due to the property of the COBRA magnetic field of sweeping out positrons emitted with small longitudinal momentum and the first two segments are parts of a complete turn within the chamber volume. Each segment is called a \lq\lq half turn\rq\rq~and corresponds, wire by wire, to a rather broad distribution of crossing angles around a central value. The most favourable situation for an efficient alignment takes place when for an individual wire there is a good angular separation (tens of degrees) between the central values of two or three segments, especially when two of the segments are almost perpendicular one to the other.

Some wires are located in regions of the chamber which are crossed by one \lq\lq half turn\rq\rq~only (usually the second one), so that the distribution of the wire crossing angles is limited, making the wire alignment achievable in all angles except that perpendicular to the typical track angle.  However, for the same reason we are not sensitive to this misalignment, the track quality should not be degraded by these types of misalignments. 

The track angular coverage depends on the radial and azimuthal position of the wires; the alignment is better for wires located at the center of the CDCH (small radii) and at azimuth angles far from the borders of the trigger acceptance and worse for wires at large radii or at the borders of the acceptance. 

The MEG~II trigger requires that positron and photon are not only almost backward in angle, but also in time coincidence within a pre-selected window. If this condition is released, one can use segments of tracks which are out-of-time with respect to the photon signal; these segments have a larger angular coverage and using them the alignment sensitivity can be improved in the CDCH regions outside or at the borders of the acceptance for MEG~II triggers. 

Another point to be stressed is that a global alignment fit requires that all types of tracks involved in the alignment share some common wires. Tracks with all hits deeply in the upstream or downstream section of the chamber $\left( \left| z \right| > 30~{\rm cm}\right)$ can't be used to align simultaneously the two chamber ends because of the lack of information connecting them. If the hits of all tracks were within one of the two chamber ends, the ends could be aligned only separately. However, since the distribution of the hits is centered at $z=0$, the relative alignment between upstream and downstream sections can be obtained by the using the central part $\left( \left| z \right| < 30~{\rm cm}\right)$ of the chamber as a common link. 

If this recovery procedure works well for the global upstream/downstream hemispheres, the situation is more complicated for individual wires or sectors. We examined the relative coupling $\left( B|A \right)$ between the wires, i.e. the probability that one track crossing wire $A$ crosses also wire $B$ and we found, as expected, that the coupling is maximum for wires more frequently crossed by the three \lq\lq half turn\rq\rq~tracks and minimum for wires on the extreme CDCH sectors, at the borders of the geometrical and trigger acceptance. Since the first and the last sectors are almost uncoupled, their relative alignment requires a second order procedure by combining information of tracks crossing the first and central sectors with that of tracks crossing the central and the last sectors.

\begin{figure}[htb]
\centering 
\includegraphics[width=0.49\textwidth]{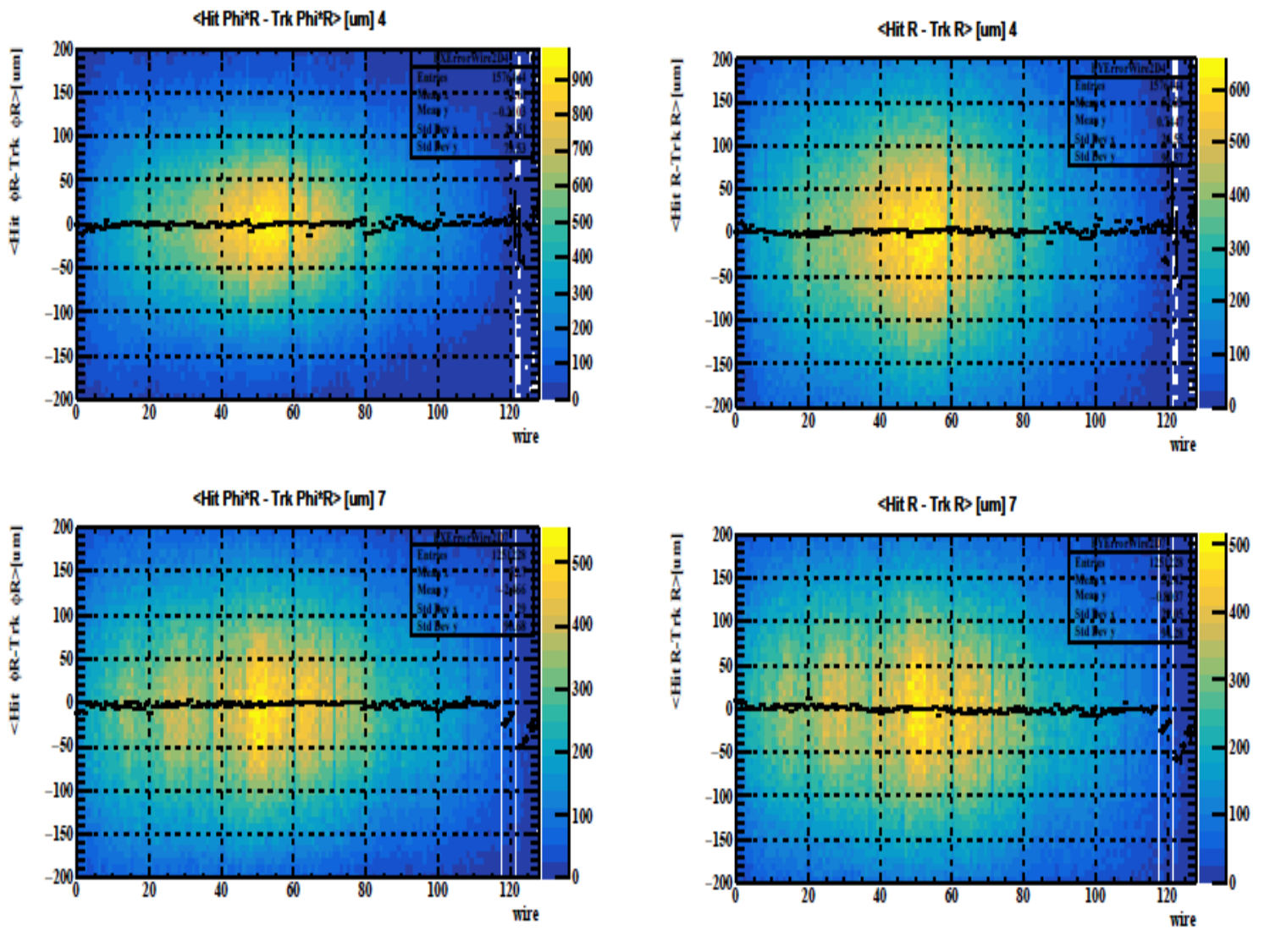}
  \caption{Examples of hit residuals for two representative layers, \num{4} and \num{7}, in the $x$ and $y$ axes of the MEG~II reference frame after \num{14} alignment iterations.}
 \label{fig:dchxyafteralign}
\end{figure}

\begin{figure}[htb]
\centering 
\includegraphics[width=0.48\textwidth]{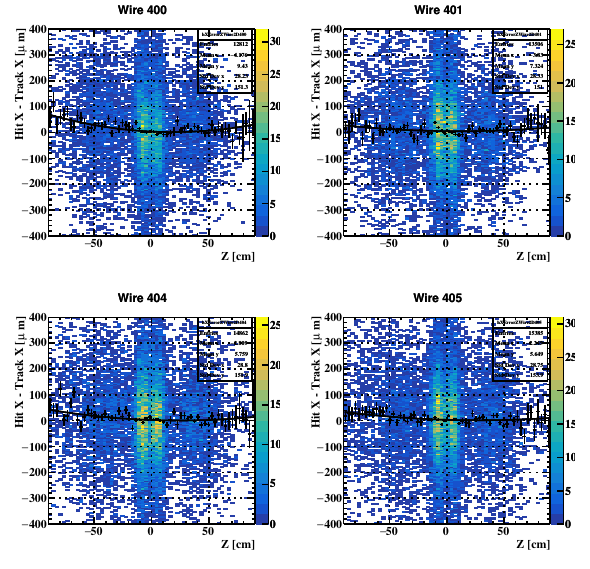}
  \caption{Examples of mean residual on the $x$ axis as a function of wire longitudinal coordinate ($z$) for the same wires of 
  Fig.~\ref{fig:Xresid} after \num{14} alignment iterations.}
\label{fig:Xresidafteralign}
\end{figure}

The alignment procedure used approximately \num{500} runs, each one with \num{2000} events; 
tracks with high number of hits (\num{40} on average per track) were selected to improve the quality of the global fit. The total number of used hits was $\sim 20M$. The alignment quality was verified by comparing the figures of merit (size of residuals and track reconstruction quality) before and after some alignment iterations. To make this comparison more meaningful, a cut was introduced removing wires with less than \num{2000} hits, mainly concentrated on the edges of each layer. Since \num{54} more wires had no hits because of electronic issues, the analysis sample was formed by \num{1000} wires.   
We show in Fig.~\ref{fig:dchxyafteralign}
the hit residuals on the $x$ and $y$ axes of the MEG~II reference frame for the same representative layers of Fig.~\ref{fig:dchxymisalign} after \num{14} alignment iterations.
The improvement with respect to Fig.~\ref{fig:dchxymisalign} is evident; the residuals are highly suppressed. As another check of the alignment quality we show in
Fig.~\ref{fig:Xresidafteralign} the residuals on the $x$ axis as a function of the wire longitudinal coordinate ($z$) for the same wires of Fig.~\ref{fig:Xresid}. 
Again, the improvement is shown by the reduction in size of the residuals. Here, the remaining unaccounted for sagitta is highly suppressed.  

The fit parameters $p_{0}$ and $p_{2}$, corresponding to the wire translation and sagitta, have a strong linear correlation, as expected because it implies a good alignment and centering of the wire at zero error;  
however, sagitta fluctuations cause fluctuations in the translation term too. The remaining sagitta has a typical magnitude of $\sim$ \SI{13}{\micro\meter}, but there are cases of remaining sagitta's reaching $\sim$ \SI{100}{\micro\meter}. These large values are associated with wires with only few thousands of hits; a reduction in sagitta uncertainty is possible by a significant increase in hit statistics, particularly close to the wire edges. Most of these wires are concentrated on the extreme sectors of CDCH, as expected.

\begin{figure}[htb]
\centering 
\includegraphics[width=0.44\textwidth]{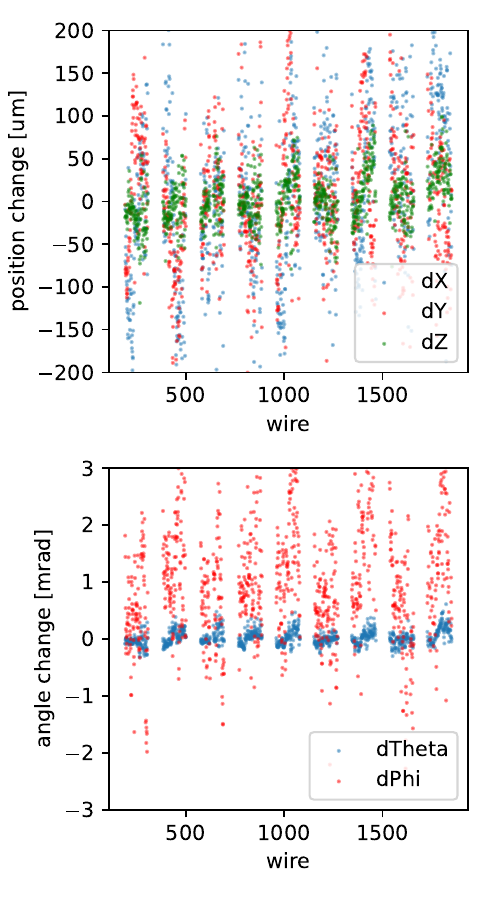}
  \caption{Top: the $x$, $y$ and $z$ components of the global wire displacement as a function of the wire number at the end of alignment procedure. Bottom: the polar angle rotations $\theta$ and $\phi$ as a function of the wire number.}
 \label{fig:p0p1final}
\end{figure}

\begin{figure}[htb]
\centering 
\includegraphics[width=0.44\textwidth]{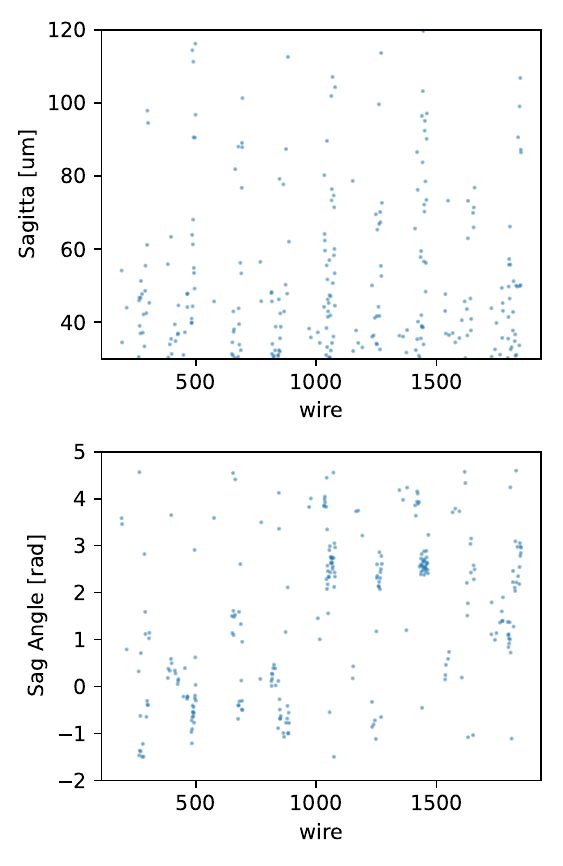}
  \caption{Top: the sagitta amplitude $p_{2}$ as a function of the wire number. Bottom: the sagitta angle as a function of the wire number. Only wires with final sagitta larger than \SI{30}{\micro \meter} are included in these plots.}
 \label{fig:sagittafinal}
\end{figure}

The distributions of the values of the final translation and rotation of the wire as a function of the wire number are shown in Fig.~\ref{fig:p0p1final}.
In the top plot we report the three components of the wire global displacement, in the bottom plot the final rotation angles $\theta$ and $\phi$. All the components of the global translation vector are centered at zero, but there is a clear correlation between such components and the layer number, not yet explained. The spreads of the three components of the global translation amount to tens of microns. 
The final global rotation is $< 0.1$ \unit{\milli\radian} for the $\theta$ angle and $\sim 1.1$ \unit{\milli\radian} for the $\phi$ angle; the corresponding spreads are $\sim 0.1$ \unit{\milli\radian} and $\sim 1.2$ \unit{\milli\radian} for $\theta$ and $\phi$ respectively. The plot on the bottom of Fig.~\ref{fig:p0p1final} shows that the distribution of $\phi$ is biased towards positive values, implying the need of a net global rotation; on the other hand the dependence of the final angular corrections on the wire and layer number is rather weak. The distribution of the sagitta value (which is positively defined) reaches its maximum around \SI{13}{\micro \meter}, with a long right tail, so that the mean is   
\SI{26}{\micro \meter} and the standard deviation is \SI{27}{\micro \meter}. The distribution of the sagitta amplitude $p_{2}$ and angle as a function of wire number are shown in Fig.~\ref{fig:sagittafinal} for sagittas larger than \SI{30}{\micro \meter}.
%
The plot on the bottom shows an interesting but unexpected feature: the sagitta angles are centered at $\sim 0^{\circ}$ in the external layers (wire number $< 1000$) and at $\sim 180^{\circ}$ in the internal layers. This is due to the interplay between the gravitational and the electrostatic sagitta’s. The gravitational sagitta is constant for all layers and always directed along the $y$-axis (the sagitta angle varies from $-90^{\circ}$ to $+90^{\circ}$, including the missing sectors), whereas the electrostatic sagitta, because of the slightly trapezoidal shape of the drift cell, will always point toward the center of CDCH (sagitta angle $=180^{\circ}$) and will be maximum at the inner layer, exceeding the value of the gravitational sagitta, and decreasing with the square of the ratio between high voltage and cell width, to the outer layers, where the gravitational sagitta dominates.

\begin{figure}[p]
\centering 
\includegraphics[width=0.4\textwidth]{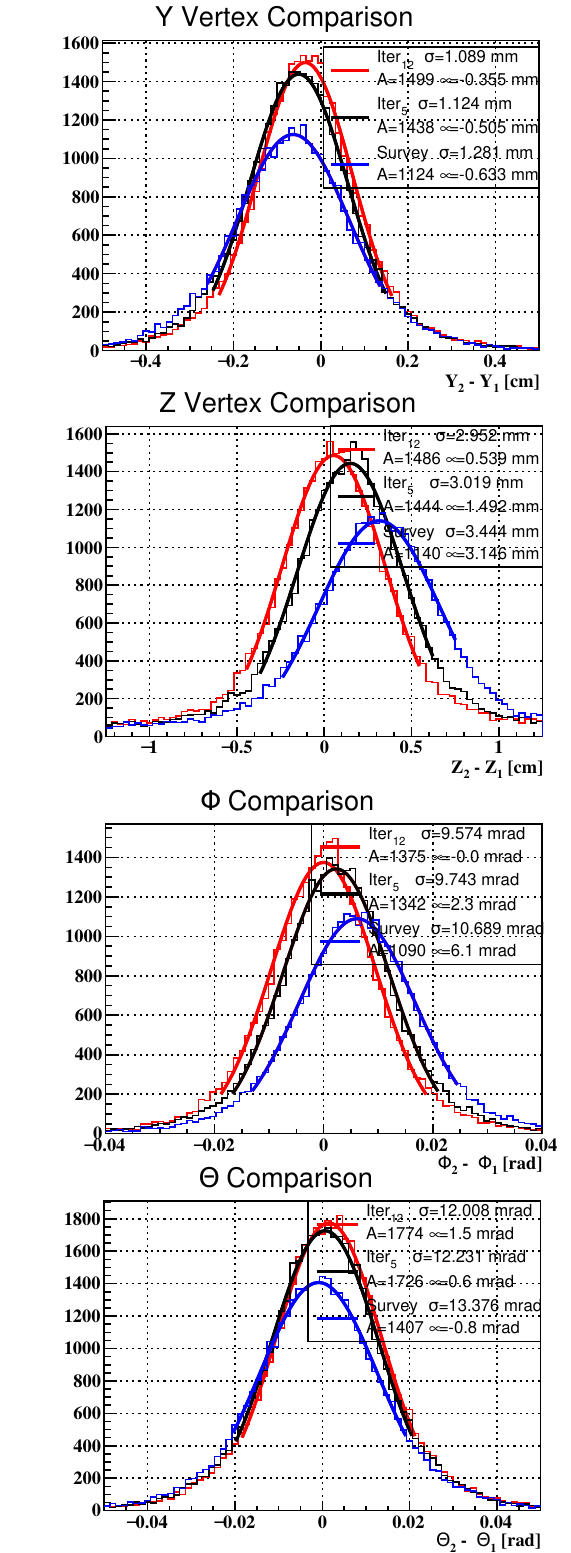}
  \caption{Results of the double turn method to estimate the resolutions on positron vertex coordinates 
  $Y_{\rm e}$ and $Z_{\rm e}$ and on track polar angles $\phi$ and $\theta$ at the target location. The blue, black and red curves correspond to reconstructions based on alignments obtained using the survey and after \num{5} and \num{12} iterations. 
 \label{fig:doubleturnalign}}
\end{figure}

The most important effect of a good wire alignment is an improvement of the position and angular resolutions of the tracking. In order to evaluate this improvement we used the double turn method, which will be briefly explained in section~\ref{sec:performances}. This technique uses tracks crossing the chamber on two complete \lq\lq turns\rq\rq, which can be reconstructed separately by the tracking algorithm providing two independent estimates of tracking parameters. The distributions of the differences between these estimates provide a measurement (after including some corrections which will be explained later on) of the resolutions of the individual parameters. Fig.~\ref{fig:doubleturnalign} shows the results of the double turn analysis for (from top to bottom) $y_{\rm e}$ and $z_{\rm e}$ coordinates of the muon decay vertex (i.e. of positron production point), $\phi$ and $\theta$ angles of positron track at the target location where the positron is produced. 
For each plot three distributions and the corresponding fitting curves are compared: the blue one was obtained using the survey based alignment, the black and the red curves using the iterative alignment after \num{5} and \num{12} steps respectively. Note that the absolute values of the standard deviations are not yet a measurement of the MEG~II resolution on these kinematic variables, but the positive effects of the alignment are shown by significant reductions of the systematic biases and of the widths of all plots. 

\subsection{Global alignment and magnetic field corrections} 
\label{sec:magfield}

The determination of the relative positron-photon angles relies on the relative alignment of the CDCH with respect to the LXe detector and the muon stopping target. Moreover, a possible misalignment of the CDCH with respect to the magnetic field produces non-uniformity in the energy scale as a function of the positron emission angle, due to the gradient of the magnetic field.

The initial global alignment of the CDCH is based on an optical survey, performed by means of a laser tracker and a group of corner cubes. The laser tracker system is placed one after the other on the upstream and the downstream side of the detector and measures the locations of the corner cube reflectors mounted on dedicated positions on the CDCH, the pTC and the muon stopping target. The location of the laser tracker itself is derived from fiducial marks on the concrete walls and in the concrete floor of the experimental area. The overall precision of the survey method can be estimated to the level of a few hundreds of microns. The results of this survey have been used to fix the position of the CDCH with respect to the MEG II reference frame, and shifts are applied to the magnetic field, the target and the LXe detector to recover the correct relative alignments, making the reconstruction of the relative angles independent of the initial assumption.

\begin{figure}[htbp]
  \centering
  \includegraphics[width=\linewidth]{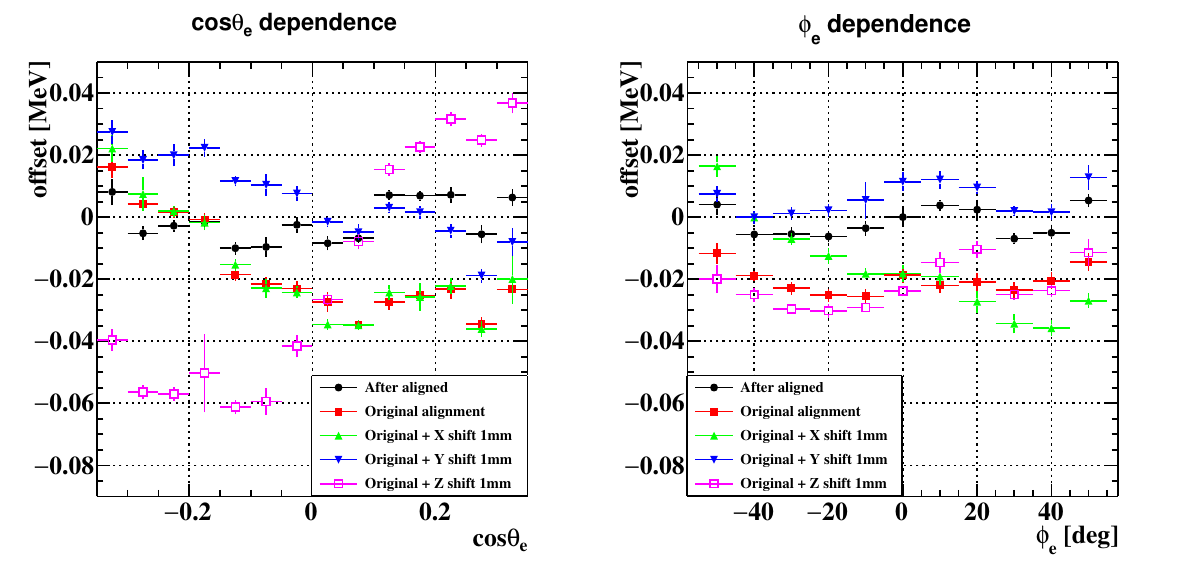}
  \caption{Angle dependence of the energy scale before and after the alignment. The impact of shifting the magnetic field by \SI{1}{mm} is also shown for illustration.}
  \label{fig:BFieldAlignment}
\end{figure}

The relative alignment of the CDCH with respect to the LXe detector and the target are exhaustively described in \cite{megiidetector}. Concerning the misalignment of the CDCH with respect to the magnetic field, the impact is illustrated in Fig.~\ref{fig:BFieldAlignment}, which shows the $\theta_\mathrm{e}$ and $\phi_\mathrm{e}$ dependence of the energy scale evaluated from the Michel edge fitting (described in section~\ref{sec:trackres}) to sliced data samples.
The misalignment in the positive $x$ direction results in a decreasing energy scale with an increasing $\phi_\mathrm{e}$.
The misalignment in the positive $y$ direction results in an energy scale maximized at $|\phi_\mathrm{e}| = 0$. 
The misalignment in the positive $z$ direction results in an increasing energy scale with an increasing $\cos\theta_\mathrm{e}$.

The magnetic field is thus aligned to minimize the observed energy scale dependence on the emission angle.
We start from the magnetic field that is calculated from the COBRA design, which is then shifted by \SI{100}{\micro \meter} in $x$, \SI{700}{\micro \meter} in $y$, and \SI{300}{\micro \meter} in $z$, where the estimated precision in $x$ and $z$ direction is \SI{100}{\micro \meter} and that in $y$ is \SI{200}{\micro \meter}.

Although we do not consider here possible relative rotations between the CDCH and the magnetic field, the good uniformity achieved in the energy scale (one order of magnitude better than the resolution) indicates that there can be only a minor impact on the energy reconstruction.

The overall scale in the magnetic field strength is also calibrated from Michel edge fitting and scaled by 0.9991 with a \SI{0.01}{\percent} precision.

\section{Performances}
\label{sec:performances}
\subsection{Tracking resolutions}
\label{sec:trackres}

\begin{figure}[htb]
\centering
  \includegraphics[width=0.5\textwidth,angle=0] {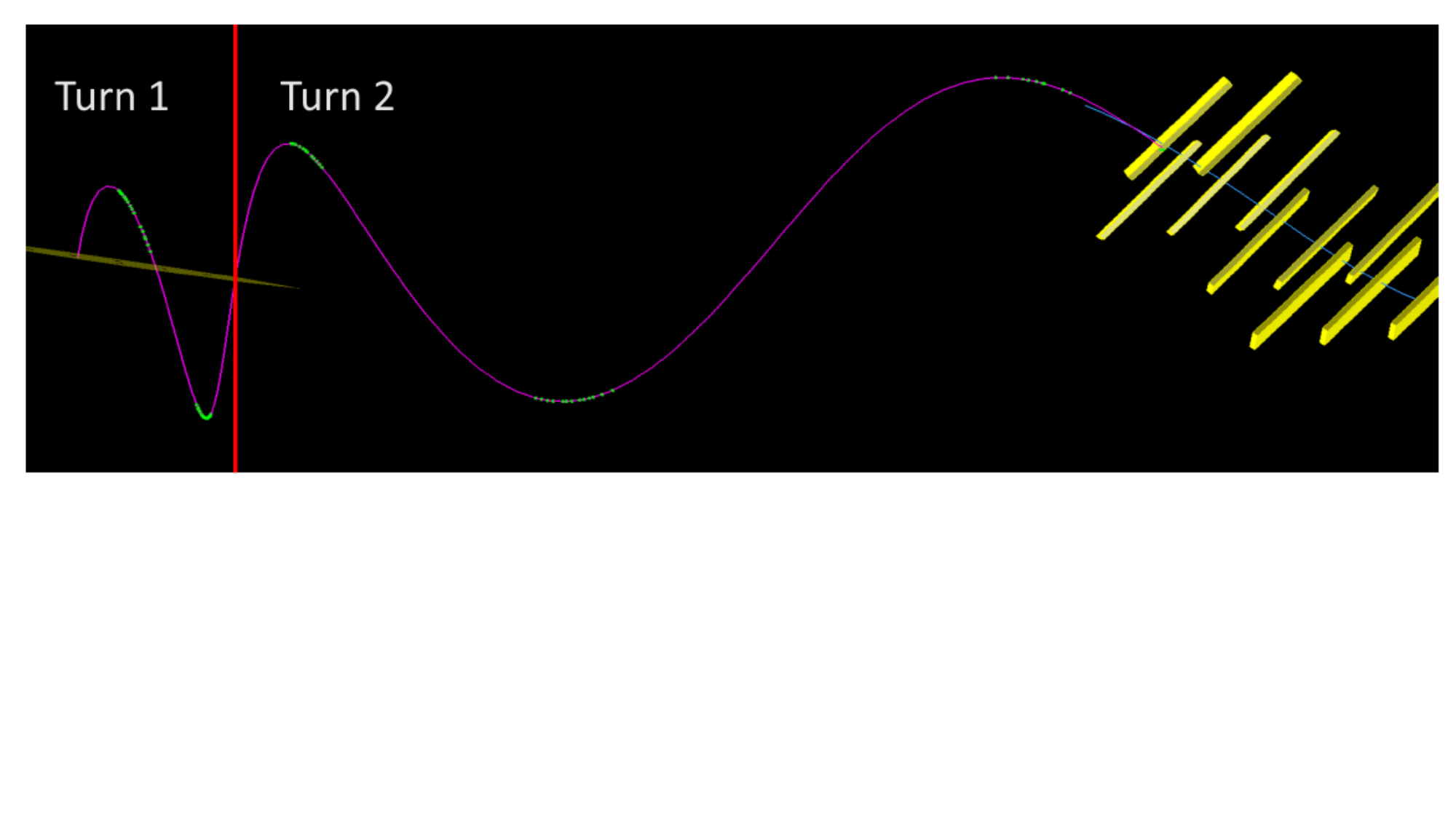}
  \caption{Example of a double turn positron track in the \num{2021} data-set. The green dots represent intersected wires with signal in the drift chamber; the yellow tiles represent the pTC tiles with signal.} 
  \label{fig:doubleturnevent}
\end{figure}

The tracking resolution can be evaluated on Monte Carlo events and on experimental data. The Monte Carlo events have the clear advantage that one can compare the reconstructed values of the kinematic quantities with the generated ones; obviously this method is not applicable to real data, but it can be used as a benchmark if one is able to identify a technique for estimating the resolution valid for both Monte Carlo events and experimental data. This technique is the \lq\lq double turn\rq\rq~method, already used in MEG and discussed in detail in~\cite{Baldini_2016}. The idea is to treat tracks that perform two complete turns within the chamber volume as formed by two independent tracks; an example of this type of tracks is shown in Fig.~\ref{fig:doubleturnevent}.
 
In MEG~II, \SI{\sim 15}{\percent} of positron tracks intersect the chamber volume five times, two times in the first segment of the track and three times in the second, hence crossing \numproduct{9 x 5} layers. The segment formed by the hits closer to the target is called the first turn and the segment formed by the hits closer to one of the chamber ends the second turn.
The two track segments are independently fitted and propagated (one in backward and one in forward direction) to the target plane. The distributions of the differences between the kinematic variables reconstructed by the two segments are then compared and fitted with a convolution of two double Gaussian curves. The choice of the fit curve is based on the expectation that the resolution of each turn is well described by a double Gaussian. 

The fit assumes that the two double Gaussian curves have the same resolution; this is an approximation because the first turn has a lower number of hits and thus a degraded resolution. Therefore, reliable estimates of the resolutions cannot be extracted by simply using the fit results. To account for this difference, we apply the double turn analysis to Monte Carlo Michel positron tracks and build corrections by comparing the core and tail resolutions in the double turn analysis with double Gaussian fit results of the reconstructed kinematics in the Monte Carlo. 

\begin{figure}[htb]
\centering
  \includegraphics[width=0.5\textwidth,angle=0] {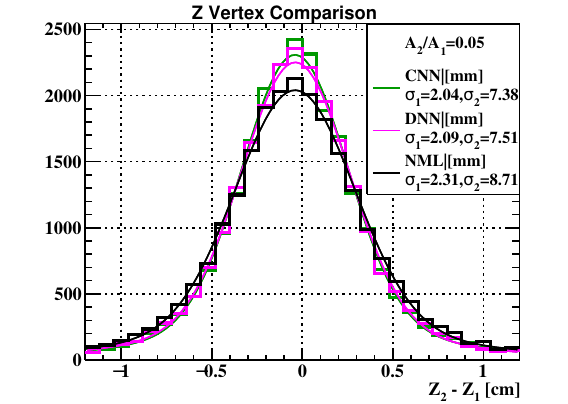}
  \caption{The experimental distribution of the double-turn difference for $z_{\rm e}$ at the positron production point in the target obtained by using the conventional and the neural network approaches for the DOCA reconstruction. A convolution of two equal double Gaussians is fitted to the distribution. 
  }
 \label{dch:doubleturn}
\end{figure}

We show in Fig.~\ref{dch:doubleturn} the double-turn $z_{\rm e}$ distribution for experimental data; the corresponding distributions for $y_{\rm e}$, $\phi_{\rm e}$ and $\theta_{\rm e}$ are similar. The quoted resolutions are the result of the convolution of two double Gaussians. 

The three distributions of Fig.~\ref{dch:doubleturn} are obtained by using DOCA based on the conventional (black lines) and on the neural network based approaches (purple and green lines). The best result is from the convolutional neural network \cite{CNN}. 

In table \ref{tab:perftab1}, we compare the core resolution from the double-turn fits on experimental data and Monte Carlo Michel positrons to the Monte Carlo simulation resolutions for signal and Michel positrons. 
\begin{table}[htb]
    \centering
    \caption{Core Gaussian $\sigma$ for double-turn fits with a convolution of two double Gaussian curves and the core Gaussian $\sigma$ for double Gaussian fits in the Monte Carlo.}
    \begin{tabular}{l c c c c } 
    \hline 
    Data 
    & $y_{\rm e}$  
    & $z_{\rm e}$ 
    & $\phi_{\rm e}$  
    & $\theta_{\rm e} $\\
    & $\left(\unit{\mm}\right)$ 
    & $\left(\unit{\mm}\right)$
    & $\left(\unit{\milli\radian}\right)$
    & $\left(\unit{\milli\radian}\right)$ \\
    \hline
    2021 Data DT & $0.77$ & $2.09$ & $5.82$ & $7.91$ \\ 
    Michel MC,DT & $0.67$ & $1.76$ & $5.27$ & $7.17$ \\ 
    Michel MC    & $0.75$ & $1.84$ & $5.39$ & $6.85$ \\
    Signal MC    & $0.73$ & $1.68$ & $5.21$ & $6.55$ \\

    \hline
    \end{tabular}
    \label{tab:perftab1}    
\end{table}
The resolutions are better for Monte Carlo events, as expected, but the differences are rather small (compare the first and the third lines): \SI{0.02}{\milli \meter} in $y_{\rm e}$, \SI{0.25}{\milli \meter} in $z_{\rm e}$, \SI{0.43}{\milli \radian} in $\phi_{\rm e}$ and \SI{1.06}{\milli \radian} in $\theta_{\rm e}$. Since the tracking algorithm and the alignments can be further optimized, this good similarity demonstrates that our detector is well under control and our knowledge of its capabilities is solid. The comparison between the third and the fourth lines shows that the Monte Carlo resolutions are better for signal positrons than for Michel positrons. This is also expected, since signal tracks have on average a higher number of hits than Michel tracks; so, if one wants to determine the resolution on signal positrons he needs a set of Michel-to-signal correction factors. 

The resolutions on Monte Carlo events can be compared with the double turn results on the corresponding sample of Monte Carlo events; in this way one obtains the correction factor needed to convert the $\sigma$'s of the double-turn distributions, variable by variable, into the corresponding resolutions. We indicate with $\sigma_\mathrm{S/M}$ the (single Gaussian) resolutions determined by fitting the (reconstructed $-$ generated) distributions on Monte Carlo events for signal (S) and Michel (M) positron tracks and with $\sigma_\mathrm{MC/Data,DT}$ that obtained with the double-turn method on Monte Carlo events of Michel positron tracks or on experimental data. The latter values are listed in 
Table~\ref{tab:perftab1} in the first and second lines. The ratio $\sigma_\mathrm{S}/\sigma_\mathrm{M}$ allows to convert a resolution obtained on Michel tracks in the corresponding for signal tracks and the ratio 
$\sigma_\mathrm{M}/\sigma_\mathrm{MC,DT}$ is needed to convert the $\sigma$'s of the double-turn distributions in the resolution of the corresponding variables for Michel positrons. The correction factors are reported in Table~\ref{tab:perftab2}.
\begin{table}[htb]
\centering
   \caption{Correction factors to be inserted to convert single Gaussian $\sigma$'s of double-turn distributions into effective resolutions of kinematic variables for experimental data. We report in the first line the ratios between the single Gaussian $\sigma$'s for signal ($\mathrm{S}$) and Michel ($\mathrm{M}$) positron Monte Carlo events obtained by comparing the reconstructed and the generated values of the kinematic variables; in the second line the ratios between the single Gaussian $\sigma$'s extracted with the (reconstructed-generated) and with the double-turn method on simulated Michel positron events; in the third line, the reduction factors that account for the correlations included in the PDFs.}
    \begin{tabular}{l c c c c } 
    \hline 
    Data & $y_{\rm e}$ & $z_{\rm e}$ 
         & $\phi_{\rm e}$  & $\theta_{\rm e}$\\
         & $\left( \unit{mm} \right)$ 
         & $\left( \unit{mm} \right)$ 
         & $\left( \unit{\milli\radian} \right)$
         & $ \left( \unit{\milli\radian} \right)$
         \\
    \hline
    $\sigma_\mathrm{S}/\sigma_\mathrm{M}$
    & $0.97$ & $0.91$ & $0.97$ & $0.96$ \\
    $\sigma_\mathrm{M}/\sigma_\mathrm{MC,DT}$
    & $1.12$ & $1.05$ & $1.02$ & $0.96$ \\
    $\prod_i \sqrt{1-\rho_i^2}$ & 0.88 & 1.00 & 0.71 & 1.00 \\
    \hline
    \end{tabular}
    \label{tab:perftab2}
\end{table}

The resolutions estimated in this way do not account for the correlations existing among the positron kinematical observables. Indeed, when the target is propagated to the muon stopping plane, the extrapolation of the positron trajectory to a point, that is constraint to lie on a plane, introduces relevant correlations among the measurement errors in energy, angles and position. For signal events, being the true energy known, the correlations can be used to apply a correction to the measured angles and positions. As a consequence, the effective resolution on angles and positions is reduced by a factor $\sqrt{1-\rho_i^2}$ for each correlation (with correlation factor $\rho_i$) that is taken into account.~\footnote{In the likelihood analysis for the search of \megp\, we do not correct explicitly the angles and positions, to not deform the background distributions, but the correlations are accounted for in the construction of the probability density functions, so producing on the analysis power the same effect of a reduction of the resolutions.} 

In summary, the effective resolutions $\sigma_{\mathrm{Data}}$ determining the sensitivity of the experiment can be calculated from the formula:
\begin{align}
\sigma_\mathrm{Data} = \sigma_\mathrm{Data,DT} \times \left( \frac{\sigma_\mathrm{M}}{\sigma_\mathrm{MC,DT}} \right) \times \left( \frac{\sigma_\mathrm{S}}{\sigma_\mathrm{M}} \right) \prod_i \sqrt{1-\rho_i^2}
\label{eq:dchcorrections}
\end{align}
%
The results are summarized in Table~\ref{tab:resolutions}.
\begin{table}[tb]
    \centering
    \caption{Effective resolutions (single Gaussian $\sigma$'s) on the \num{2021} experimental data obtained by combining the double-turn results reported in Table~\ref{tab:perftab1} with the Monte Carlo correction factors listed in Table~\ref{tab:perftab2}. The $\phi$ resolution is for $\phi=0$ and correction for any correlations are not applied.}
    \begin{tabular}{c c c c } 
    \hline 
      $y_{\rm e} \left( \unit{mm} \right)$ & $z_{\rm e} \left( \unit{mm} \right)$ & $\phi_{\rm e} \left( \unit{\milli\radian} \right)$ & $\theta_{\rm e} \left( \unit{\milli\radian} \right)$\\
    \hline
    $0.74$ & $2.0$ & $4.1$ & $7.2$ \\
    \hline
    \end{tabular}
    \label{tab:resolutions} 
\end{table}

The positron momentum resolution is measured by fitting the experimental spectrum of positron momentum for Michel events close to its upper edge. In principle the double-turn method can be used for the positron momentum too (it gives comparable results), but since the value of the upper edge is theoretically known (\SI{52.83}{\mega \eV}), the fit of the positron momentum spectrum provides a direct evaluation of the resolution, without needs of Monte Carlo based correction factors. 

\begin{figure}[htb]
\centering
  \includegraphics[width=0.5\textwidth,angle=0] {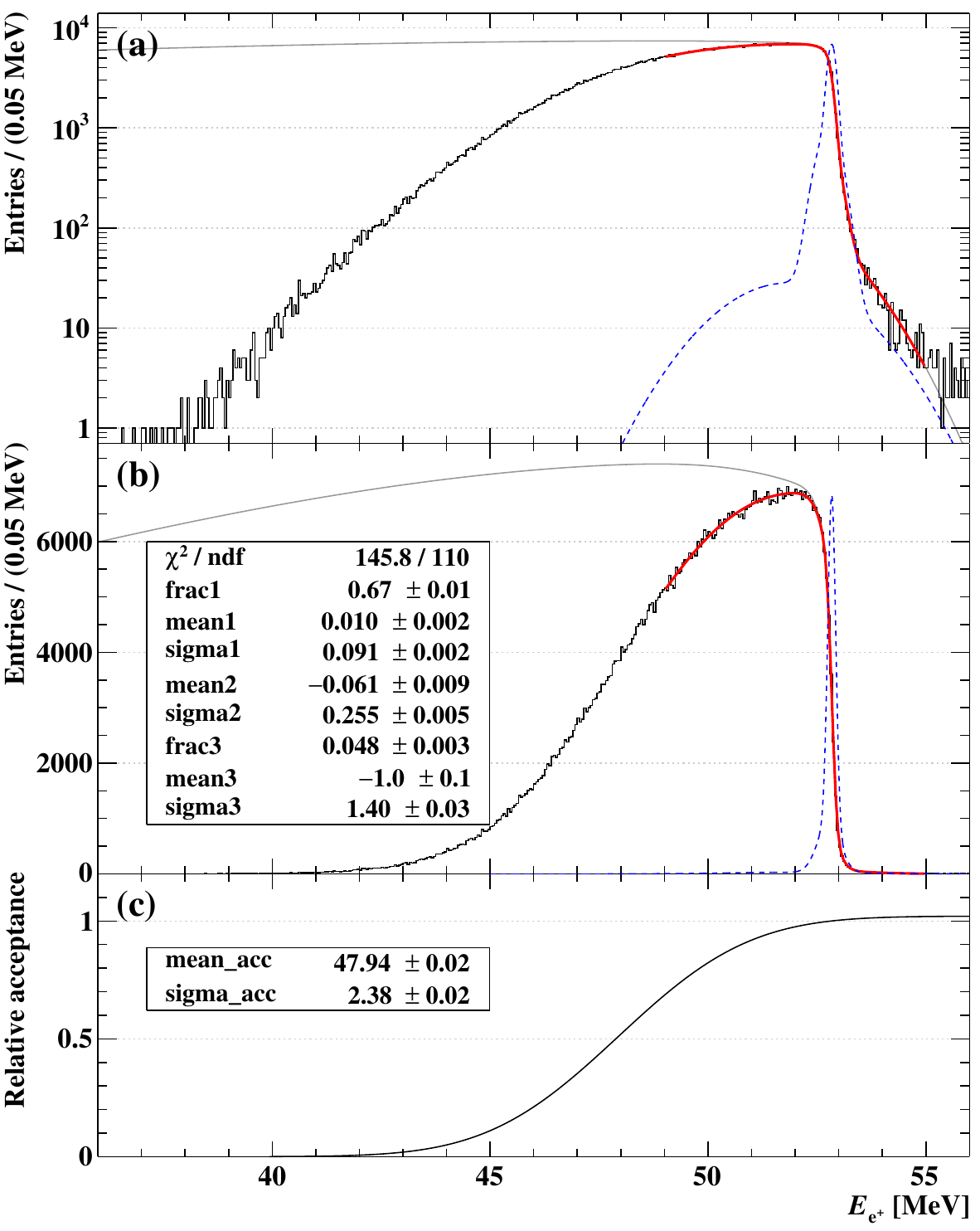}
  \caption{Fit of Michel positron spectrum for a beam intensity of \SI{3e7} stopped muons per second. The top panel shows the spectrum in logarithmic scale, the central panel in linear scale. In both panels the red continuous curve is the global fit and the blue dotted curve is the resolution, modelled with a three Gaussian shape. The bottom panel shows the spectrometer acceptance function, which reaches \num{0.5} around \SI{48}{\mega\eV}}. 
 \label{dch:michel}
\end{figure}

The positron momentum spectrum is the result of three components: 1) the theoretical Michel spectrum, including the radiative corrections \cite{kinoshita_1959}; 2) the spectrometer acceptance, which rejects most of the low energy events ($E < \SI{45}{\mega \eV}$) and selects preferentially the positron tracks emitted in opposite direction to the photon detector; 3) the spectrometer resolution, which is phenomenologically modeled with a triple Gaussian shape. The Gaussian shape which accounts for the largest fraction of the resolution curve integral is called the \lq\lq core\rq\rq~and measures the spectrometer resolution at energies close to that of the signal. The positron momentum spectrum at a beam intensity of \SI{3e7} stopped muons per second is shown in Fig.~\ref{dch:michel}, in logarithmic (top panel) and in linear (central panel) scales.  
In the bottom panel of the same figure we show the spectrometer acceptance function, which reaches \num{0.5} at about \SI{48}{\mega \eV}.
The core Gaussian fraction is $\approx 0.67$ and its $\sigma$ is \SI{91}{\keV}, better by \SI{40}{\keV} of the value quoted in the MEG~II proposal. The corresponding value for MEG was \SI{320}{\keV}; so, the resolution of CDCH is almost a factor \num{4} better than that of the MEG segmented drift chamber. Going to higher beam intensities, up to \SI{5e7} stopped muons per second, the resolution worsens by no more than \SI{10}{\percent}.
\subsection{Efficiency}
\label{sec:dchperformeff}
The CDCH efficiency is defined for \SI{52.8}{MeV} positrons emitted in the opposite of the LXe acceptance and detected by the pTC.
The efficiency is evaluated with data samples with a trigger that requires only at least one pTC trigger.
The total number of efficient positron tracks during the whole data taking are obtained by counting the number of reconstructed positrons with the energy larger than \SI{50}{MeV}.
This is then divided by the expected number of positrons, which is a product of the beam rate, the branching ratio of the Michel decay with the \SI{50}{MeV} cut (0.101), the geometrical acceptance of LXe detector (0.11), the pTC detection efficiency (0.91), and the efficiency for the pTC-only trigger ($\sim0.9$ depending on the beam rate).
The energy dependence of the efficiency is also corrected to match the definition at \SI{52.8}{MeV} ($\sim\SI{9}{\percent}$ effect) according to the result of Michel spectrum fitting.

\begin{figure}[htb]
\centering
  \includegraphics[width=0.5\textwidth,angle=0] {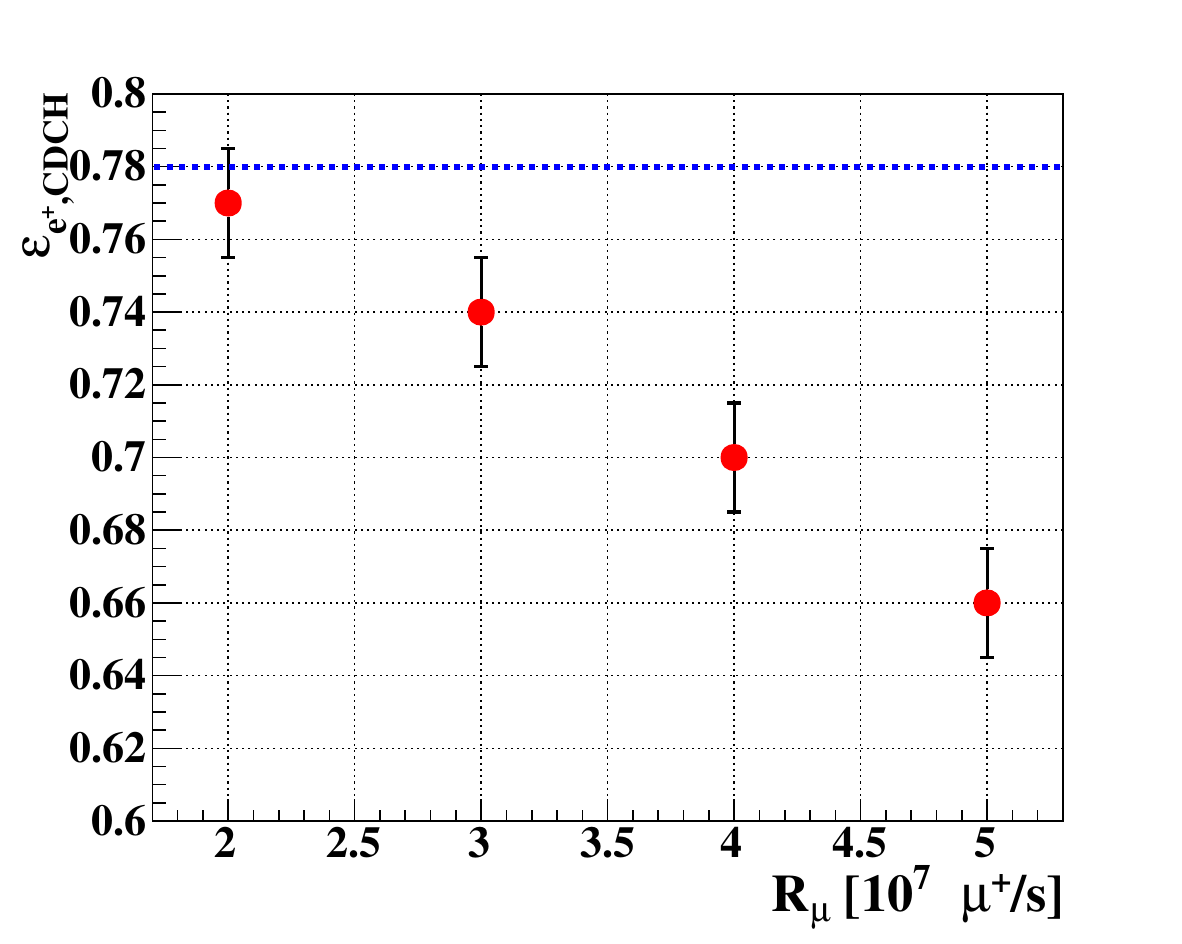}
    \caption{CDCH efficiency as a function of the beam rate intensity $R_{\mu}$.}  
\label{dch:cdcheff} 
\end{figure}

Fig.~\ref{dch:cdcheff} shows the CDCH tracking efficiency as a function of $R_{\mu}$. 
The efficiency decreases from \SI{77}{\percent} to \SI{66}{\percent} when the beam rate increases from $R_{\mu} = \SI{2e7}{\per\second}$ to $R_{\mu} = \SI{5e7}{\per\second}$. This is not surprising, since when the beam intensity increases the probability of accidental superimposition of hits coming from different tracks also increases, making the track finder algorithm less effective in singling out the hits belonging to each individual track. 

Nevertheless, the efficiency reduction from the smallest to the highest beam intensity is not dramatic, \SI{11}{\percent}. The blue dotted line at \SI{78}{\percent} represents the MEG~II design value, which has been almost reached at the smallest beam intensity and is not far from being reached also at higher intensities. We are confident that further improvements in hit reconstruction and tracking algorithms could allow to move closer to the design value also at higher beam intensities.

The beam intensity of the MEG experiment was $R_{\mu} = \SI{3e7}{\per\second}$, with a positron efficiency of \SI{\sim 40}{\percent}. This number includes the pTC detection, which is \SI{\sim 91}{\percent} for MEG~II. So, the positron efficiency for MEG~II is $0.74 \times 0.91 = 0.67$. i.e. \SI{\sim 67}{\percent}. 
The new chamber is more than \num{1.5} times more efficient than the original one, even though the CDCH performances are not yet optimized.

\begin{figure}[htb]
\centering  
  \includegraphics[width=0.5\textwidth,angle=0] {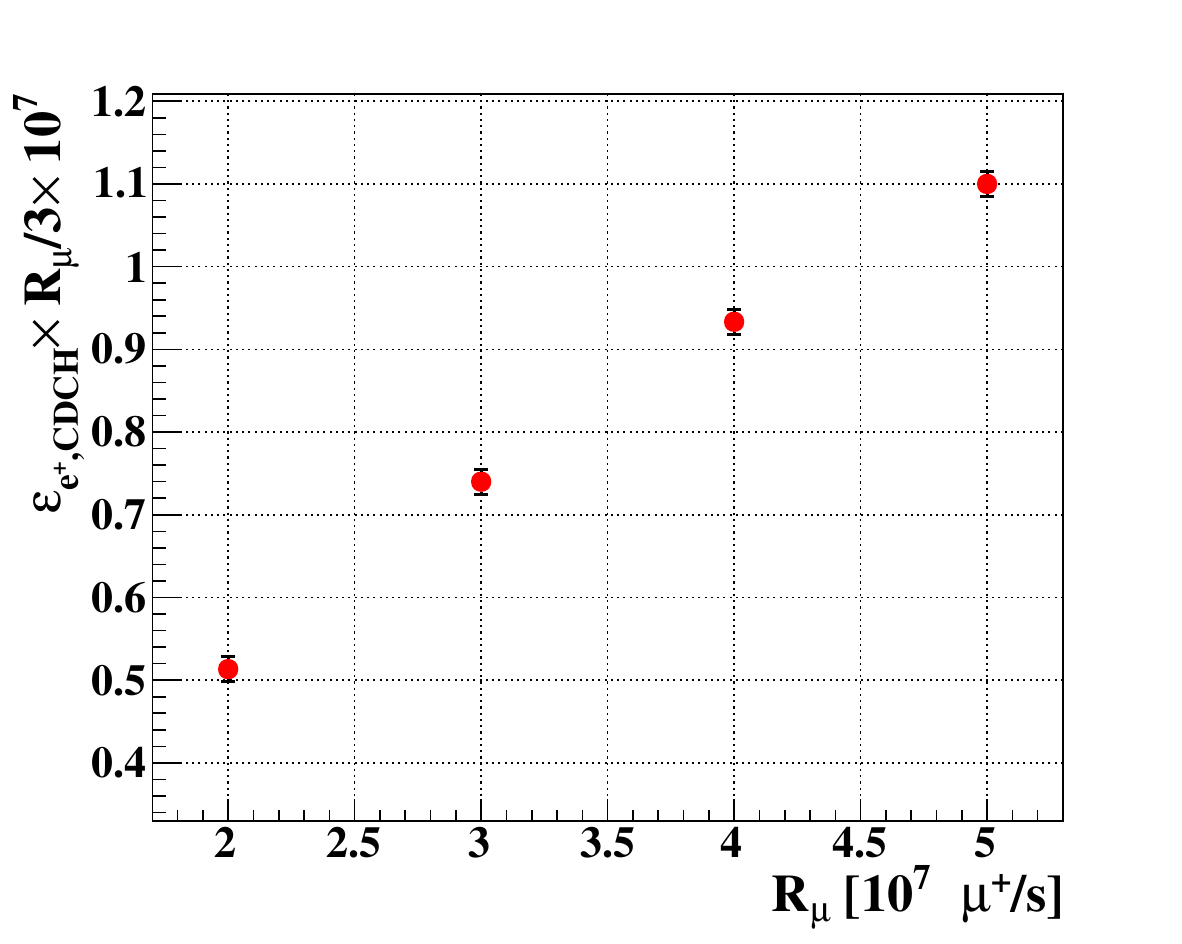}
  \caption{Product between the CDCH efficiency at a given beam rate and the beam rate intensity $R_{\mu}$ divided by \SI{3e7}{\per\second} 
  as a function of the beam rate. The value at $R_{\mu} = \SI{3e7}{\per\second}$ is the efficiency at the same beam intensity used in MEG.}
 \label{dch:cdcheffbeam}
\end{figure}

Moreover, we remind that the gain in positron statistics is determined by the product of two factors, the CDCH efficiency and the beam rate. We show in Fig.~\ref{dch:cdcheffbeam} this product as a function of the beam rate, normalized to the benchmark intensity $R_{\mu} = \SI{3e7}{\per\second}$, the same used in MEG.
This figure clearly shows that higher intensity muon beams are preferable: the small decrease in efficiency is largely compensated by the higher number of stopped muons, so that the number of positron tracks increases by a factor \num{1.5} from $R_{\mu} = \SI{3e7}{\per\second}$ to $R_{\mu} = \SI{5e7}{\per\second}$. Compared with MEG performances, the global gain increases by almost a factor \num{3} at fixed data taking time. A higher beam intensity is the favoured choice for the next years of data taking. 

\section{Cosmic ray tracks}  

During the initial phases of the detector setup, cosmic ray hits and tracks were extensively employed for analysis both with and without the application of magnetic fields since in the assembly phase of the MEG~II detector there were certain limitations, such as the unavailability of the muon beam and the inability to energize the magnets. Consequently, the only viable option for signal checks and partial tracking reconstruction was to expose the detector to cosmic rays without the application of a magnetic field. As a result, numerous events of this nature were recorded, and subsequently, a robust reconstruction method was developed to analyze and interpret these events. This approach proved essential in obtaining valuable insights and calibration for the detector's performance under different conditions.

\begin{figure}[htb]
\centering 
\includegraphics[width=0.48\textwidth]{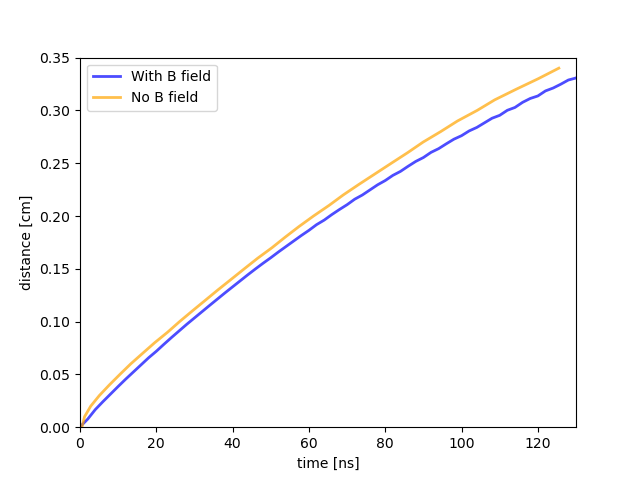}
  \caption{Comparison of distance vs drift time relationships in the case of magnetic field switched on (blue line) or off (orange line).}
 \label{fig:txyBnoBcomp}
\end{figure}

Since the cosmic ray data were collected with the magnetic field switched off, a set of TXY tables without magnetic field was prepared. The differences with the corresponding TXY tables with magnetic field on are not large, but clearly visible, especially close to the borders of the cells. Fig.~\ref{fig:txyBnoBcomp} shows an example of the distance vs drift time average relationships for layer \num{9} and $z = 0$ with (blue line) and without (orange line) magnetic field switched on.    

\begin{figure}[htb]
\centering
  \includegraphics[width=0.5\textwidth,angle=0] {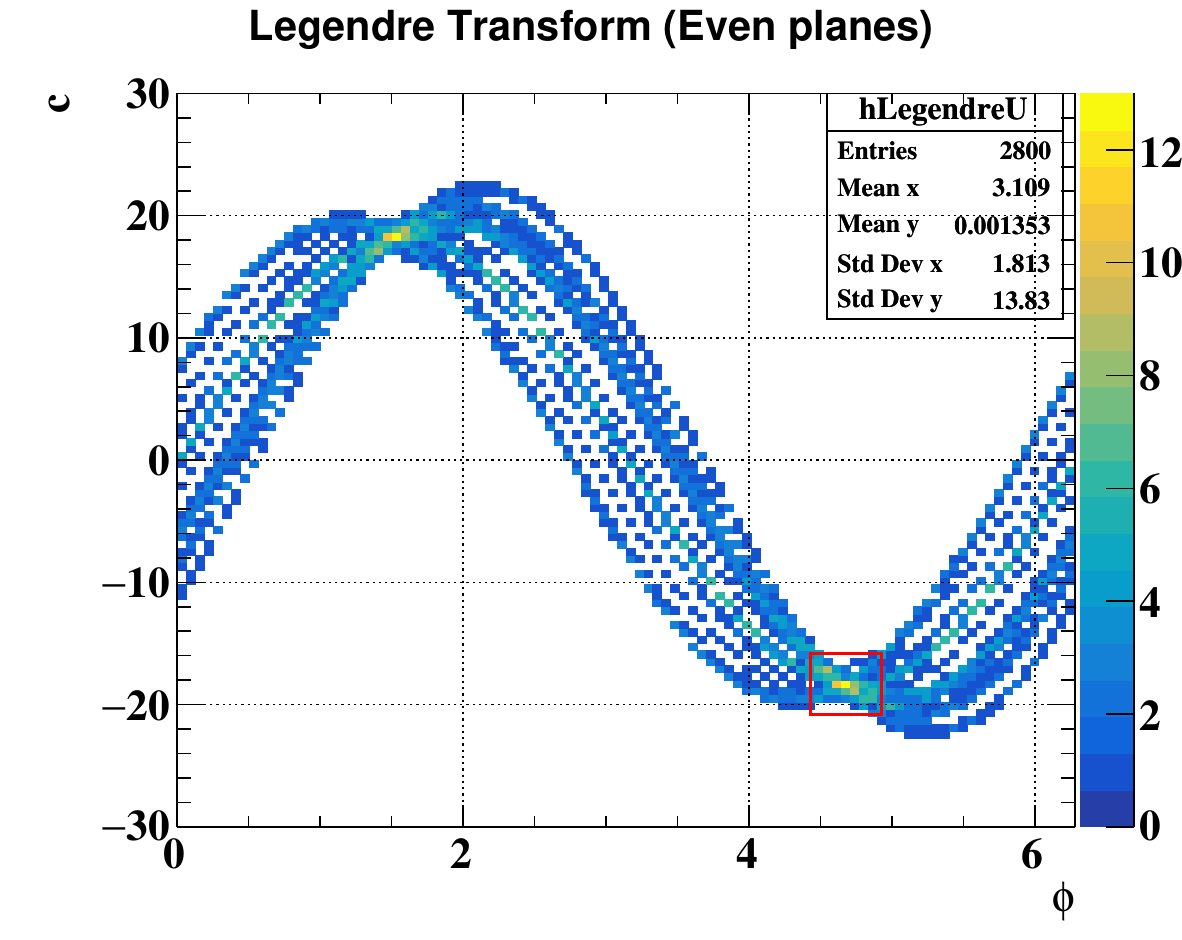}
\caption{Example of identification of the best track by using a fine grid in the Legendre plane $\left( c, \phi \right)$. The best track corresponds to the bin in this plane where the number of intersecting curves is maximum and is highlighted by the red square.}
 \label{dch:legplane}
\end{figure}

The track finding algorithm for cosmic rays' tracks is based on the Legendre transform method \cite{fruhwirth2021pattern}.
In the Legendre transform the tangent lines to a circle of radius $\rho_{i}$ centered in $\left( x_{i}, y_{i} \right)$ (the coordinates of the center of the $i^{th}$ wire) are mapped in the 2-D parameter space which we call the $\left(c, \phi \right)$ plane ($c$ is the line intercept and $\phi$ is the angle between the track and the horizontal axis).
The mapping equation from cartesian coordinates to the parameter space is the following: 
\begin{align}
c \pm \rho_i = x_i \sin{\phi} + y_i \cos{\phi}
\label{eq:dchlegendre}
\end{align}
The radius $\rho_{i}$ corresponds to the DOCA of the $i^{th}$ hit from the track.
Given a set of measured hit points in CDCH, each point defines two groups of curves in the Legendre plane, according to equation~\ref{eq:dchlegendre}; the curves are sampled in a fine grid and the bins where they intersect more densely are searched for; the best track is identified by the bin where the number of intersecting tracks is maximum, as shown in Fig.~\ref{dch:legplane}. 
The Legendre transform is applied to the two stereo views independently and the curves which intersect in the bin of maximum content in the Legendre plane single out the hits belonging to the track.
The algorithm is iterated to find all possible track segment candidates.
The size of the CDCH's drift cells sets the order of magnitude of the bin sizes. The optimal bin size has been chosen
testing the performances on Monte Carlo simulations. We observed that the pattern recognition performances in our case
are stable in a wide range of grid bin sizes.

\begin{figure}[htb]
  \includegraphics[width=0.5\textwidth,angle=0] {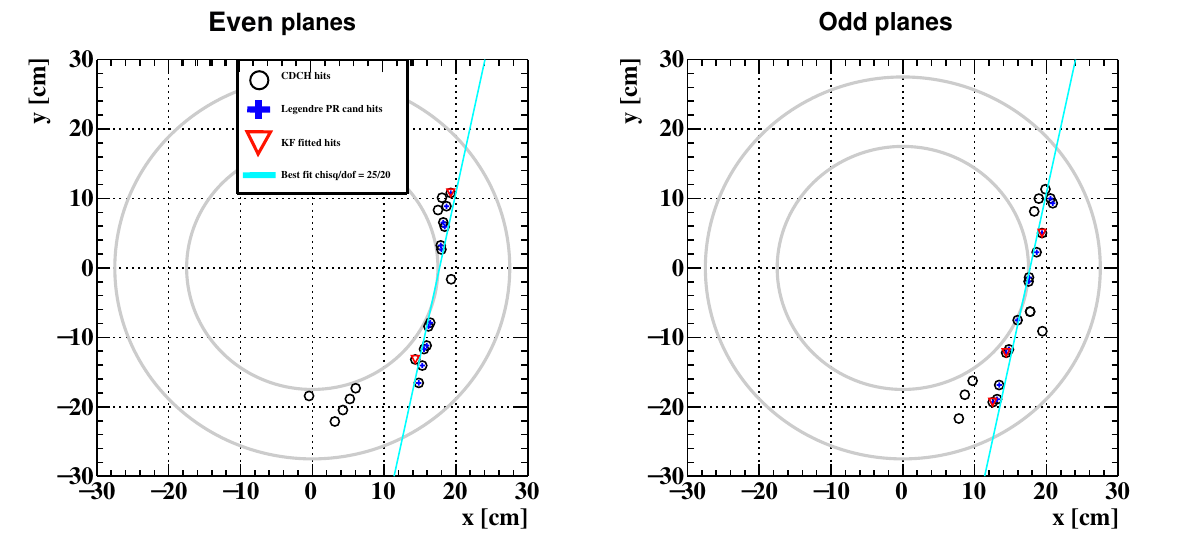}
  \caption{Example of track reconstruction in the two stereo views performed by using the Legendre transform in the $z = 0$ plane. The black circles represent the hit wires, the red triangles the hits selected by the Kalman Filter fitting (but no optimized pattern recognition) and the blue crosses the hits selected by the Legendre transform pattern recognition, as explained in the legend in the left plot.}
 \label{dch:legvskalman}
 \end{figure}

This pattern recognition algorithm has been used in place of a preliminary fitting algorithm which consisted
simply of feeding all CDCH hits to the Kalman filter fitting routine complemented with the DAF hit rejection.
Fig.~\ref{dch:legvskalman} shows an example of comparison of the pattern recognition results in the $z=0$ plane in the two stereo views obtained with the Legendre transform and with the Kalman filter. 
The yellow circles identify the hit wires, the red triangles the hits used by the Kalman filter fit and the blue crosses the hits selected by the Legendre transform pattern recognition. The light blue line is an eye guide which shows the quality of the hit selection.
It looks clear that the Legendre transform has a good efficiency in identifying the hits which really belong to each straight track. This efficiency has been evaluated on Monte Carlo simulations to be above 90\%.
Monte Carlo studies show that the hit sample selected for each track by the Legendre transform pattern recognition is almost free from contamination of hits belonging to a different track (contamination below 1\%).

\subsection{Cosmic rays for detector internal alignment}

In addition to the detector commissioning, straight tracks are also used in physics analysis. Cosmic rays tracks can pass through both the CDCH and the LXe detectors, providing a unique opportunity to examine and refine their relative alignment using real data. It is the only physics signal with these characteristics.

To investigate the alignment between the CDCH and LXe detectors, we employ a two-step process. First, the CDCH reconstructs the cosmic ray's track, allowing us to extrapolate its path to the LXe detector. Then, we compare the measurements obtained from both detectors to assess any deviations or misalignment. By studying cosmic rays and their tracks, we can gain valuable insights into the alignment accuracy of the CDCH and LXe detectors, helping us to improve the precision and reliability of our data analysis.

\subsection{Future developments: alignment with straight tracks and MillePede}

\label{sec:alicr}

Muon tracks offer an excellent alternative sample for assessing alignment procedures and comparing them with the current benchmark, as outlined in Section~\ref{sec:alimich}. A straight track-based alignment method is currently in development, leveraging the well-known MillePede~\cite{millipede} algorithm.

The MillePede algorithm is data-driven and concurrently determines the alignment parameters of the tracking system along with the optimal track parameters. This is achieved by minimizing the $\chi^{2}$ function, given by:

\begin{align}
\chi^2 = \sum_j^{N_{tracks}}\sum_i^{N_{hits}}
\frac{\left( m_{ij} - f(m_{ij}, \Vec{\tau}_j, \Vec{p}_i \right)^2}
{\sigma_{ij}^2}
\label{eq:dchmille}
\end{align}

Here, $m_{ij}$ represents the coordinates of the $i^{th}$ hit in the $j^{th}$ track, $\vec{p}_{i}$ is a vector containing the parameters required to align the $i^{th}$ wire hit, and $\vec{\tau}_{j}$ is a vector containing the local parameters of the $j^{th}$ track. The function $f$ calculates the expected hit coordinates, taking into account the alignment and track parameter vectors. 
The MillePede algorithm is applied on cosmic rays data recorded with the COBRA magnetic field turned off to have an estimate of the alignment parameters independent of the magnetic field calibration and because analytical tracks
allow to determine analytical derivatives used
for the $\chi^2$ minimization.

The $\vec{p}_{i}$ vector is formed, wire by wire, incorporating the wire coordinates at the CDCH center $X_{C}$ and $Y_{C}$, the wire tilt angles $\theta$ and $\phi$, the wire sagitta $s$, and a rotation angle $\gamma$ defining the plane in which the sagitta lies.

In the future, we anticipate comparing the performance of the two alignment methods and, if feasible, combining their results to further enhance the reconstruction performance
(a promising method is described in \cite{Bilka:2021rqj}). By leveraging both cosmic ray and Michel tracks, we can refine the alignment accuracy, leading to more reliable and precise data analysis.

\section{Conclusions} 
\label{sec:conclusions}
In this paper we discussed the performances of the MEG~II cylindrical drift chamber CDCH. This detector was designed to overcome the problems of the old MEG drift chamber, where the segmented structure, the presence of passive materials along the positron trajectories and the formation of localized spatial charge excesses on the cathodic strips caused significant efficiency losses and resolution degradation. The unique volume, the high granularity, the extremely light mechanical structure and the large radiation length of the gas mixture of CDCH, together with its cluster counting capabilities for particle identification, form the basis of an ideal solution for tracking detectors at future $\mathrm{e^+e^-}$ machines, like FCC-ee \cite{FCC-ee}
and CEPC \cite{CEPC:2018}. The measured single hit and tracking resolutions are just a bit worse than the predictions of the Monte Carlo simulations and we are pretty confident that further improvements will be possible since the calibration and alignment procedures are not yet completely optimized. The positron detection efficiency is rapidly approaching the design value and seems only weakly depending on the muon stopping rate. The CDCH is able to sustain high stopping muon rates, up to $\SI{5e7}{\per\second}$, corresponding to hit rates per cell of \SI{3}{\mega\hertz\per\square\cm} at the innermost layers, which ensure a global gain in positron statistics despite an unavoidable (but small) reduction of tracking efficiency. Taking into account the expected performances of the other MEG~II subdetectors we estimated a final sensitivity of our experiment in the search for $\mu^{+} \rightarrow {\rm e^{+}} \gamma$ decay of \num{\approx 6e-14} in a data taking period of about \num{100} weeks.

\section*{Acknowledgments}
We are grateful for the support and co-operation provided  by PSI as the host laboratory and to the technical and engineering staff of our institutes and particularly to G.Balestri, A.Bianucci, M.D'Elia, A.Innocente, A.Miccoli, C.Pinto, G.Petragnani and A.Tazzioli for their invaluable work throughout all the assembling period of CDCH.

This work is supported by DOE DEFG02-91ER40679 (USA); INFN (Italy); H2020 Marie Skłodowska-Curie ITN Grant Agreement 858199;
JSPS KAKENHI numbers JP26000004, 21H04991, 21H00065, 22K21350 and JSPS Core-to-Core Program, A. Advanced Research Networks JPJSCCA20180004 (Japan);
Schweizerischer Nationalfonds (SNF) Grants 206021\_177038,
206021\_157742, 200020\_172706, 200020\_162654 and 200021\_137738 (Switzerland); the Leverhulme Trust, LIP-2021-01 (UK).

\bibliographystyle{elsarticle-my}
\bibliography{MEG}

\begin{thebibliography}{10}
\expandafter\ifx\csname url\endcsname\relax
  \def\url#1{\texttt{#1}}\fi
\expandafter\ifx\csname urlprefix\endcsname\relax\def\urlprefix{}\fi
\expandafter\ifx\csname href\endcsname\relax
  \def\href#1#2{#2} \def\path#1{#1}\fi

\bibitem{baldini_2018}
A.M. Baldini, E.~Baracchini, C.~Bemporad, et~al., {The design of the MEG II experiment}, Eur.\ Phys.\ J.\ C 78~(5).
\newblock \href {http://dx.doi.org/10.1140/epjc/s10052-018-5845-6} {\path{doi:10.1140/epjc/s10052-018-5845-6}}.

\bibitem{megdet}
J.~Adam, X.~Bai, A.M. Baldini, et~al., {The MEG detector for $\mu^+ \to \mathrm{e}^+ \gamma$ decay search}, Eur. Phys. J. C 73~(4) (2013) 2365.
\newblock \href {http://dx.doi.org/10.1140/epjc/s10052-013-2365-2} {\path{doi:10.1140/epjc/s10052-013-2365-2}}.

\bibitem{Baldini_2016}
A.M. Baldini, Y.~Bao, E.~Baracchini, et~al., {Search for the lepton flavour violating decay \megp\ with the full dataset of the MEG experiment}, Eur.\ Phys.\ J.\ C 76~(8) (2016) 434.
\newblock \href {http://dx.doi.org/10.1140/epjc/s10052-016-4271-x} {\path{doi:10.1140/epjc/s10052-016-4271-x}}.

\bibitem{kuno_2001}
Y.~Kuno, Y.~Okada, {Muon decay and physics beyond the standard model}, Rev. Mod. Phys. 73~(1) (2001) 151--202.
\newblock \href {http://dx.doi.org/10.1103/RevModPhys.73.151} {\path{doi:10.1103/RevModPhys.73.151}}, \href {http://arxiv.org/abs/hep-ph/9909265} {\path{arXiv:hep-ph/9909265}}.

\bibitem{megiidetector}
K.~Afanaciev, A.M. Baldini, S.~Ban, et~al., Operation and performance of {MEG II} detector, Eur.\ Phys.\ J.\ C to be submitted.

\bibitem{Baldini:2018ing}
A.M. Baldini, E.~Baracchini, G.~Cavoto, et~al., Gas distribution and monitoring for the drift chamber of the {MEG II} experiment, J.\ Instrum. 13~(06) (2018) P06018.
\newblock \href {http://dx.doi.org/10.1088/1748-0221/13/06/P06018} {\path{doi:10.1088/1748-0221/13/06/P06018}}, \href {http://arxiv.org/abs/1804.08482} {\path{arXiv:1804.08482}}.

\bibitem{Adinolfi:2002}
M.~Adinolfi, F.~Ambrosino, A.~Andryakov, et~al., The tracking detector of the {KLOE} experiment, Nucl.\ Instrum.\ Methods A 488 (2002) ~51--73.
\newblock \href {http://dx.doi.org/10.1016/S0168-9002(02)00514-4} {\path{doi:10.1016/S0168-9002(02)00514-4}}.

\bibitem{Cascella:2014}
M.~Cascella, F.~Grancagnolo, G.~Tassielli, Cluster counting/timing techniques for drift chambers, Nucl. Phys. B (Proc.Suppl.) 248-250 (2014) 127--130.
\newblock \href {http://dx.doi.org/10.1016/j.nuclphysbps.2014.02.025} {\path{doi:10.1016/j.nuclphysbps.2014.02.025}}.

\bibitem{BlumRolandi}
W.~Blum, W.~Riegler, L.~Rolandi, {Particle detection with drift chambers; 2nd ed.}, Springer, Berlin, 2008.
\newblock \url{https://cds.cern.ch/record/1105920}, \href {http://dx.doi.org/10.1007/978-3-540-76684-1} {\path{doi:10.1007/978-3-540-76684-1}}.

\bibitem{Panareo:2020}
M.~Panareo, M.~Chiappini, G.~Chiarello, et~al., The front end electronics for the drift chamber readout in {MEG} experiment upgrade, J.\ Instrum. 15~(C07009).
\newblock \href {http://dx.doi.org/10.1088/1748-0221/15/07/C07009} {\path{doi:10.1088/1748-0221/15/07/C07009}}.

\bibitem{Galli:2019}
L.~Galli, A.M. Baldini, F.~Cei, et~al., {WaveDAQ}: An highly integrated trigger and data acquisition system, Nucl.\ Instrum.\ Methods A 936 (2019) 399--400.
\newblock \href {http://dx.doi.org/10.1016/j.nima.2018.07.067} {\path{doi:10.1016/j.nima.2018.07.067}}.

\bibitem{francesconi2023wavedaq}
M.~Francesconi, A.~Baldini, H.~Benmansour, et~al., The {WaveDAQ} integrated trigger and data acquisition system for the {MEG II} experiment, Nucl.\ Instrum.\ Methods A 1045 (2023) 167542.
\newblock \href {http://dx.doi.org/10.1016/j.nima.2022.167542} {\path{doi:10.1016/j.nima.2022.167542}}.

\bibitem{ritt_2004_nim}
S.~Ritt, The {DRS} chip: Cheap waveform digitizing in the {GHz} range, Nucl.\ Instrum.\ Methods A 518 (2004) 470--471.
\newblock \href {http://dx.doi.org/10.1016/j.nima.2003.11.059} {\path{doi:10.1016/j.nima.2003.11.059}}.

\bibitem{unet}
O.~Ronneberger, P.~Fischer, T.~Brox, {U-Net}: Convolutional networks for biomedical image segmentation, in: Nassir Navab, Joachim Hornegger, WilliamM. Wells, et~al. (Eds.), Medical Image Computing and Computer-Assisted Intervention -- MICCAI 2015, pp. 234--241.
\newblock \href {http://dx.doi.org/10.1007/978-3-319-24574-4\_28} {\path{doi:10.1007/978-3-319-24574-4\_28}}.

\bibitem{garfield++}
H.~Schindler, {Garfield++ user guide (Version 2023.1)}.
\newblock \url{https://garfieldpp.web.cern.ch/documentation/}.

\bibitem{CNN}
D.~Palo, W.~Molzon, Neural network applications to improve drift chamber track position measurement, Nucl.\ Instrum.\ Methods A Forthcoming.

\bibitem{Alme:2010ke}
J.~Alme, Y.~Andres, H.~Appelsh{\"a}user, et~al., {The ALICE TPC, a large 3-dimensional tracking device with fast readout for ultra-high multiplicity events}, Nucl.\ Instrum.\ Methods A 622 (2010) 316--367.
\newblock \href {http://dx.doi.org/10.1016/j.nima.2010.04.042} {\path{doi:10.1016/j.nima.2010.04.042}}, \href {http://arxiv.org/abs/1001.1950} {\path{arXiv:1001.1950}}.

\bibitem{BelleIITrackingGroup:2020hpx}
V.~Bertacchi, T.~Bilka, N.~Braun, et~al., {Track finding at Belle II}, Comput. Phys. Commun. 259 (2021) 107610.
\newblock \href {http://dx.doi.org/10.1016/j.cpc.2020.107610} {\path{doi:10.1016/j.cpc.2020.107610}}, \href {http://arxiv.org/abs/2003.12466} {\path{arXiv:2003.12466}}.

\bibitem{Fruhwirth:1987fm}
R.~Fr{\"u}hwirth, Application of {Kalman} filtering to track and vertex fitting, Nucl.\ Instrum.\ Methods A 262 (1987) 444--450.
\newblock \href {http://dx.doi.org/10.1016/0168-9002(87)90887-4} {\path{doi:10.1016/0168-9002(87)90887-4}}.

\bibitem{genfit1}
C.~Höppner, S.~Neubert, B.~Ketzer, et~al., A novel generic framework for track fitting in complex detector systems, Nucl.\ Instrum.\ Methods A 620~(2) (2010) 518--525.
\newblock \href {http://dx.doi.org/https://doi.org/10.1016/j.nima.2010.03.136} {\path{doi:https://doi.org/10.1016/j.nima.2010.03.136}}.

\bibitem{daf}
R.~Frühwirth, A.~Strandlie, Track fitting with ambiguities and noise: A study of elastic tracking and nonlinear filters, Computer Phys. Communications 120~(2) (1999) 197--214.
\newblock \href {http://dx.doi.org/https://doi.org/10.1016/S0010-4655(99)00231-3} {\path{doi:https://doi.org/10.1016/S0010-4655(99)00231-3}}.

\bibitem{kinoshita_1959}
T.~Kinoshita, A.~Sirlin, Radiative corrections to {F}ermi interactions, Phys. Rev. 113~(6) (1959) 1652--1660.
\newblock \href {http://dx.doi.org/10.1103/PhysRev.113.1652} {\path{doi:10.1103/PhysRev.113.1652}}.

\bibitem{fruhwirth2021pattern}
R.~Fr{\"u}hwirth, A.~Strandlie, Pattern Recognition, Tracking and Vertex Reconstruction in Particle Detectors, Springer Cham, 2021.
\newblock \href {http://dx.doi.org/10.1007/978-3-030-65771-0} {\path{doi:10.1007/978-3-030-65771-0}}.

\bibitem{millipede}
V.~Blobel, Software alignment for tracking detectors, Nucl.\ Instrum.\ Methods A 566 (2006) 5--14, https://www.terascale.de/wiki/millepede\_ii/.
\newblock \href {http://dx.doi.org/10.1016/j.nima.2006.05.157} {\path{doi:10.1016/j.nima.2006.05.157}}.

\bibitem{Bilka:2021rqj}
T.~Bilka, J.~Kandra, C.~Kleinwort, et~al., Simultaneous global and local alignment of the {Belle II} tracking detectors, EPJ Web Conf. 251 (2021) 03028.
\newblock \href {http://dx.doi.org/10.1051/epjconf/202125103028} {\path{doi:10.1051/epjconf/202125103028}}.

\bibitem{FCC-ee}
A.~Abada, M.~Abbrescia, S.S. AbdusSalam, et~al., {FCC}-ee: The lepton collider, Eur.\ Phys.\ J.\ Spec.\ Top. 228 (2019) 261--623.
\newblock \href {http://dx.doi.org/10.1140/epjst/e2019-900045-4} {\path{doi:10.1140/epjst/e2019-900045-4}}.

\bibitem{CEPC:2018}
{The CEPC Study} Group, {CEPC} conceptual design report: Volume 1 -- accelerator (2018).
\newblock \href {http://dx.doi.org/10.48850/arXiv.1809.00285} {\path{doi:10.48850/arXiv.1809.00285}}, \href {http://arxiv.org/abs/1809.00285} {\path{arXiv:1809.00285}}.

\end{thebibliography}

\end{document}